\documentclass[aps,prd,multicol,groupedaddress]{revtex4}
\usepackage{graphicx}

\begin{document}

\title{Search for the possible S=+1 Pentaquark states in 
Quenched Lattice QCD}

\author{Toru T. Takahashi, Takashi Umeda, Tetsuya Onogi and Teiji Kunihiro}

\affiliation{Yukawa Institute for Theoretical Physics, Kyoto University,
Kitashirakawa-Oiwakecho, Sakyo, Kyoto 606-8502, Japan}

\date{\today}

\begin{abstract}
We study spin $\frac12$ hadronic states in
quenched lattice QCD to search for a possible $S=+1$ pentaquark resonance.
Simulations are carried out
on $8^3\times 24$, $10^3\times 24$, $12^3\times 24$
and $16^3\times 24$ lattices at $\beta$=5.7 at the quenched level
with the standard plaquette gauge action and the Wilson quark action.
We adopt a Dirichlet boundary condition in the time direction 
for the quark to circumvent
the possible contaminations due to the (anti)periodic boundary
condition for the quark field, which are peculiar to the pentaquark.
By diagonalizing the $2\times 2$ correlation matrices
constructed from two independent operators with 
the quantum numbers $(I,J)=(0,\frac12)$,
we successfully obtain the energies of the lowest state and 
the 2nd-lowest state in this channel.
The analysis of 
the volume dependence of the energies and spectral weight factors
indicates that a resonance state 
is likely to exist slightly above
the NK threshold in $(I,J^P)=(0,\frac12^-)$ channel.
\end{abstract}

\maketitle

\section{Introduction}\label{Intro}
After the discovery~\cite{Netal03} of $\Theta^+(1540)$
followed by the other 
experiments~\cite{DIANA03,CLAS03,SAPHIR03,ZEUS04,ADK04,HERMES04,COSY04,CLAS04},
identifying the properties of the particle is one of the central problems
in hadron physics.
While the isospin of $\Theta^+$ is likely to be zero~\cite{SAPHIR03},
the spin and the parity and the origin
of its tiny width still remain open questions~\cite{O04,J04}.
In spite of many theoretical studies on 
$\Theta^+$~\cite{DPP97,P03,EMN04,H03,SDO04,O04,J04},
the nature of this exotic particle,
including the very existence of the particle,
is still controversial.
Among theoretical approaches,
the lattice QCD calculation is considered as one of the
most reliable {\it ab initio} methods
for studying the properties of hadronic states,
which has been very successful
in reproducing the non-exotic hadron mass spectra~\cite{CPPACS01}.
Up to now, several lattice QCD studies have been reported,
which aim to look for pentaquarks in various different ways.
However, the conclusions are unfortunately contradictory with each other.
On one hand, the authors in Refs.~\cite{CFKK03,S03,TTTT04} conclude
that the parity of $\Theta^+$ is likely to be negative, 
while in Ref.~\cite{CH04} the state with the similar mass to
$\Theta^+$ in the positive parity channel is reported.
In Refs.~\cite{Metal04,IDIOOS04}, the absence of $\Theta^+$ is suggested.

One of the difficulties in the spectroscopy calculation with 
lattice QCD arises from
the fact that the hadron masses suffer from systematic 
errors due to the discretization, the chiral extrapolation, the 
quenching effect, the finite volume effect and the contaminations 
from higher excited-states. The difficulty specific to the 
present problem is that the signal of $\Theta^+$ is embedded 
in the discrete spectrum of NK scattering states in finite volume. 
In order to verify the existence of a resonance state, one needs 
to isolate the first few low energy states including the lowest 
NK scattering state, identify a resonance state and study its 
volume dependence which can distinguish itself from other scattering 
states. Therefore, ideally one should extract  multistates from a high 
statistics unquenched calculation for several different physical 
volumes, where both the continuum and the chiral limits are taken.
However, due to enormous computational costs, so far there are no 
lattice QCD study  which performs all these steps.  

In the present situation where even the very existence of the 
resonance state is theoretically in dispute, the primary task is 
to provide evidences which distinguish the candidate resonance 
state from a scattering state. 
As long as other systematic errors only affect the numerical 
values of the masses but not the characteristic evidences of 
the resonance state, they may be neglected.
Even so, the isolation 
of the first few low energy states and the study of the volume 
dependence is a minimum requisite. 

Therefore, at this stage as a first step towards a more complete 
analysis, we propose to focus only on analyses using rather heavy 
quarks on coarse quenched lattices
but with a good statistics. By such a strategy, 
we can afford taking 
several different lattice volumes with thousands 
of gauge configurations so that the careful separation of states 
and the studies of volume dependence are possible. Although
giving well controlled continuum and chiral extrapolations may be 
important, 
we simply assume that the contents in spectra would not be drastically 
changed, although there are some cases where level crossings of resonance
states occur as the quark masses decrease~\cite{Metal05}.
Even with such a compromise, we can hopefully learn about the 
existence and much of the qualitative properties of $\Theta^+$.

In this paper, we study $(I,J)=(0,\frac12)$ channel in quenched 
lattice QCD to search for possible resonance states.
We adopt two independent operators with $I=0$ and $J=\frac12$
and diagonalize the $2\times 2$ correlation matrices by the 
variational method for all the combinations of lattice sizes 
and quark masses to extract the 2nd-lowest state slightly 
above the NK threshold in this channel.
After the careful separation of the states, we investigate the 
volume dependence of the energy as well as the spectral 
weight~\cite{Metal04} of each state so that we can distinguish 
the resonance state from simple scattering states.

The paper is organized as follows.
We present the formalism used in the analysis in Sec.~\ref{Formalism}
and show the simulation conditions in Sec.~\ref{Setup}.
The process of the analysis is shown in Sec.~\ref{Data}.
Secs.~\ref{Negative1}-\ref{Positive} are devoted to 
the interpretation of the obtained results
and the verification 
of the existence of a resonance state,
as well as some checks on the consistency and
the reliability of the obtained data.
In Sec.~\ref{Discussion},
we discuss the operator dependence of the results and 
compare our results with the previous works.
We finally summarize the paper in Sec.~\ref{Summary}.
In Appendix, we show the result of another trial
to estimate the volume dependences of the spectral weights,
which requires no multi-exponential fit.

\section{Formalism}\label{Formalism}

As $\Theta^+$ lies above the NK threshold, any hadron correlators, 
which have the $\Theta^+$ signal, also contain the discrete-level
NK scattering states in a finite volume lattice. In order to isolate 
the resonance state from the scattering states, one needs to extract 
{\it at least} two states before anything else.

Since a double-exponential fit of a single correlator 
becomes numerically ambiguous,
we adopt the variational method using 
correlation matrices constructed from independent 
operators~\cite{PM90,LW90,TS04,CH04}.
A set of independent operators 
,\{$O_{\rm snk}^I$\} for sinks and\{$O_{\rm src}^{I\dagger}$\} for sources,
is needed to construct correlation matrices 
${\cal C}^{IJ}(T)\equiv\langle O_{\rm snk}^I(T) 
O_{\rm src}^{J\dagger}(0)\rangle$,
which can be 
decomposed into the sum over the energy eigenstates $|i \rangle$ as
\begin{eqnarray}
{\cal C}_{IJ}(T)=
\langle O_{\rm snk}^I(T) O_{\rm src}^{J\dagger}(0)\rangle
=\sum_{i} \sum_{j} C^\dagger_{{\rm snk}Ii}\Lambda(T)_{ij} C_{{\rm src}jJ},
= (C^\dagger_{\rm snk} \Lambda(T) C_{\rm src})_{IJ},
\end{eqnarray}
with the general matrices which depend on the operators as 
\begin{eqnarray}
C^\dagger_{{\rm snk} Ii}\equiv \langle 0 | O_{\rm snk}^I | i \rangle, 
C_{{\rm src} jI}\equiv \langle j | O_{\rm src}^{J \dagger} | 0 \rangle, 
\end{eqnarray}
and the diagonal matrix
\begin{equation}
\Lambda(T)_{ij}\equiv \delta_{ij} e^{- E_iT}.
\end{equation}
From the product 
\begin{equation}
{\cal C}^{-1}(T+1){\cal C}(T)
=C_{\rm src}^{-1}\Lambda(-1)C_{\rm src},
\end{equation}
we can extract the energies \{$E_i$\} as
the logarithm of eigenvalues \{$e^{E_i}$\} of the matrix 
${\cal C}^{-1}(T+1){\cal C}(T)$.

While there are $N$ independent operators for the 
correlation matrix, the number of the intermediate states $|i \rangle$ 
which effectively contribute to this matrix may differ from $N$ 
in general. Let us call this number as $N_{\rm eff}$.  If $N_{\rm eff}$ is larger 
than $N$, the higher excited-states are non negligible and their 
contaminations  give rise to a $T$-dependence of eigenvalues 
as \{$e^{E_i(T)}$\}. If on the other hand $N_{\rm eff}$ is smaller 
than $N$, ${\cal C}$ becomes non-invertible 
so that the extracted energies become numerically fairly unstable
and we cannot extract all the $N$ eigenvalues.
In order to have a reliable extraction of states, 
we therefore need to find an appropriate window of 
$T$ ($T_{\rm min}\leq T\leq T_{\rm max}$) so that $N_{\rm eff} = N$.
(Of course, even in the case when $N > N_{\rm eff}$,
we can extract $N_{\rm eff}$ eigenvalues
with the reduced $N_{\rm eff}\times N_{\rm eff}$ correlation matrices.)
The stability of \{$e^{E_i(T)}$\} against $T$ is expected in 
this $T$ range and we can obtain $N$ eigenenergies \{$E_i$\} 
($0\leq i\leq N-1$) by fitting the eigenvalues $e^{E_i(T)}$ as 
$E_i=E_i(T)\equiv\ln (e^{E_i(T)})$ in $T_{\rm min}\leq T\leq T_{\rm max}$.
Since finding the stability of the energies against $T$ in noisy data
may suffer from uncontrollable biases,
the result could be quite subjective.
In order to avoid such biases, one should impose some concrete 
criteria to judge the stability as will be explained in later sections 
and select only those data which satisfy the criteria.
After the separation of the states,
we need to distinguish a possible resonance state from NK scattering
states by the volume dependence of each state. It is expected that 
the energies of resonance states have small volume dependence,
while the energies of NK scattering states are expected to scale as 
$\sqrt{M_N^2+|\frac{2\pi}{L}\vec{\bf n}|^2}+
\sqrt{M_K^2+|\frac{2\pi}{L}\vec{\bf n}|^2}$ according to
the relative momentum $\frac{2\pi}{L}\vec{\bf n}$ between N and K on
a finite periodic lattice, 
provided that the NK interaction is weak and 
negligible which is indeed the case for the leading order in chiral 
perturbation theory. 

Although the variational method is powerful for extracting the 
energy spectrum, one can obtain only part of the information on
the spectral weights $C$. In order to extract the spectral weights,
we also perform constrained double exponential fits using the energies 
from the variational method as inputs.

\section{Lattice set up}\label{Setup}

We carry out simulations on four different sizes of lattices,
$8^3\times 24$, $10^3\times 24$, $12^3\times 24$ and $16^3\times 24$
with 2900, 2900, 1950 and 950 gauge configurations using the standard 
plaquette (Wilson) gauge action at $\beta=5.7$ and the Wilson quark action. 
The hopping parameters for the quarks are 
$(\kappa_{u,d},\kappa_{s})$=$(0.1600,0.1650)$, $(0.1625,0.1650)$,
$(0.1650,0.1650)$, $(0.1600,0.1600)$ and $(0.1650, 0.1600)$,
which correspond to the current quark masses 
$(m_{u,d},m_{s})\sim(240,100)$, $(170,100)$,
$(100,100)$, $(240,240)$ and $(100,240)$, respectively in the unit of 
MeV~\cite{BCSVW94}.
The lattice spacing $a$ from the Sommer scale
is set to be 0.17 fm, which implies the physical lattice sizes are
$1.4^3\times 4.0$ fm$^4$, $1.7^3\times 4.0$ fm$^4$, 
$2.0^3\times 4.0$ fm$^4$ and $2.7^3\times 4.0$ fm$^4$.

We adopt the following two operators used in
Ref.~\cite{CFKK03} for the interpolating operators at the sink
$\{O_{\rm snk}^I\}$;
\begin{eqnarray}
\Theta^1(x)\equiv
\varepsilon^{abc}[u_a^{\rm T}(x)C\gamma_5d_b(x)]
\{u_e(x)[{\overline{s_e}(x)\gamma_5d_c(x)}]
-(u\leftrightarrow d)\},
\label{theta1}
\end{eqnarray}
which is expected to have a larger overlap with $\Theta^+$ state,
and 
\begin{eqnarray}
\Theta^2(x)\equiv
\varepsilon^{abc}
[u_a^{\rm T}(x)C\gamma_5d_b(x)]
\{u_c(x)[{\overline{s_e}(x)\gamma_5d_e(x)}]
-(u\leftrightarrow d)\},
\label{theta2}
\end{eqnarray}
which we expect to have larger overlaps with NK scattering states.
Here, the Dirac fields $u(x)$, $d(x)$ and $s(x)$ are up, down and 
strange quark fields, respectively and the Roman alphabets
\{a,b,c,e\} denote color indices.
For measuring the energy spectrum, 
the two operators at the source $\{O_{\rm src}^{I \dagger}\}$ are chosen to be 
$\Theta^1_{\rm wall}(t)$ and 
$\Theta^2_{\rm wall}(t)$
defined using spatially spread quark fields
$\sum_{\vec{\bf x}}q(x)$
with the Coulomb gauge:
\begin{eqnarray}
\Theta^1_{\rm wall}(t)&\equiv&\left(\sqrt{\frac{1}{V}}\right)^5
\sum_{\vec{\bf x_1}\sim \vec{\bf x_5}}
\varepsilon^{abc} 
[u_a^{\rm T}(x_1)C\gamma_5
d_b(x_2)]\{u_e(x_3)[{\overline{s_e}(x_4)
\gamma_5 d_c(x_5)}]
-(u\leftrightarrow d)\},
\end{eqnarray}
and
\begin{eqnarray}
\Theta^2_{\rm wall}(t)\equiv\left(\sqrt{\frac{1}{V}}\right)^5
\sum_{\vec{\bf x_1}\sim \vec{\bf x_5}}
\varepsilon^{abc} 
[u_a^{\rm T}(x_1)C\gamma_5
d_b(x_2)]\{u_c(x_3)[{\overline{s_e}(x_4)
\gamma_5 d_e(x_5)}]
-(u\leftrightarrow d)\}.
\end{eqnarray}
The above operators give a $2\times 2$ correlation matrix in 
the channel with the quantum number of $(I,J)=(0,\frac12)$.
We note here that the baryonic correlators have the spinor indices,
which we omit in the paper, and they contain the propagations of both 
the positive and negative parity particles. For the parity projection,
we simply multiply the correlators by $\frac12(1\pm\gamma_0)$
and extract the contributions proportional to $(1\pm\gamma_0)$
from the negative-parity and positive-parity particles, respectively.

We fix the source operator $\overline{\Theta}_{\rm wall}(t)$
on $t=t_{\rm src}\equiv 6$ plane to reduce the effect of the Dirichlet boundary
on $t=0$ plane~\cite{CPPACS04,CPPACS95}. We adopt the operators
$\sum_{\vec{\bf x}}\Theta^I(\vec{\bf x},t)$
as sink operators, which is summed over all space to project out 
the zero-momentum states.
We finally calculate 
\begin{equation}
{\cal C}^{IJ}(T)=
\sum_{\vec{\bf x}} \langle
\Theta^I(\vec{\bf x},T+t_{\rm src}) 
\overline{\Theta^J}_{\rm wall}(t_{\rm src})\rangle.
\label{correlation_matrix}
\end{equation}

Using two independent operators,
we can extract the first {\it two} states, 
namely the lowest and the next-lowest states.
The lowest state is considered to be the ``lowest'' NK scattering state.
In order to extract a possible resonance state with controlled
systematic errors, we need to choose the physical volume of the lattice 
in an appropriate range. If we choose $L$ to be too large, the resonance
state becomes heavier than the 2nd-lowest NK scattering state whose energy is naively 
expected to scale as 
$\sqrt{M_N^2+(\frac{2\pi}{L})^2}+\sqrt{M_K^2+(\frac{2\pi}{L})^2}$ 
according to the spatial lattice extent $L$. In this case we need to 
extract the 3rd state using a $3\times 3$ correlation matrix, 
which requires more computational time. The energy difference between
the lowest and the next-lowest NK scattering states 
$\sqrt{M_N^2+(\frac{2\pi}{L})^2}+\sqrt{M_K^2+(\frac{2\pi}{L})^2}-
(M_N+M_K)$, for example, ranges from 180 MeV to 860 MeV in 
$1.4 {\rm fm}\leq L\leq 3.5 {\rm fm}$.
Taking into account that $\Theta^+$ lies about 100 MeV
above the NK threshold, we take 3.5 fm as the upper limit of $L$.
On the other hand, if we choose $L$ too small, unwanted
finite-volume artifacts from the finite sizes of particles 
become non-negligible. It is however difficult to estimate the 
lower limit of $L$, because the finite-volume effect is rather 
uncontrollable. We shall take the spatial extents $L=8,10,12,16$ at 
$\beta =5.7$ as a trial.

We take periodic boundary conditions in all directions for the gauge field,
whereas we impose periodic boundary conditions on the spatial directions
and the Dirichlet boundary condition on the temporal direction 
for the quark field
in order to avoid possible contaminations from those propagating
beyond the boundary at $t=0$ in (anti)periodic boundary conditions. 
Since the source of possible contaminations is peculiar to the pentaquark
and has not been properly noticed in previous studies, 
it is worthwhile to 
dwell on this problem for a moment. 

\begin{figure}[h]
\includegraphics[scale=0.4]{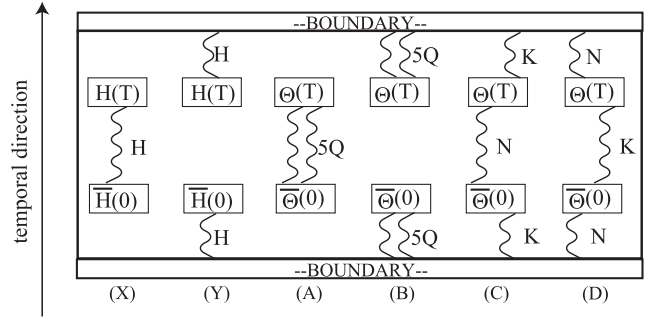}
\caption{
\label{hadronic_corr}
Schematic figure for the explanation on the possible contaminations
of the particles propagating over the temporal boundary.
$H(t)$ and $\Theta(t)$ are the interpolating operators
and their arguments are the distances from the source point.
$H(t)$ is a generic hadronic operator, which
creates and annihilates the particles which cannot decay in quenched QCD.
The wave lines represent the propagations of states.
Resonance states like $\Theta^+$ are represented 
by ``5Q'' in the figure, as well as NK scattering states.
{\it 
The five-quark state can dissociate into 
forward-propagating nucleon (Kaon) 
and backward-propagating Kaon (nucleon).
}
}
\end{figure}

Let us denote the correlators in the pentaquark channel with
periodic/antiperiodic boundary conditions and Dirichlet boundary
conditions as ${\cal C}_{P/AP}(T)$ and ${\cal C}_{D}(T)$, respectively.
Inserting complete set of states these correlators read
\begin{eqnarray}
{\cal C}_{P/AP}(T) 
&=& \sum_{m,n} (-)^{\epsilon_n} 
\langle n | \Theta | m \rangle \langle m | \overline{\Theta} |n\rangle
e^{- E_m T - E_n (N_t -T)}, 
\nonumber\\
{\cal C}_{D}(T) 
&=& \sum_{m,n_f,n_i} 
\langle D | n_f \rangle \langle n_f | 
\Theta | m \rangle \langle m |\overline{\Theta} 
| n_i \rangle \langle n_i |  D \rangle 
e^{ - E_m T - E_{n_f} (N_t -(T+t_{\rm src})) - E_{n_i} t_{\rm src}},
\end{eqnarray}
where the states 
$|m\rangle,\ |n\rangle,\ |n_i\rangle,\ |n_f\rangle$ 
are the eigenstates with energies
$E_{m}$, $E_{n}$, $E_{n_i}$, $E_{n_f}$ 
respectively. $ | D \rangle$ is the state which 
corresponds to the Dirichlet boundary condition and $(-)^{\epsilon_n}$ 
is the factor which represents the $\pm$ sign with antiperiodic boundary
condition. The factor $(-)^{\epsilon_n}$ is equal to $1(-1)$
when $|n\rangle$ contains an even(odd) number of valence quarks.
It should be noted that the up, down, and strange quark 
numbers $U,D,S$ for the state $|D\rangle$  are restricted to zero since all 
the quark fields $\psi({\bf x},0)$ are set to zero in Dirichlet
boundary condition, while there is no restriction in the quark sector for 
the states $|n\rangle$ which appear in periodic/antiperiodic boundary 
conditions. This means that the following states can contribute 
in the correlators 
\begin{eqnarray}
|n\rangle = & \mbox{any states including  } 
|0\rangle, |\bar{K}\rangle, |\bar N\rangle , |\bar N\bar{K}\rangle
\cdots & \mbox{for periodic/antiperiodic b.c.}, 
\nonumber\\
|n_{i,f} \rangle  = & \mbox{any states(with $U=D=S=0$) including  } |0\rangle
\cdots   &\mbox{for Dirichlet b.c.},
\end{eqnarray}
where $|\bar{K}\rangle$ and $|\bar N\rangle$ are the antiparticle states of 
$|K\rangle$ and $|N\rangle$, respectively.
Therefore, the correlators have the contributions in the long
range limit,
\begin{eqnarray}
{\cal C}_{P/AP}(T) 
&\sim & \langle 0 | \Theta | NK \rangle 
\langle  NK | \overline{\Theta} |0\rangle
e^{- E_{NK} T} 
+ \langle 0 | \Theta | \Theta \rangle 
\langle \Theta | \overline{\Theta} |0\rangle
e^{- E_{\Theta} T} 
+\langle \bar{K} | \Theta | N \rangle 
\langle  N | \overline{\Theta} |\bar{K} \rangle
e^{- E_{N} T - E_{\bar{K}}(N_t-T)}
\nonumber\\
&  &  \pm \langle \bar N | \Theta | K \rangle 
\langle  K | \overline{\Theta} | \bar N \rangle
e^{- E_{K} T - E_{\bar N}(N_t-T)}
\pm \langle \bar N \bar{K}| \Theta | 0 \rangle 
\langle  0 | \overline{\Theta} | \bar N \bar{K}\rangle
e^{ -E_{\bar N \bar K}(N_t-T)}
\nonumber\\
&  &  + \mbox{ and excited states }, 
\label{CPAP}
\\
{\cal C}_{D}(T) 
&\sim& 
\langle D | 0 \rangle \langle 0 | 
\Theta |  NK  \rangle \langle NK |\overline{\Theta} 
| 0 \rangle \langle 0 |  D \rangle 
e^{ - E_{NK} T }
+ \langle D | 0 \rangle \langle 0 | 
\Theta | \Theta  \rangle \langle \Theta |\overline{\Theta} 
| 0 \rangle \langle 0 |  D \rangle 
e^{ - E_{\Theta} T }
\nonumber\\
& & + \mbox{ and excited states }.
\label{CD}
\end{eqnarray}
In  Fig.~\ref{hadronic_corr}, we give a schematic picture of 
the contributions to the correlators. The first two terms in
Eq.(\ref{CPAP}) are the contributions from the five quark states 
as given in diagram (A).  The third, the fourth and the fifth 
terms 	in Eq.(\ref{CPAP}) which correspond to diagrams (C),(D) and (B) 
are hadronic contributions which propagate beyond the boundary. 
As a result, the correlation $\langle \Theta(T+t_{\rm src})\overline{\Theta} 
(t_{\rm src})\rangle$ inevitably contains  unwanted contributions such as
\begin{equation}
\langle \bar K|\Theta(T+t_{\rm src})|N\rangle \langle N|\overline{\Theta}
 (t_{\rm src})|\bar K\rangle
\sim e^{-E_NT+E_{\bar K}(T-N_t)}.
\end{equation}
In this case, the effective mass plot approaches
$E_N-E_{\bar K}$ below the NK threshold as $T$ is increased.
On the other hand, 
the contributions corresponding to diagrams (C),(D) and (B) 
do not exist with Dirichlet boundary conditions (Eq.(\ref{CD})).
Therefore, we find that it would be safest to impose the Dirichlet 
boundary condition on the temporal direction, since no quark can go
over the boundary on t=0 in the temporal direction. 
Although the boundary is transparent for the particles composed only 
by gluons; {\it i.e.} glueballs, due to the periodicity of the gauge
action, 
it would be however safe to neglect these gluonic particles going 
beyond the boundary since these particles are rather heavy. 
Then, the correlation 
$\langle \Theta(T+t_{\rm src})\overline{\Theta} (t_{\rm src})\rangle$
mainly contains only such terms ((A) in Fig.~\ref{hadronic_corr}) as 
\begin{equation}
\langle {\rm vac}|\Theta(T+t_{\rm src})|{\rm 5Q}\rangle \langle
{\rm 5Q}|\overline{\Theta} (t_{\rm src})|{\rm vac}\rangle
=
(1-\gamma_0)\sum_i W^+_ie^{-E^+_iT}+
(1+\gamma_0)\sum_i W^-_ie^{-E^-_iT}
\end{equation}
with $W^\pm_i$ the weight factor and $E^\pm_i$ the eigenenergy
of $i$-th state in positive/negative parity channel, respectively.
One sees that  one can now apply
the prescription mentioned in the last section.

One may wonder if these contaminations can be discarded with the  
parity projection of the correlators by taking linear combinations 
with periodic and antiperiodic boundary conditions. 
This method indeed works for ordinary three quark states where one can 
single out one of the two contributions diagram (X) and (Y) in 
Fig.~\ref{hadronic_corr}. 
However even if one takes such linear combinations,
one cannot make the contributions from diagram (C) seen in
Eq.(\ref{CPAP}) cancel out
as opposed to the contributions from diagram (B) and (D).
It is because of the fact that the factor $(-)^{\epsilon_n}$ 
for the contribution (C) is always equal to $1$.
(We note here that we can avoid these contaminations 
using the ``averaged quark propagator''~\cite{SS05}.)
Some of the previous lattice QCD studies on $\Theta^+$ adopted a
parity projection method using the combination with periodic 
and antiperiodic boundary conditions~\cite{CFKK03,Metal04}.
We stress that one should in principle 
be careful whether the result is free from the contamination 
owing to the boundary condition which is peculiar to the pentaquark
and can mimic a fake plateau in the 
propagator.

After obtaining the energy spectrum, we carry out a study of the
spectral weight for $(\kappa_{u,d},\kappa_{s})$=$(0.1600,0.1600)$.
Introducing two smeared operators $\overline{\Theta^1}_{\rm smear}$, 
$\overline{\Theta^2}_{\rm smear}$ we compute the following correlators
\begin{equation}
{\cal C}^{IJ}(T)=
\sum_{\vec{\bf x}} \langle
\Theta^I(\vec{\bf x},T+t_{\rm src}) 
\overline{\Theta^J}_{\rm smear}(t_{\rm src})\rangle,
\end{equation}
from which we extract the spectral weights using a constrained double 
exponential fit. The details will be explained in Sec.~\ref{Negative2-3}.

\section{Lattice QCD data}\label{Data}
\begin{figure}[h]
\includegraphics[scale=0.32]{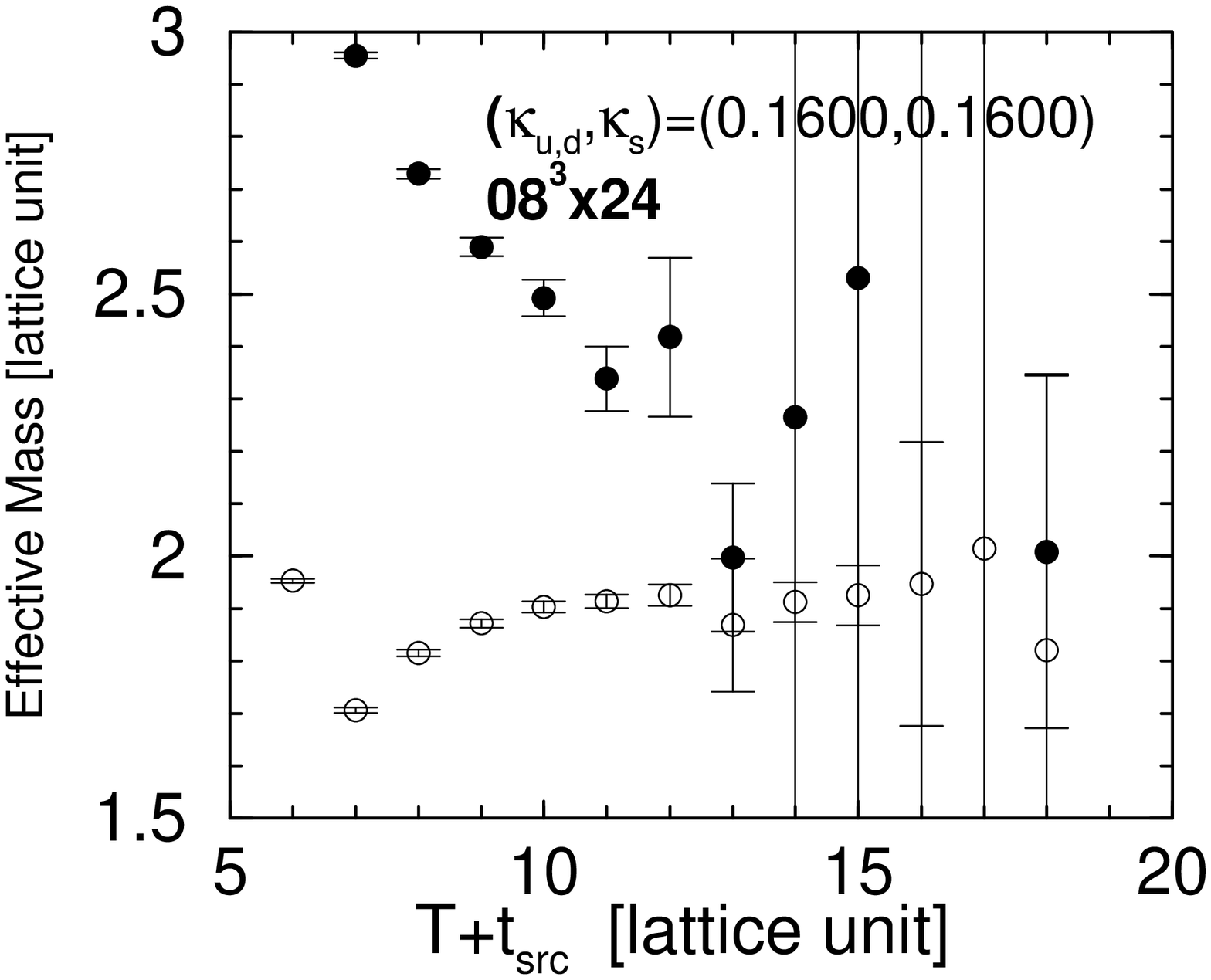}
\includegraphics[scale=0.32]{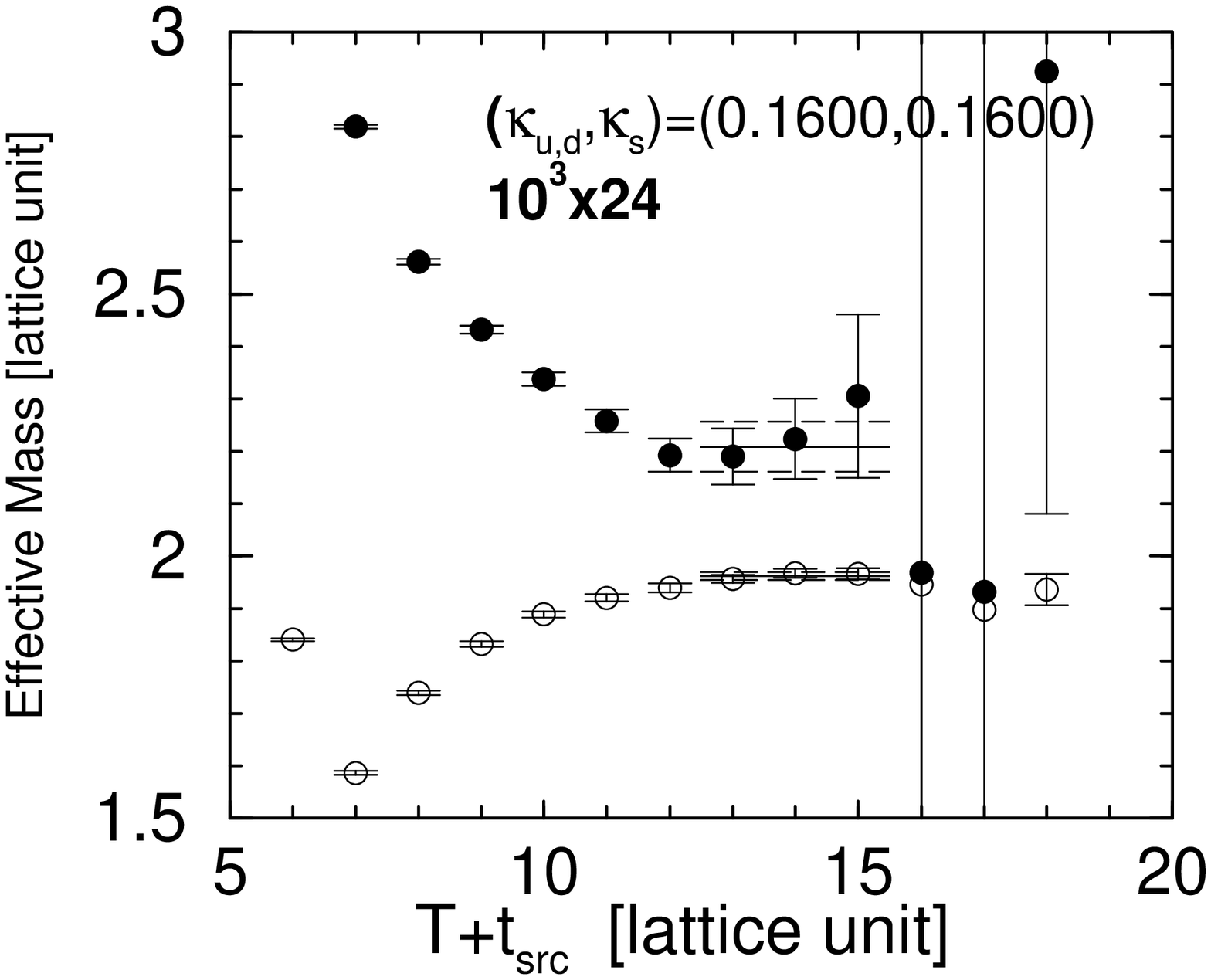}\\
\includegraphics[scale=0.32]{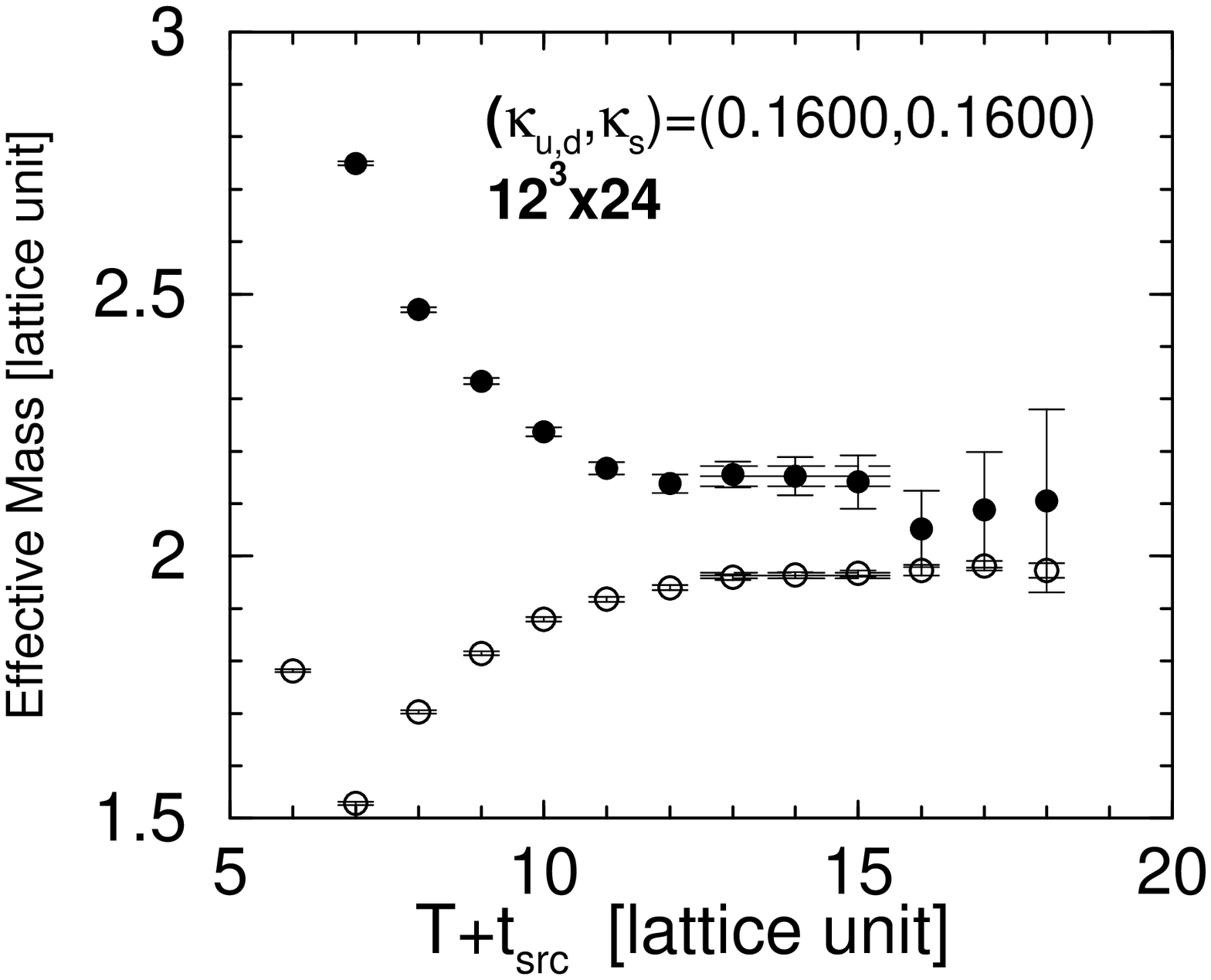}
\includegraphics[scale=0.32]{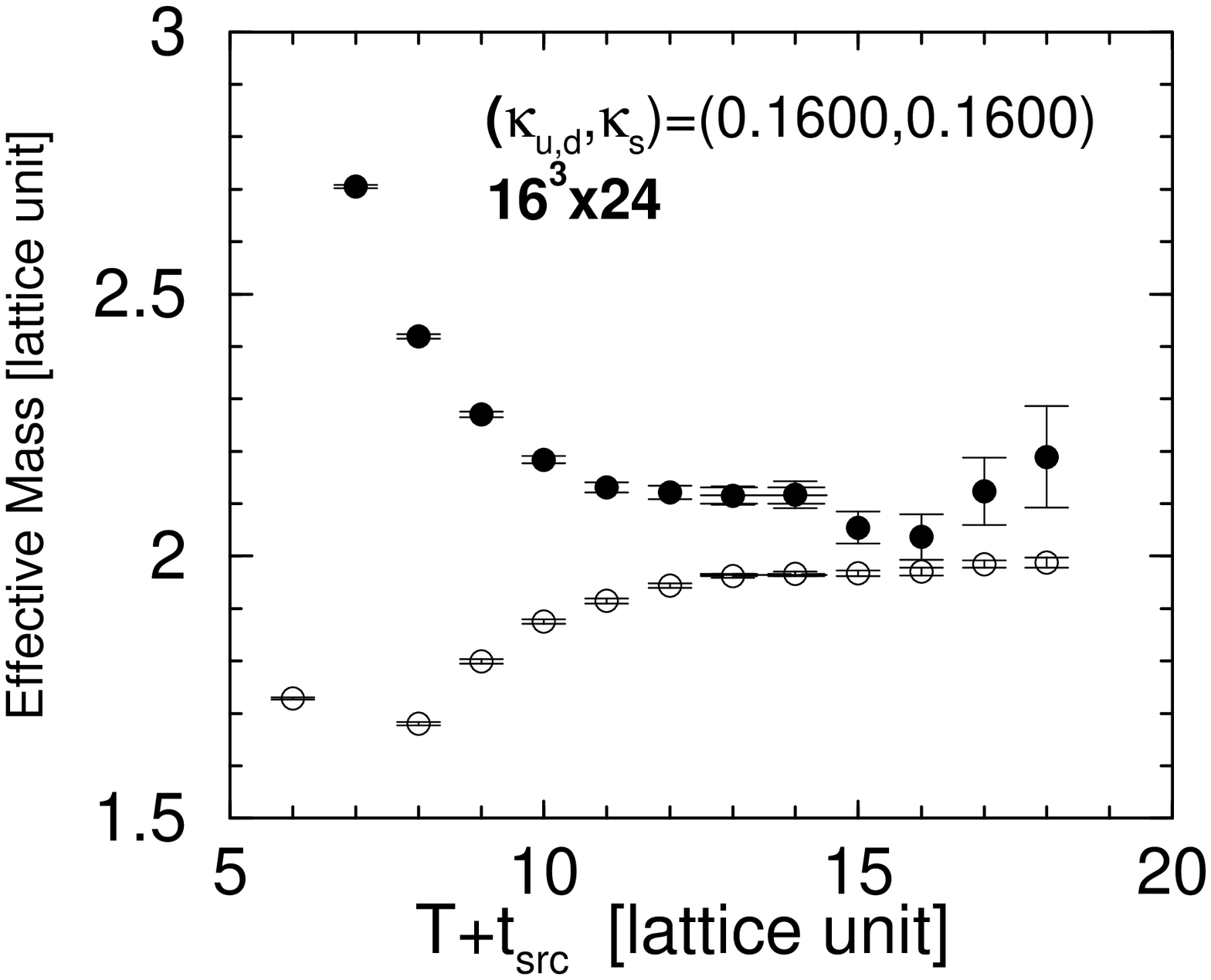}
\caption{
\label{effective_mass}
The ``effective mass'' plot $E_i(T)$ as the function of $T$,
the separation between source operators and sink operators,
in $(I,J^P)=(0,1/2^-)$ channel 
with the hopping parameters $(\kappa_{u,d},\kappa_s)=(0.1600.0.1600)$
on $8^3\times 24$, $10^3\times 24$, $12^3\times 24$,$16^3\times 24$  
lattice at $\beta =5.7$.
The stability of each $E_i(T)$ against $T$
means the smallness of the unwanted higher excited-state contaminations.
The solid line and the dashed lines represent
the central value and the error of the fitted masses $E_0^-$ and $E_1^-$.
}
\end{figure}
\begin{figure}[h]
\includegraphics[scale=0.32]{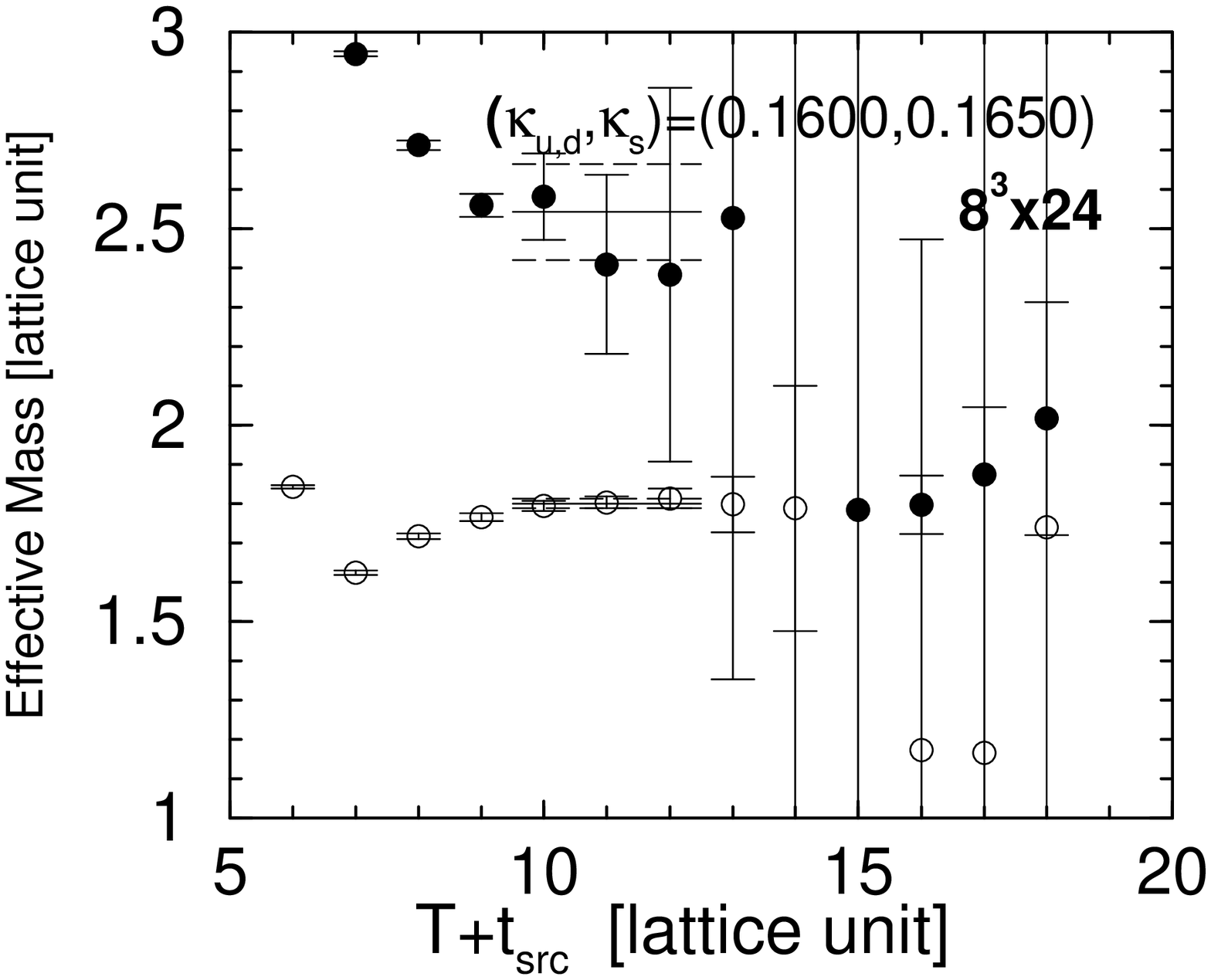}
\includegraphics[scale=0.32]{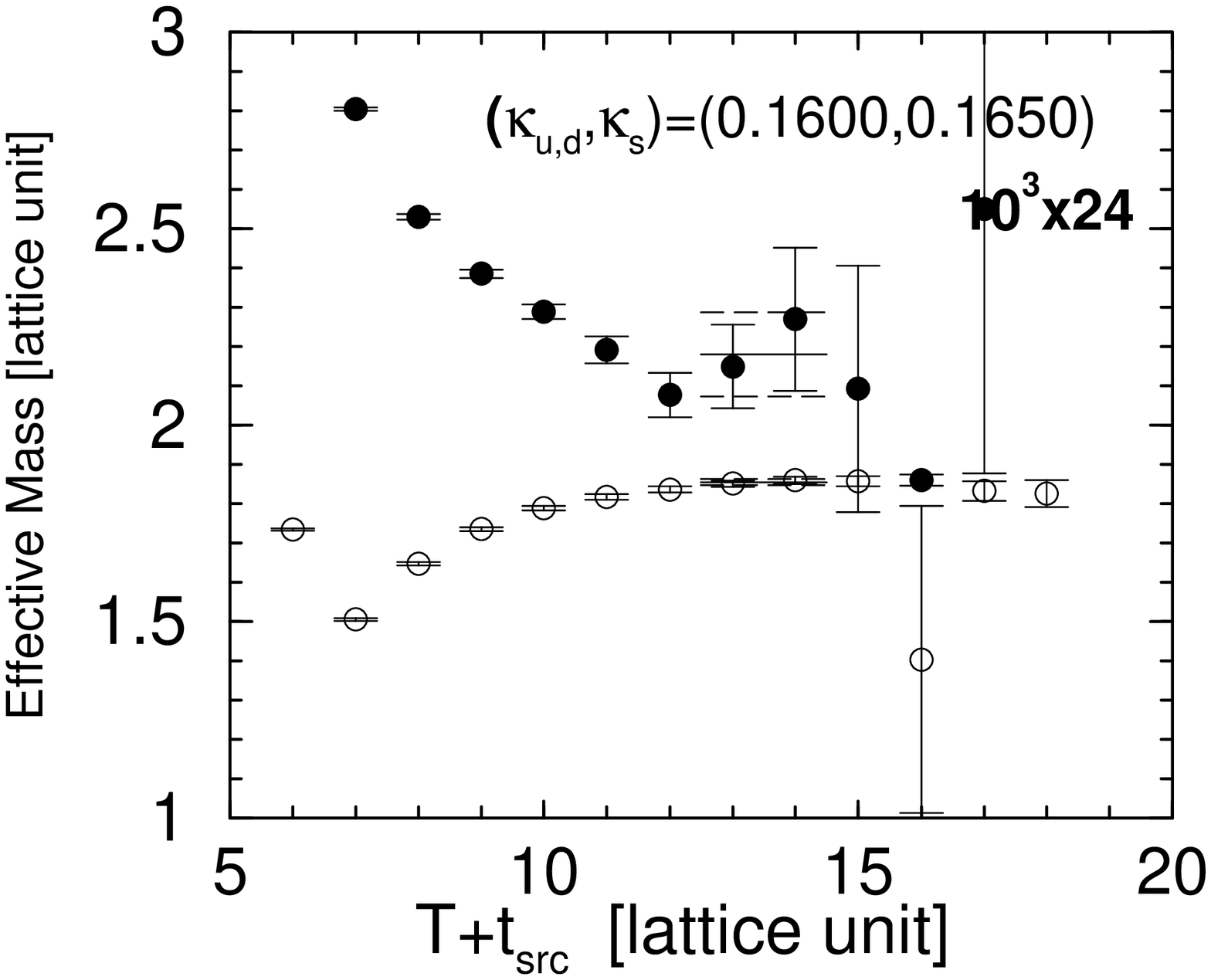}\\
\includegraphics[scale=0.32]{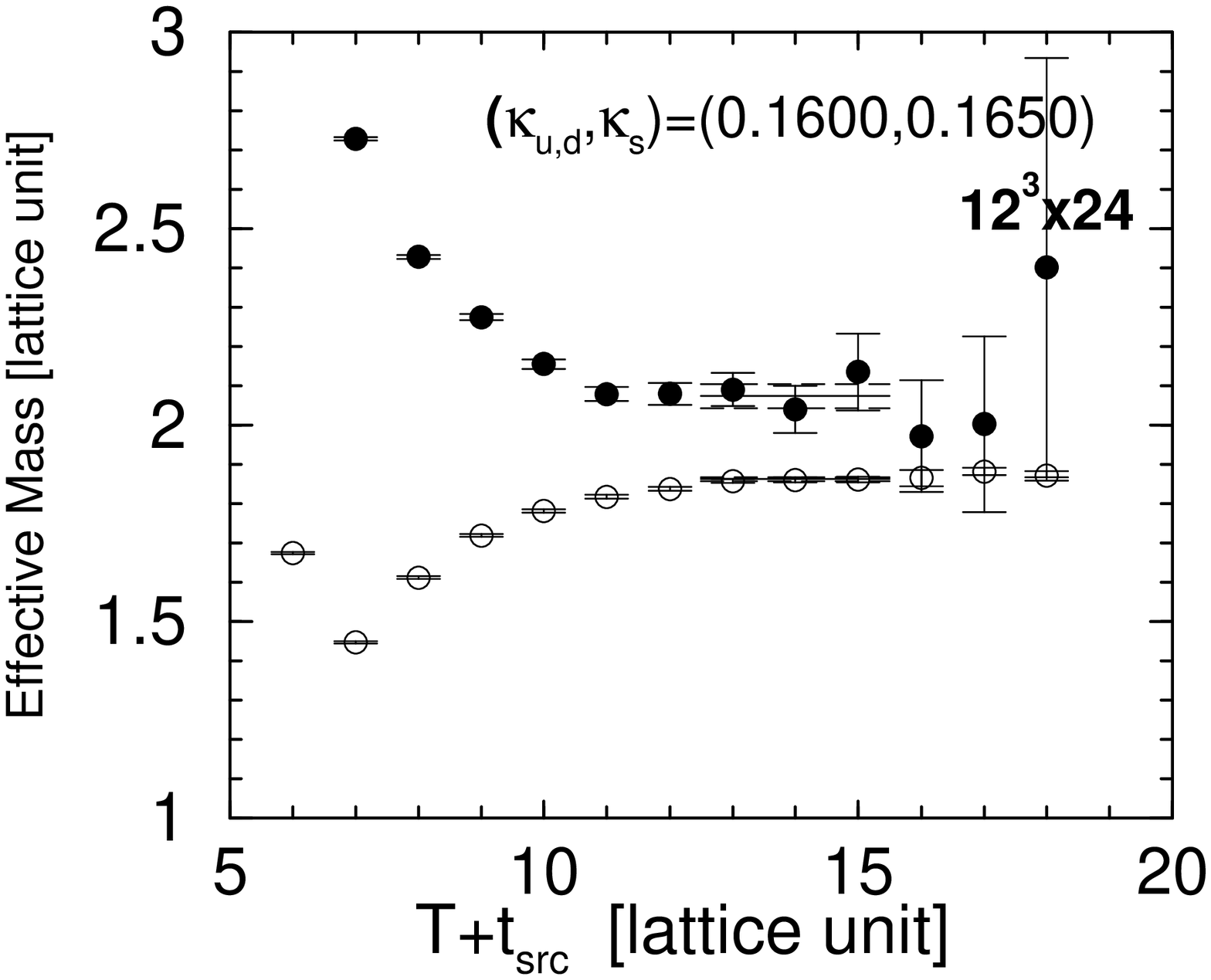}
\includegraphics[scale=0.32]{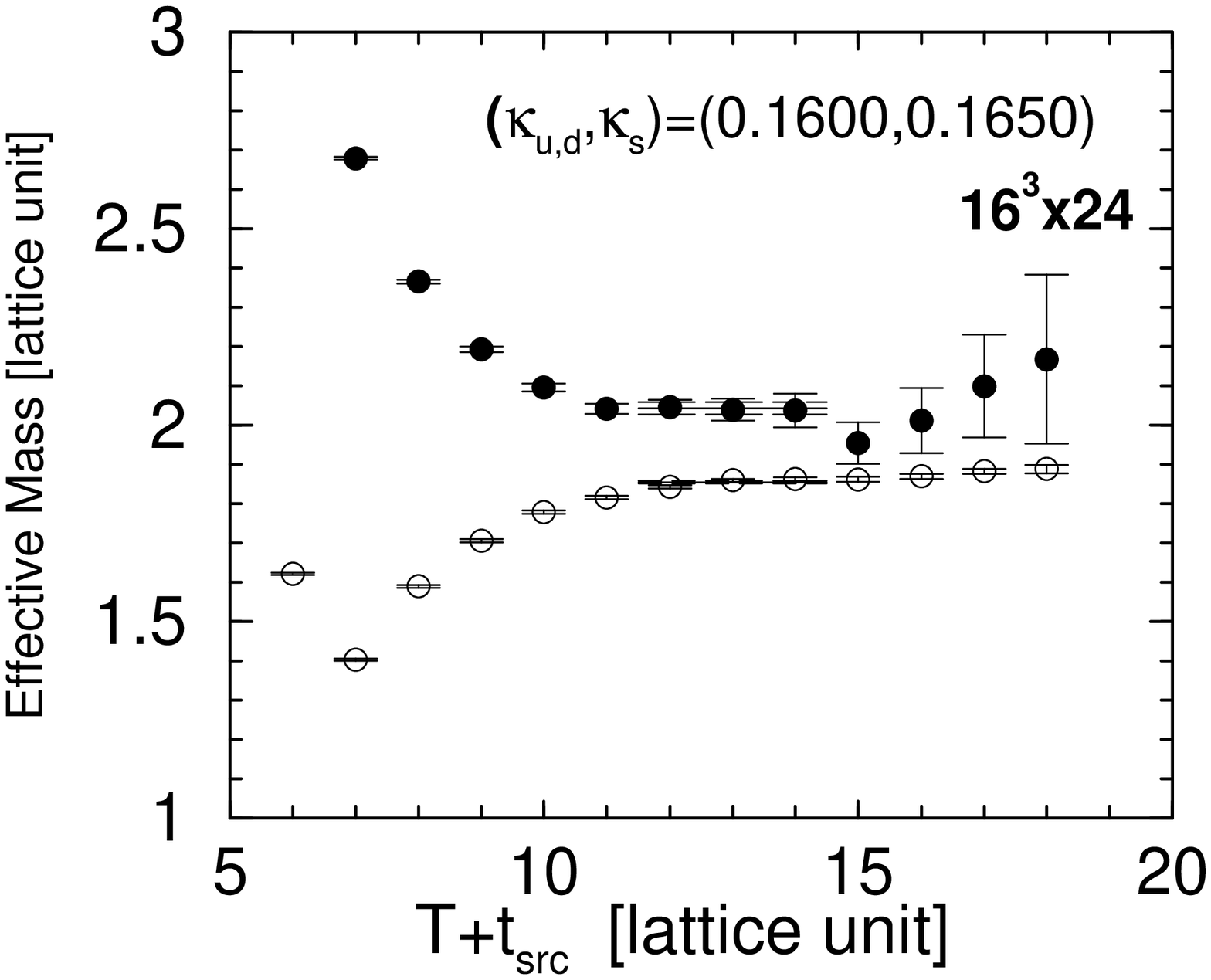}
\caption{
\label{meff221}
The ``effective mass'' plot $E_i(T)$ as the function of $T$,
the separation between source operators and sink operators,
in $(I,J^P)=(0,1/2^-)$ channel 
with the hopping parameters $(\kappa_{u,d},\kappa_s)=(0.1600.0.1650)$
on $8^3\times 24$, $10^3\times 24$, $12^3\times 24$,$16^3\times 24$  
lattice at $\beta =5.7$.
}
\includegraphics[scale=0.32]{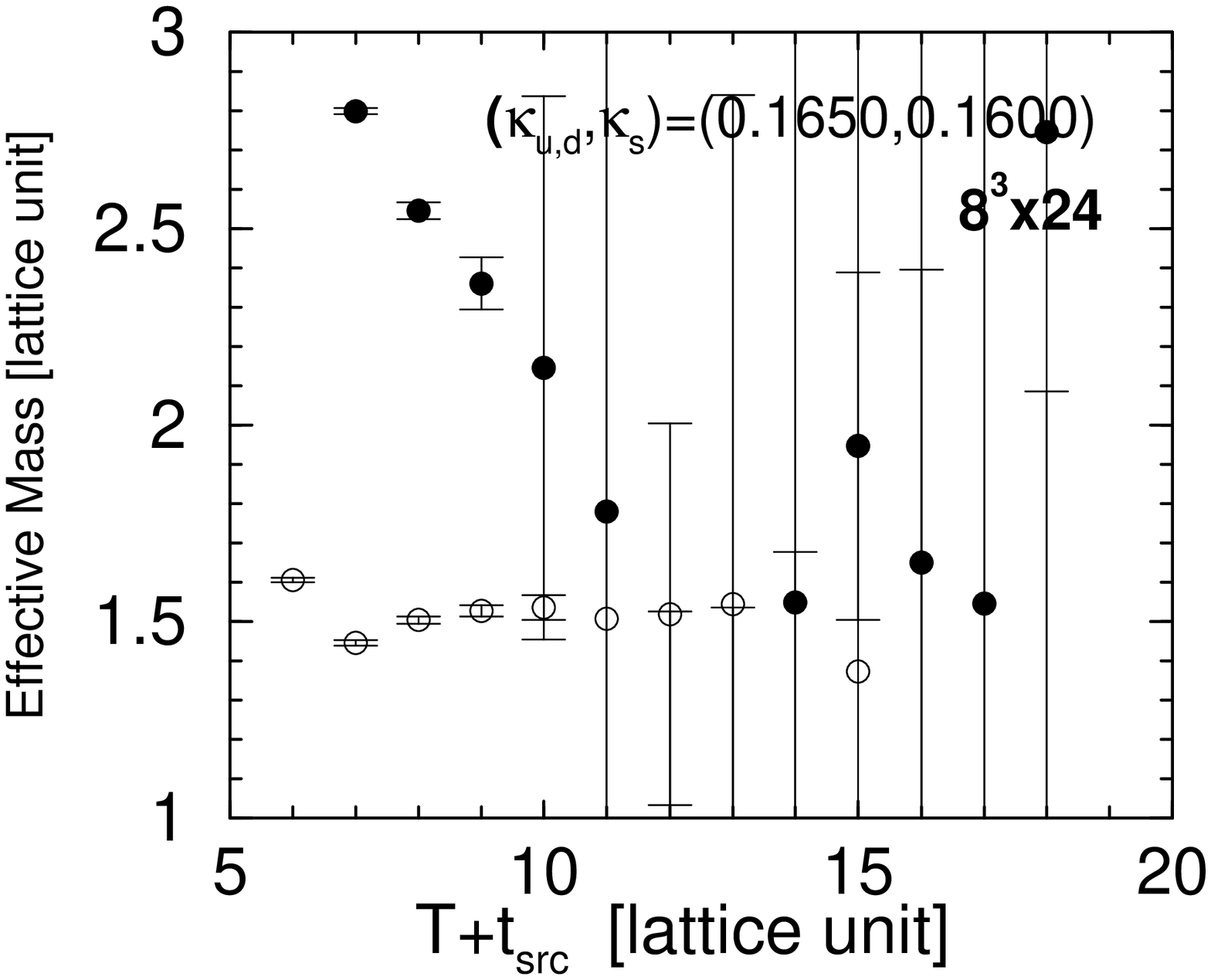}
\includegraphics[scale=0.32]{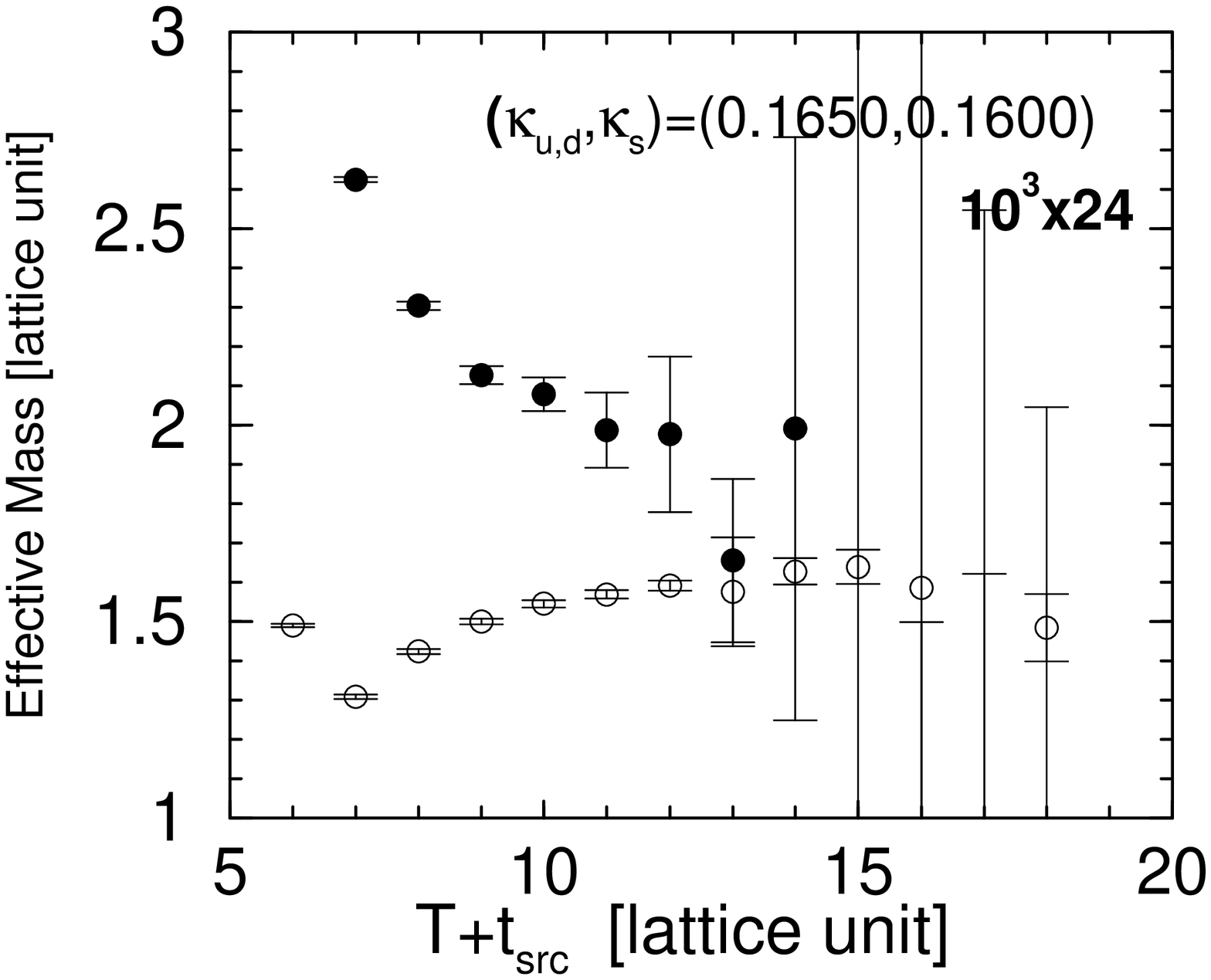}\\
\includegraphics[scale=0.32]{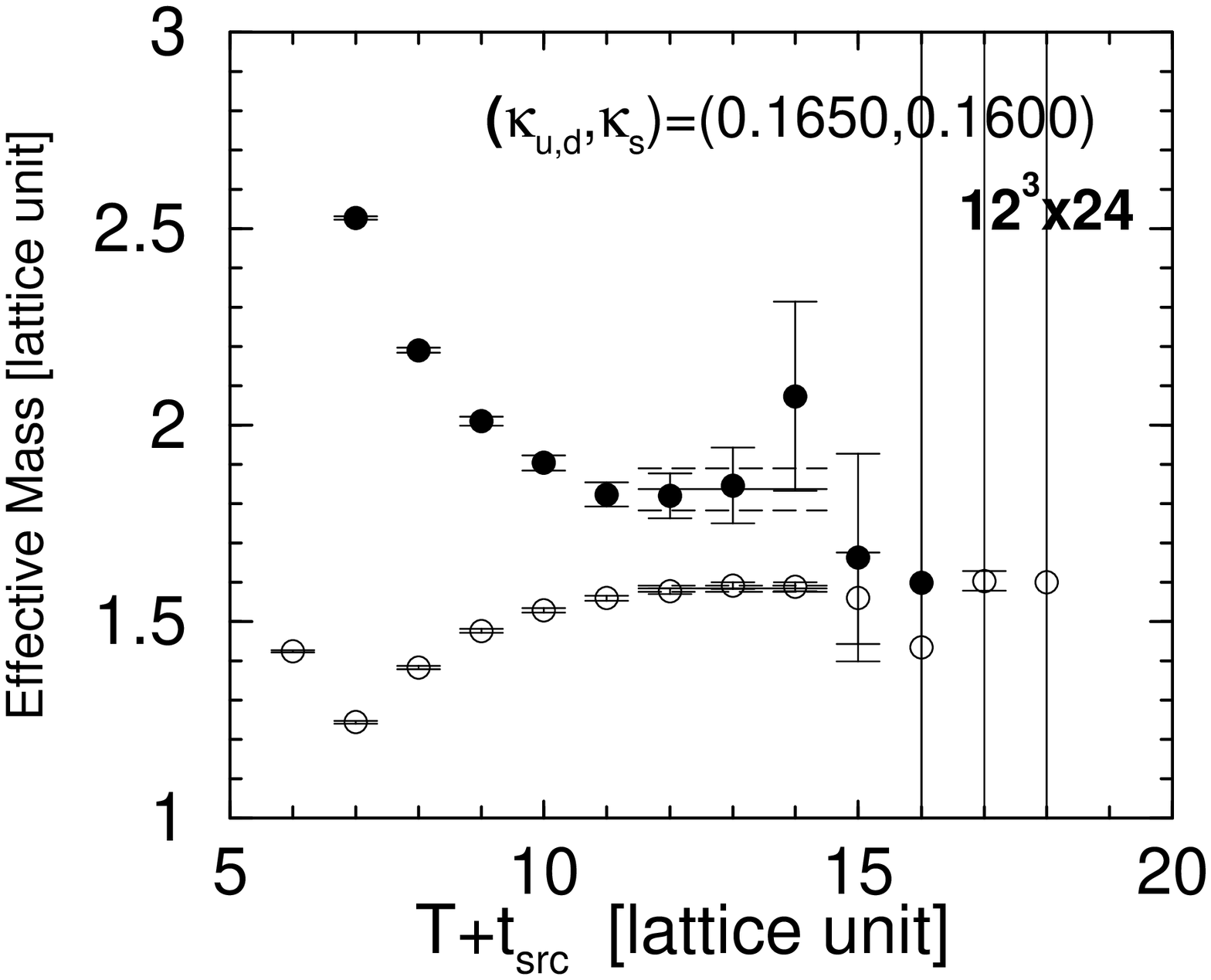}
\includegraphics[scale=0.32]{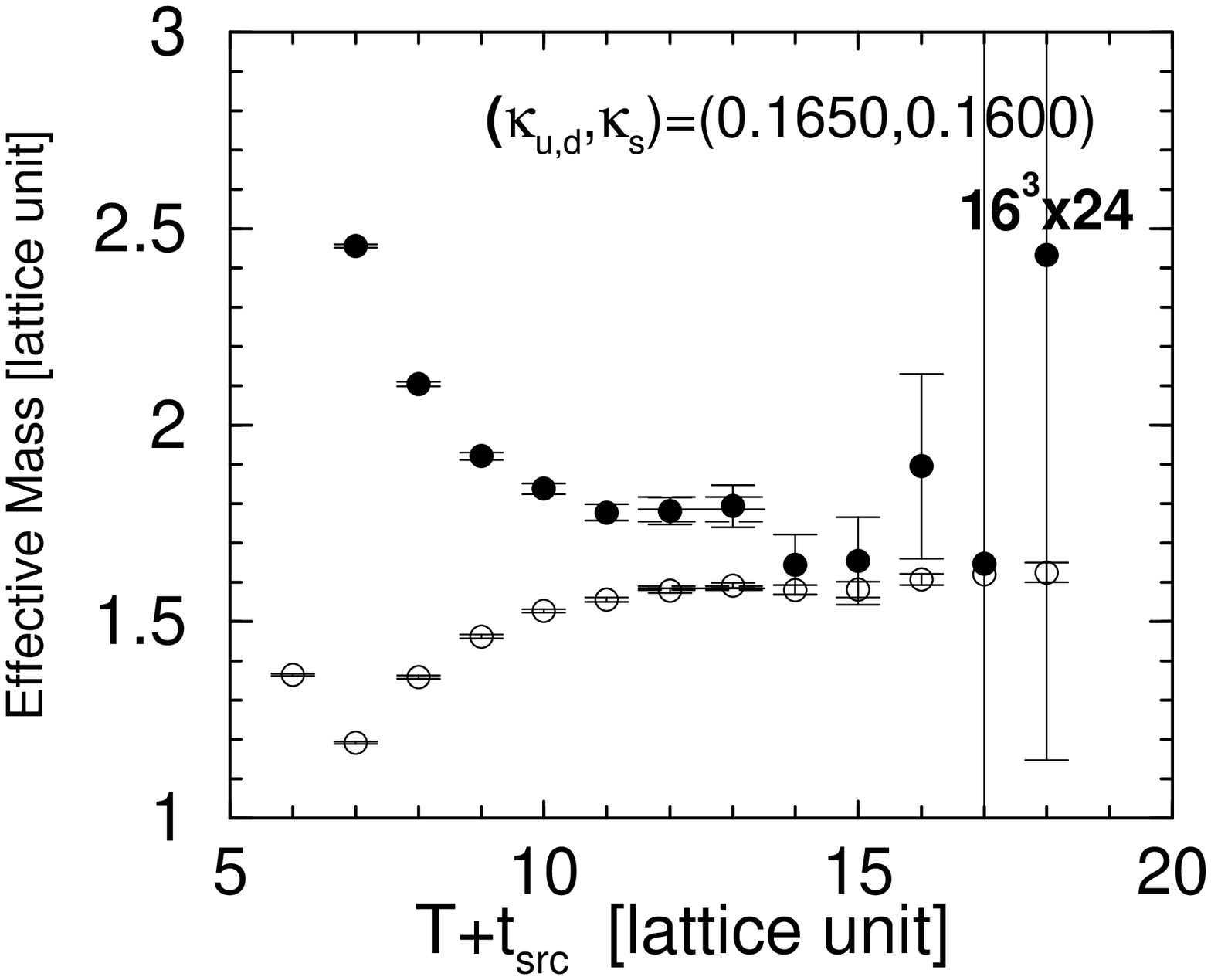}
\caption{
\label{meff112}
The ``effective mass'' plot $E_i(T)$ as the function of $T$,
the separation between source operators and sink operators,
in $(I,J^P)=(0,1/2^-)$ channel 
with the hopping parameters $(\kappa_{u,d},\kappa_s)=(0.1650.0.1600)$
on $8^3\times 24$, $10^3\times 24$, $12^3\times 24$,$16^3\times 24$  
lattice at $\beta =5.7$.
}
\end{figure}
\begin{figure}[h]
\includegraphics[scale=0.32]{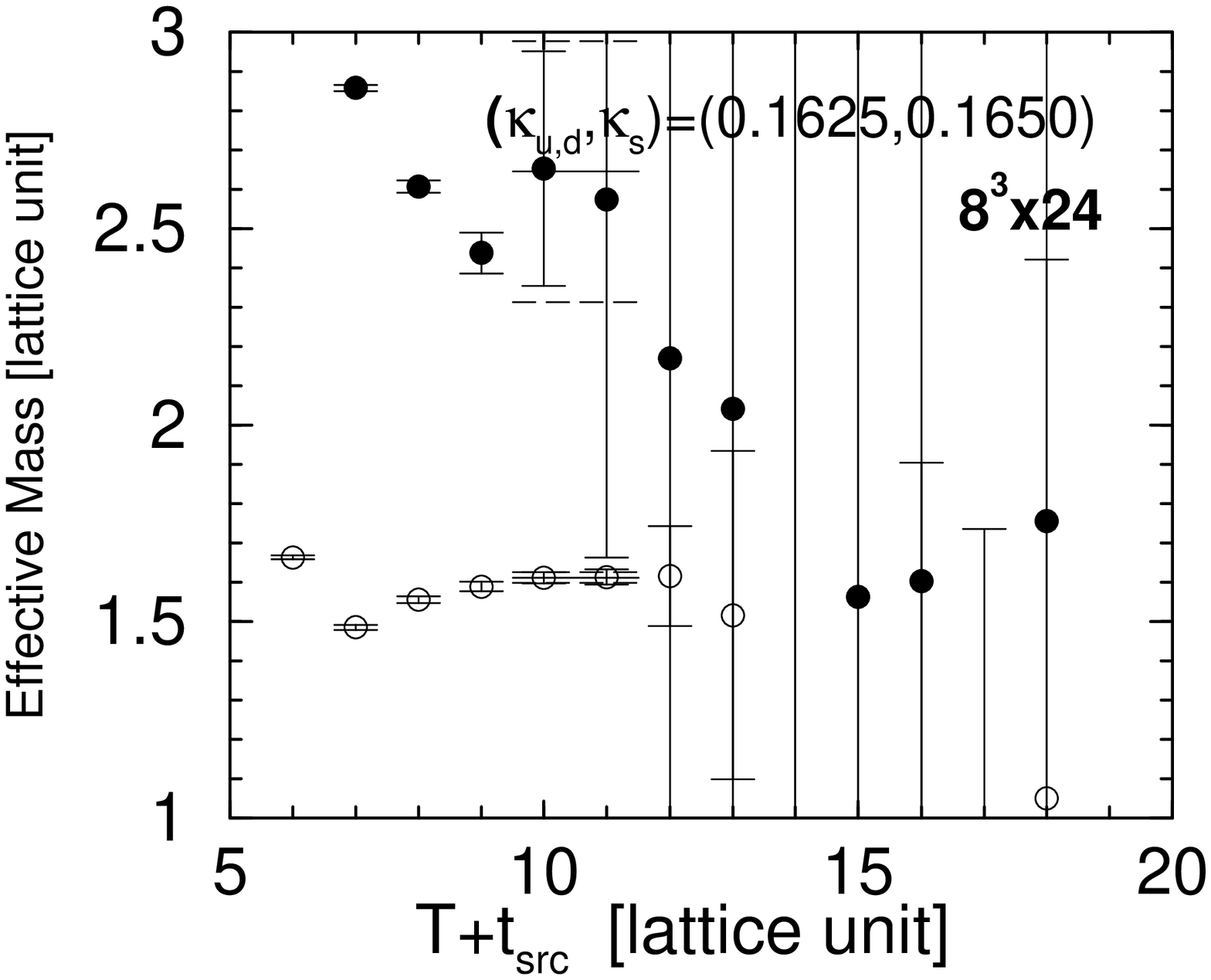}
\includegraphics[scale=0.32]{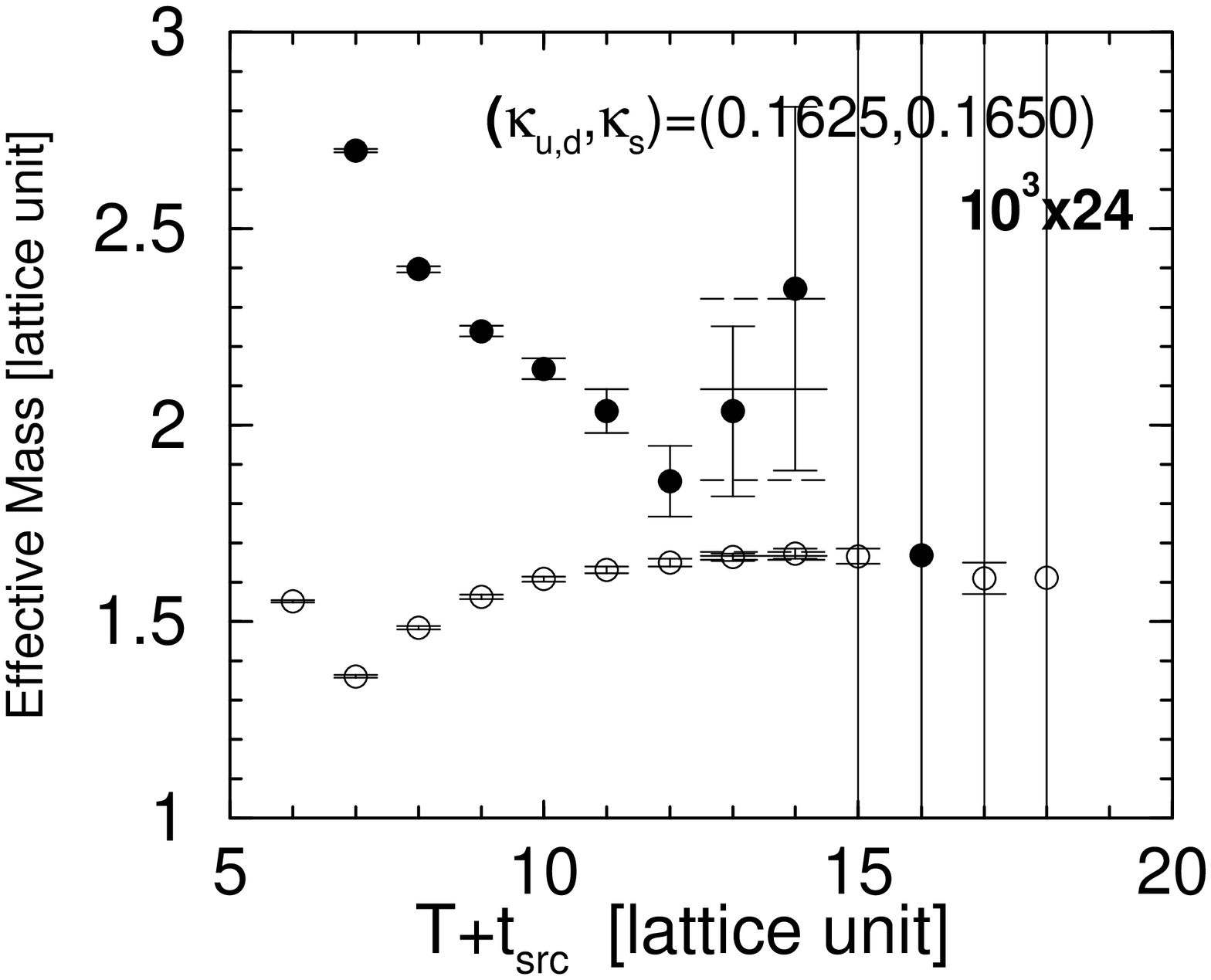}\\
\includegraphics[scale=0.32]{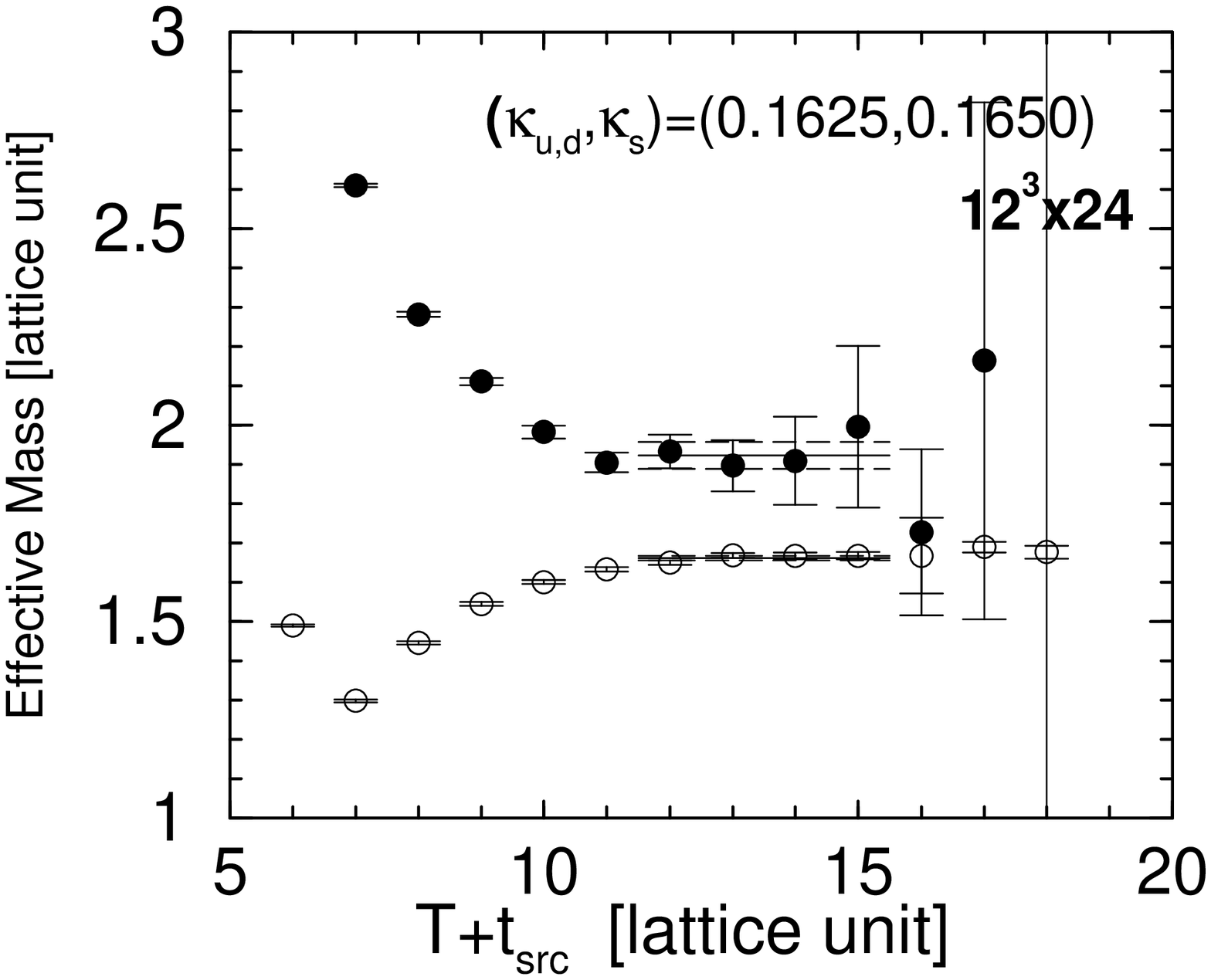}
\includegraphics[scale=0.32]{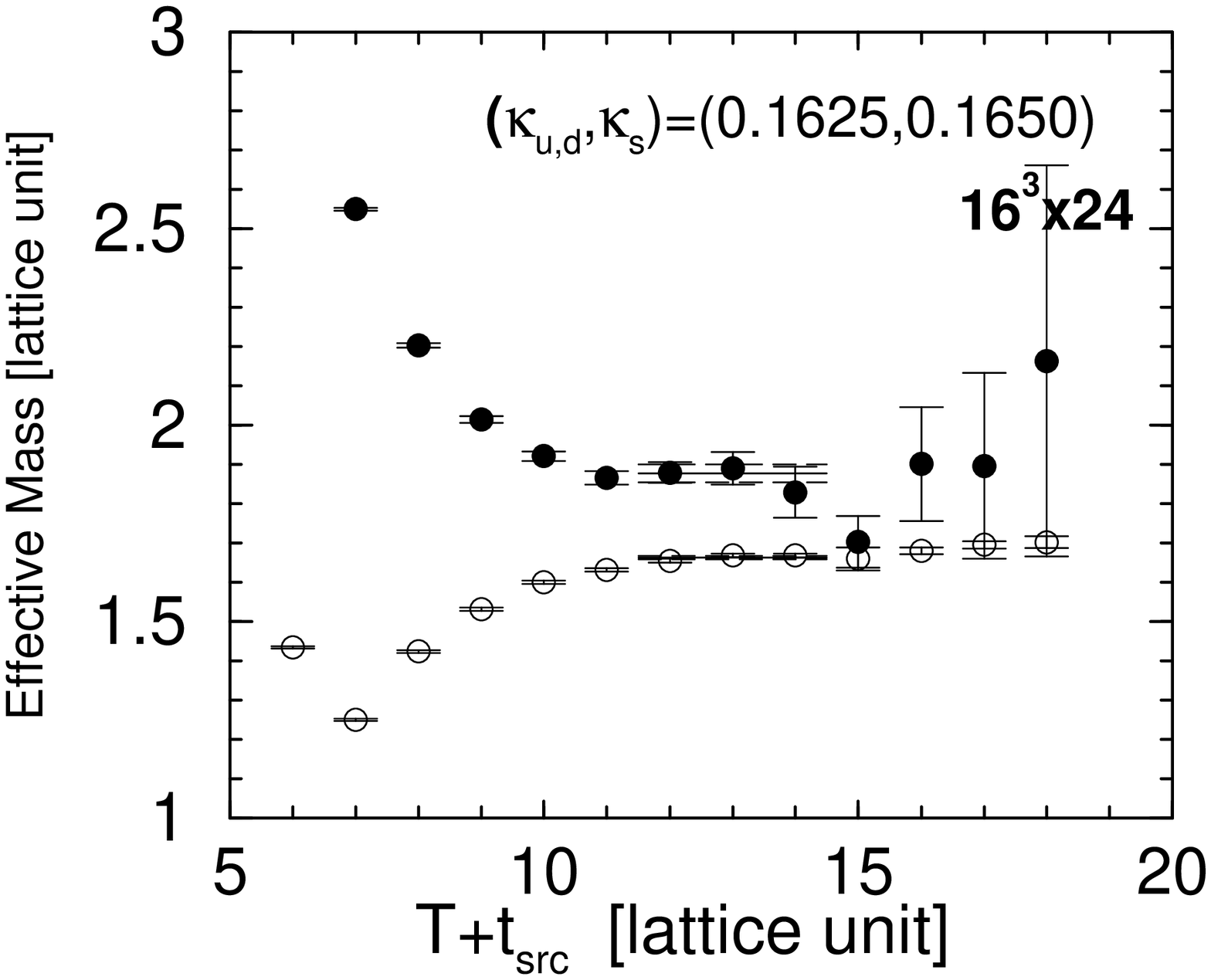}
\caption{
\label{meff331}
The ``effective mass'' plot $E_i(T)$ as the function of $T$,
the separation between source operators and sink operators,
in $(I,J^P)=(0,1/2^-)$ channel 
with the hopping parameters $(\kappa_{u,d},\kappa_s)=(0.1625.0.1650)$
on $8^3\times 24$, $10^3\times 24$, $12^3\times 24$,$16^3\times 24$  
lattice at $\beta =5.7$.
}
\includegraphics[scale=0.32]{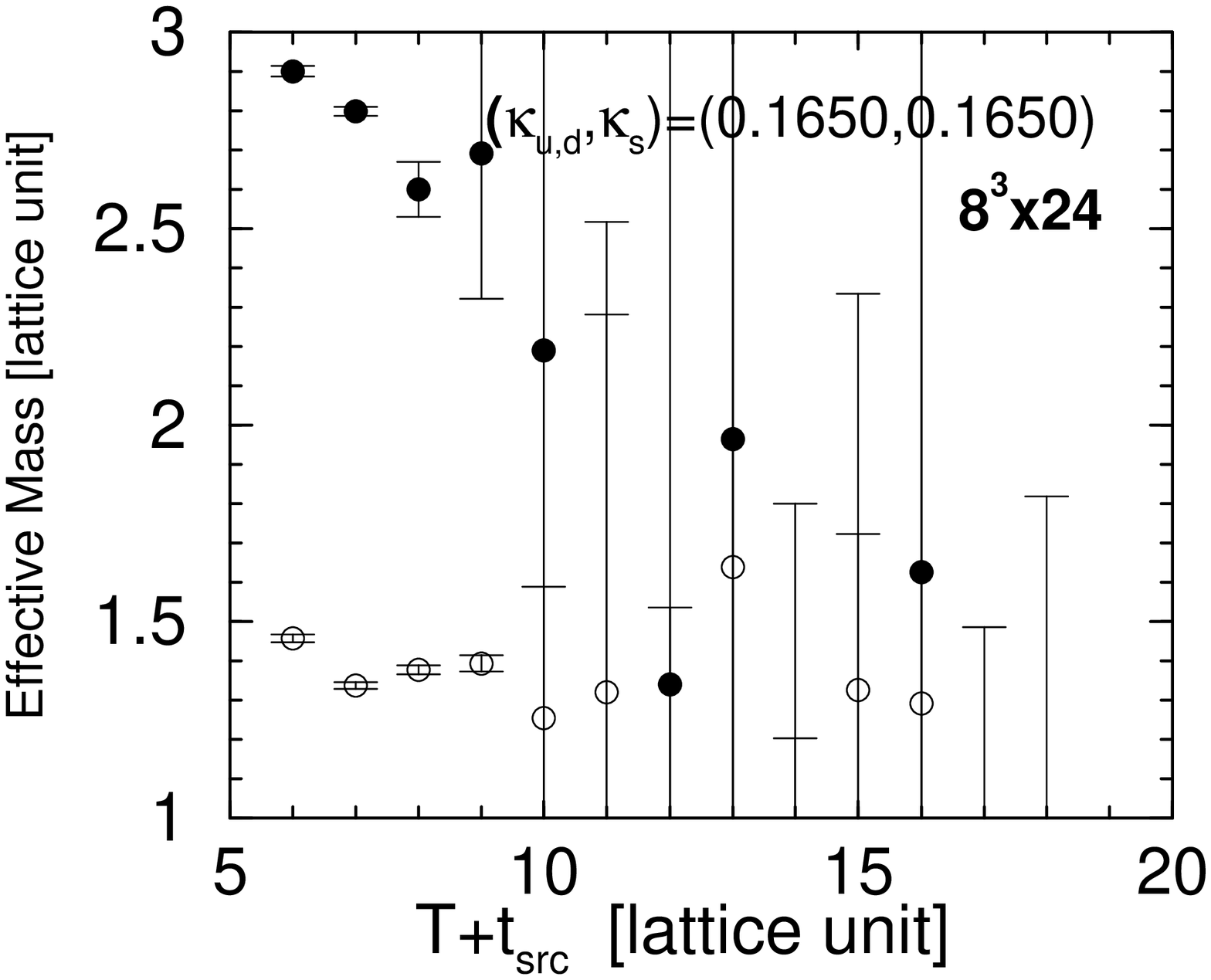}
\includegraphics[scale=0.32]{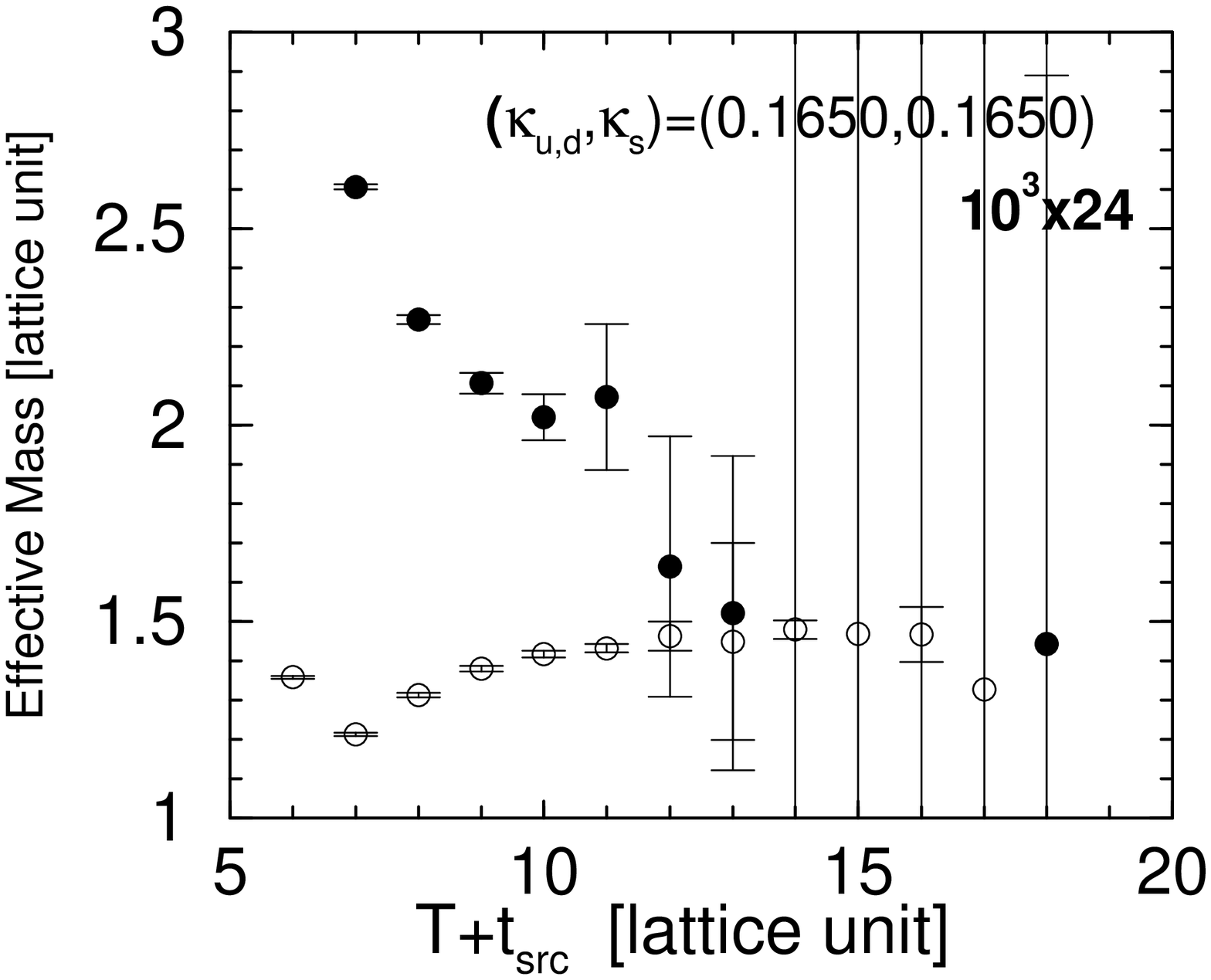}\\
\includegraphics[scale=0.32]{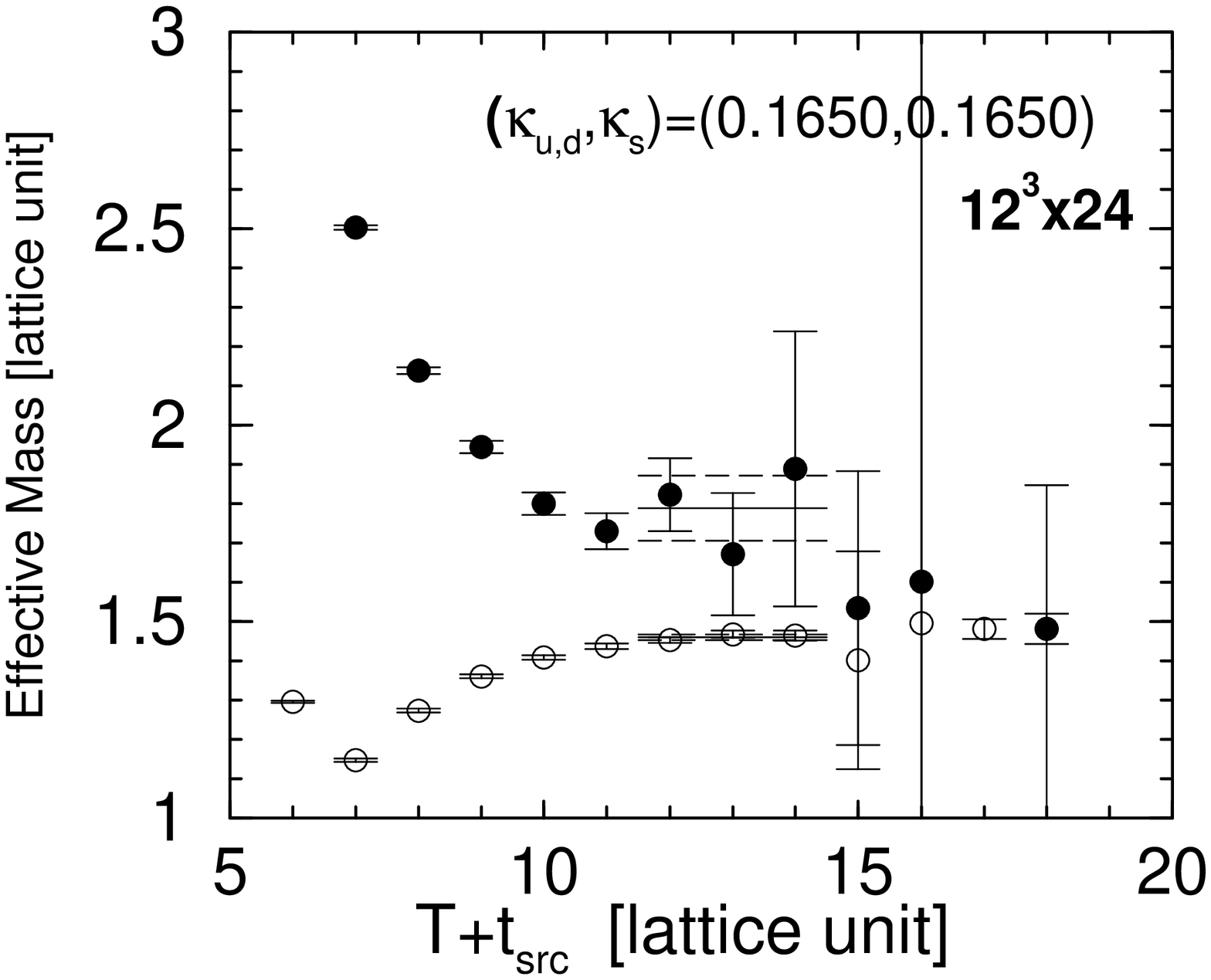}
\includegraphics[scale=0.32]{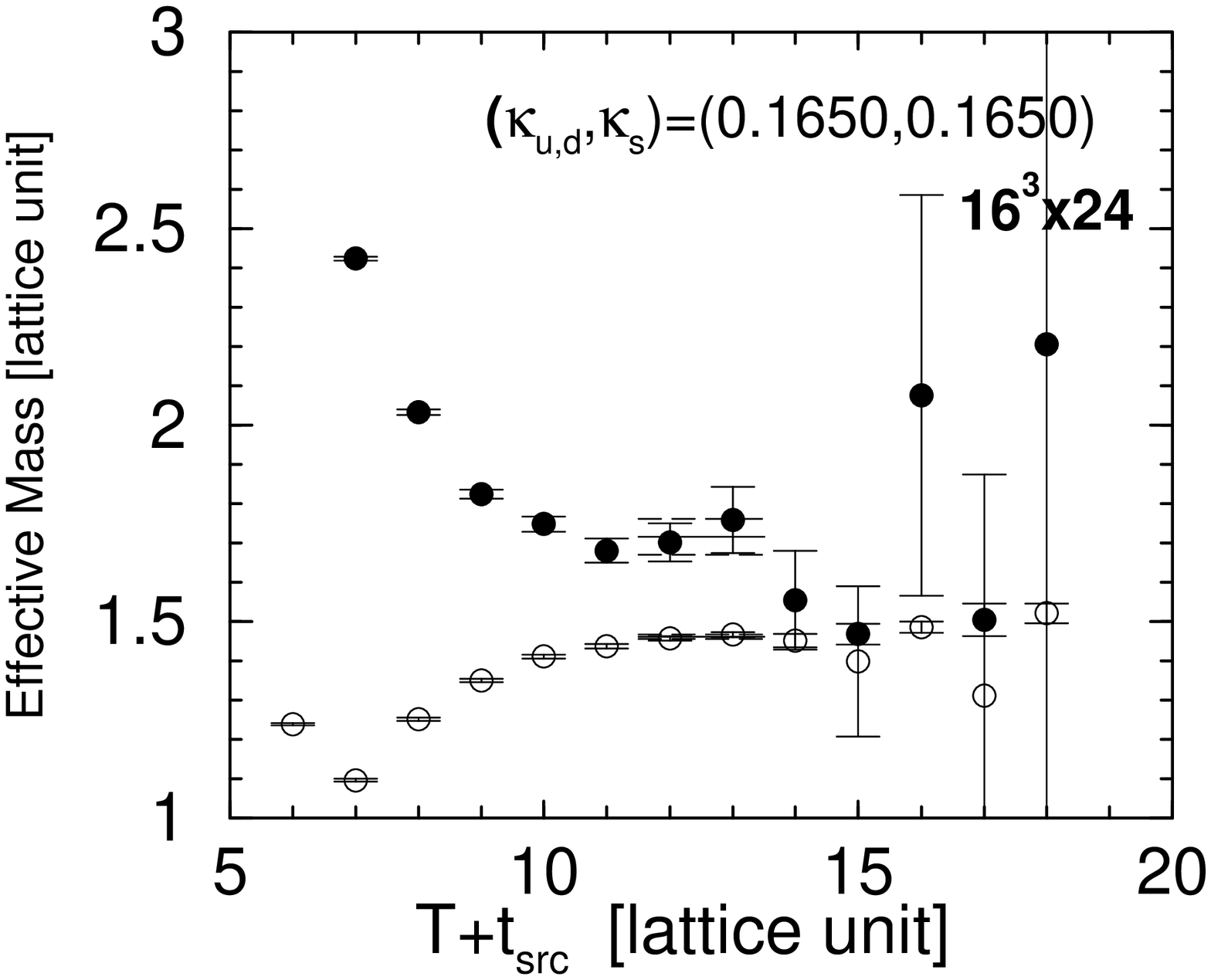}
\caption{
\label{meff111}
The ``effective mass'' plot $E_i(T)$ as the function of $T$,
the separation between source operators and sink operators,
in $(I,J^P)=(0,1/2^-)$ channel 
with the hopping parameters $(\kappa_{u,d},\kappa_s)=(0.1650.0.1650)$
on $8^3\times 24$, $10^3\times 24$, $12^3\times 24$,$16^3\times 24$  
lattice at $\beta =5.7$.
}
\end{figure}
Before obtaining the energies of 
the lowest state and the 2nd-lowest state, there are 
only a few simple steps.
First, we calculate the $2\times 2$ correlation matrix ${\cal C}(T)$
defined in Eq.(\ref{correlation_matrix})
and obtain the ``energies'' \{$E_i(T)$\} as the logarithm of 
eigenvalues \{$e^{E_i(T)}$\}
of the matrix product ${\cal C}^{-1}(T+1){\cal C}(T)$.
After finding the $T$ range ($T_{\rm min}\leq T\leq T_{\rm max}$),
where \{$E_i(T)$\} are stable against $T$,
we can extract the energies $E_i$ by the 
least $\chi$-squared fit of the data
as $E_i=E_i(T)$ in $T_{\rm min}\leq T\leq T_{\rm max}$.

Since the volume dependence of the energy is crucial 
to judge whether the state is a resonance or not, 
a great care must be paid in extracting the energy. 
Therefore, the systematic error by the contaminations from higher 
excited-states should be avoided by a careful choice of fitting 
ranges. For this reason, we impose the following criteria 
for the reliable extraction of the energies. 
Although this set of the criteria is nothing more than just one possible 
choice, we believe it is important to impose some concrete 
criteria for the fit so that we can reduce the human bias for 
the fit, though not completely. 
\begin{enumerate}
\item The effective mass plot should have ``plateau''
      for both the lowest and the 2nd-lowest states 
      simultaneously in a fit range [$T_{\rm min}$, $T_{\rm max}$],
      where the length $N_{\rm fit}\equiv T_{\rm max}-T_{\rm min}+1$      
      should be larger than or equal to 3 ( $N_{\rm fit}\geq 3$ ).
\item In the plateau region, the signal for the lowest
      and the 2nd-lowest states should be distinguishable, 
      so that the gap between the central values of the 
      lowest and the 2nd-lowest energies should be larger than their
      errors.
\item The fitted energies should be stable against the choice 
      of the fit range; i.e. the results of the fit with 
      $N_{\rm fit}$ time slices and with $N_{\rm fit}-1$ 
      time slices should be consistent within statistical errors 
      for both the lowest and the 2nd-lowest states. 
\item The lowest state energy obtained by the diagonalization method 
      using the $2\times2$ correlation matrix 
      should be consistent with the value from a single exponential 
      fit for a sufficiently large $t$.
\end{enumerate}
If the fit does not satisfy the above conditions, we discard the
result since either the data in the fit range may be contaminated 
by higher excited-states or the 2nd-lowest-state signal is too 
noisy for a reliable fit. 

Figs~\ref{effective_mass} show the ``effective mass'' plot $E_i(T)$
for the heaviest combination of quarks
$(\kappa_{u,d},\kappa_s)=(0.1600.0.1600)$.
As is mentioned in Sec.~\ref{Formalism},
we need to find the $T$ region ($T_{\rm min}\leq T\leq T_{\rm max}$)
where each $E_i(T)$ shows a plateau.
In the case of $12^3\times 24$ lattice 
in Fig.~\ref{effective_mass}, for example,
we choose the fit range of  $T_{\rm min}=6$ and $T_{\rm max}=9$
($N_{\rm fit}=4$). 
Notice that the source operators are put on the time slice
with $t=t_{\rm src}=6$.
The plateau in this region satisfies the above
criteria so that we consider the fit $E_0^-$ and $E_1^-$ for the range 
$6\leq T\leq 9$ as being reliable. The situation is similar for 
the cases of $10^3\times 24$ and $16^3\times 24$ lattices. 
On the other hand,  in the case of $8^3\times 24$ lattice 
we do not find a plateau region satisfying the above criteria.

Figs.~\ref{meff221},\ref{meff112},\ref{meff331},\ref{meff111} shows the 
``effective mass'' plots 
for the combinations with smaller quark masses. We find that 
the signal is noisier for the lighter quarks and the fit with the smaller 
volumes $8^3\times 24$ and $10^3\times 24$ lattices do not 
satisfy the criteria.

\section{Lowest-state energy in $(I,J^P)=(0,\frac12^-)$ channel}
\label{Negative1}

\subsection{The volume dependence of the lowest-state energy}
\label{Negative1-1}

Now we show the lattice QCD results
of the lowest state in $I=0$ and $J^P=\frac12^-$ channel.
The filled circles in
Fig.~\ref{negativeGSES} show the lowest-state energies $E_0^-$
in $I=0$ and $J^P=\frac12^-$ channel on four different volumes.
Here the horizontal axis denotes the lattice extent $L$
in the lattice unit and the vertical axis is the energy of the state.
The lower line denotes the simple sum $M_N+M_K$
of the nucleon mass $M_N$ and Kaon mass $M_K$
obtained with the largest lattice.
Though $M_N$ and $M_K$ are slightly affected by finite volume effects,
the deviation of $M_N+M_K\ (L=8)$ from $M_N+M_K\ (L=16)$ is 
about a few \% (Table~\ref{results}.).
We therefore simply use $M_N+M_K\ (L=16)$ as a guideline.

At a glance, 
we find that the energy of this state takes almost constant value against
the volume variation and coincides with
the simple sum $M_N+M_K$.
We can therefore conclude that 
the lowest state in $I=0$ and $J^P=\frac12^-$ channel
is the NK scattering state with the relative momentum $|{\bf p}|=0$.
The good agreement with the sum $M_N+M_K$ implies
the weakness of the interaction between N and K. 
In fact, the scattering length in the $I=0$ channel is known to be 
tiny ( $a_0^{KN}(I=0)=-0.0075$ fm ) from compilations of 
hadron scattering experiments~\cite{Dumbrajs:1983jd,Nagels:1979xh}, 
whereas the current algebra prediction from PCAC with SU(3) symmetry 
predicts that the scattering length $a_0^{KN}(I=0)=0$.

\subsection{Comparison with the previous lattice work}

We here compare our data with the previous lattice QCD studies,
which were performed with almost the same conditions as ours,
in order to confirm the reliability of our data.

The well-known hadron masses
$m_\pi$,$m_K$ and $m_N$ listed in Table~\ref{results}
can be compared with the values in Ref.~\cite{BCSVW94}.
Our data are consistent with those in Ref.~\cite{BCSVW94}.
The lowest NK scattering state
in $(I,J^P)=(0,\frac12^-)$ channel
is carefully investigated in Ref.~\cite{CPPACS95} with almost the same parameters
as our present study.
It is worth comparing our data with them.
For the complete check of our data, we re-extract the lowest state 
by the ordinary single-exponential fit of the correlator as
$\langle\Theta(T+t_{\rm src}) 
\overline{\Theta}_{\rm wall}(t_{\rm src})\rangle
=C e^{-E_{NK}T}$ in the large-T region,  
and compare them with the present lattice data $E_0^-$
obtained by the multi-exponential method
as well as the data in Ref.~\cite{CPPACS95}.
In Table~\ref{results}, we list the data of the lowest state $E_{NK}$
obtained by the single-exponential fit.
They almost coincide with the present data $E_0^-$
extracted by the multi-exponential method with about 1\% deviations,
which may be considered as the slightly remaining contaminations
of the higher excited-state.
In Ref.~\cite{CPPACS95},
the authors extracted the energy difference
$\delta E=E_{NK}-(E_N+E_K)$ with the hopping parameters
$\kappa_{u,d,s}=0.1640$ using $12^3\times 20$ lattice.
We therefore compare our data 
$\delta E=E_{NK}-(E_N+E_K)$ obtained
with the hopping parameters 
$(\kappa_{u,d},\kappa_s)=(0.1650,0.1650)$
on $12^3\times 24$ lattice.
The energy difference $\delta E$ in our study is found to be
$\delta E=-0.0128(38)$,
which is consistent with the value of
$-0.0051(38)$ in Ref.~\cite{CPPACS95} taking into account that
this error includes only statistical one.

It is now confirmed that the lowest state extracted using
the multi-exponential method is consistent with the previous works
and that our data and method are reliable enough to investigate
the 2nd-lowest state in this channel.

\section{2nd-lowest state energy in $(I,J^P)=(0,\frac12^-)$ channel}
\label{Negative2}

The $(I,J^P)=(0,\frac12^-)$ state is one of the candidates
for $\Theta^+(1540)$.
Since $\Theta^+$ is located above the NK threshold,
it would appear as an excited state in this channel.
We show the lattice data of the 2nd-lowest state in this channel.

In order to distinguish a possible resonance state from NK scattering
states, we investigate the volume dependence of 
both the energy and the spectral weight of each state.
It is expected that the energies of resonance states 
have small volume dependence,
while the energies of NK scattering states are expected to scale as 
$\sqrt{M_N^2+|\frac{2\pi}{L}\vec{\bf n}|^2}+
\sqrt{M_K^2+|\frac{2\pi}{L}\vec{\bf n}|^2}$ according to
the relative momentum $\frac{2\pi}{L}\vec{\bf n}$ between N and K on
a finite periodic lattice,
provided that the NK interaction is weak and negligible.
We can take advantage of the above difference for the discrimination.

\subsection{
Possible corrections to the volume dependence of NK scattering states}
\label{Negative2-1}

A possible candidate
for the volume dependence of the energies of NK scattering states
is the simple formula as
$E_{NK}^{\vec{\bf n}}(L)\equiv
\sqrt{M_N^2+|\frac{2\pi}{L}\vec{\bf n}|^2}+
\sqrt{M_K^2+|\frac{2\pi}{L}\vec{\bf n}|^2}$ with
the relative momentum $\frac{2\pi}{L}\vec{\bf n}$ between N and K in
finite periodic lattices,
which is justified
on the assumption that nucleon and Kaon are point particles
and that the interaction between them is negligible.
In practice, there may be
some corrections to the volume dependence of $E_{NK}^{\vec{\bf n}}(L)$.
We therefore estimate here three possible corrections;
the existence of the NK interaction,
the application of the momenta on a finite discretized lattice and
the estimation of the implicit finite-size effects.

There can be small hadronic interactions between nucleon and Kaon,
which may lead to correction to naively expected energy spectrum
$E_{NK}^{\vec{\bf n}}(L)$
of the NK scattering states.
Using L{\" u}scher formula~\cite{L91}, one can 
relate the scattering phase shift to the energy shift 
from $E_{NK}^{\vec{\bf n}}(L)$ on finite lattices.
For example, in the case when a system belongs 
to the representation ${\cal A}_1^+$ of cubic groups,
which is relevant in the present case,
the relation between the phase shift and the
possible momentum spectra is 
\begin{equation}
e^{2i\delta_0(k)}
=\frac{{\cal Z}(1;q^2)+i\pi^{\frac23}q}
{{\cal Z}(1;q^2)-i\pi^{\frac23}q}.
\end{equation}
Here ${\cal Z}(s;q^2)$ is the Zeta function defined as
\begin{equation}
{\cal Z}(s;q^2)\equiv
\frac{1}{\sqrt{4\pi}}
\sum_{n\in Z^3}\frac{1}{(n^2-q^2)^s},
\end{equation}
with the eigenenergy $q$ on a finite lattice.
We have simply omitted the corrections from the partial waves
with angular momenta higher than the next smallest one ( $l=4$ ).
Although our current quark masses are
heavier than those of the real quarks, we use the empirical values 
of the phase shift in NK scattering in Ref.\cite{HARW92}, by
simply neglecting the quark mass dependence. 
The correction using the empirical values results in 
at most a few \% larger energy than the simple formula
$E_{NK}^{\vec{\bf n}}(L)$
within the volume range under consideration;
the energies are slightly increased by
the weak repulsive force between nucleon and Kaon.

One may claim that one has to adopt momenta on a finite 
discretized lattice:
$|\vec{\bf p}|^2=
4\sin^2 (\frac{\pi}{L})\cdot|\vec{\bf n}|^2$ for Kaon
and $|\vec{\bf p}|^2=\sin^2 (\frac{2\pi}{L})\cdot|\vec{\bf n}|^2$ for
nucleon, respectively.
This correction turns out to be within only a few \% lower energy
than $E_{NK}^{\vec{\bf n}}(L)$,
although it is not certain whether this correction is meaningful or not
for composite particles like nucleon or Kaon.

We find that these corrections lead to at most a few \% 
deviations from $E_{NK}^{\vec{\bf n}}(L)$.
We then neglect these corrections for simplicity in the following discussion
and use the simple form $E_{NK}^{\vec{\bf n}}(L)$.

So far, we have neglected the implicit finite-size effects
in $E_{NK}^{\rm \vec{\bf n}}(L)$,
other than the explicit ones due to the lattice momenta
$|\vec {\bf p}|=\frac{2\pi}{L}|\vec{\bf n}|$.
Some smart readers may suspect that the dispersions 
$\sqrt{M_N^2+|\frac{2\pi}{L}\vec{\bf n}|^2}$ and 
$\sqrt{M_K^2+|\frac{2\pi}{L}\vec{\bf n}|^2}$
may be affected by the uncontrollable finite-size effects
due to the finite sizes of N and K,
and no longer valid.
In order to make sure of the small implicit artifacts
especially with $|\vec{\bf n}|=1$,
which we are mainly interested in,
we also calculate the  
sum $E_N^{\rm |\vec{\bf n}|=1} + E_K^{\rm |\vec{\bf n}|=1}$
of energies of nucleon $E_N^{\rm |\vec{\bf n}|=1}$
and Kaon $E_K^{\rm |\vec{\bf n}|=1}$ 
with the smallest non-zero lattice momentum 
$|\vec {\bf p}|=\frac{2\pi}{L}|\vec{\bf n}|=2\pi/L$.
We extract $E_N^{\rm \vec{\bf n}}$ and $E_K^{\rm \vec{\bf n}}$
from the correlators
$\sum_{\bf \vec x}e^{i\frac{2\pi}{L}{\rm \vec{\bf n}}\cdot{\bf \vec x}}
\langle N({\bf \vec x},t+t_{\rm src}) \overline{N}(0,t_{\rm src})\rangle$
and
$\sum_{\bf \vec x}e^{i\frac{2\pi}{L}{\rm \vec{\bf n}}\cdot{\bf \vec x}}
\langle K({\bf \vec x},t+t_{\rm src}) \overline{K}(0,t_{\rm src})\rangle$.
These results are denoted by the open squares in Fig.~\ref{negativeGSES}
as the sum $E_N^{\rm |\vec{\bf n}|=1} + E_K^{\rm |\vec{\bf n}|=1}$.
The upper lines in Fig.~\ref{negativeGSES}
show $E_{NK}^{\rm |\vec{\bf n}|=1}\equiv
\sqrt{M_N^2+(2\pi/L)^2}+\sqrt{M_K^2+(2\pi/L)^2}$.
The deviations of
$E_N^{\rm |\vec{\bf n}|=1} + E_K^{\rm |\vec{\bf n}|=1}$
from $E_{NK}^{\rm |\vec{\bf n}|=1}$ are very small,
which implies the smallness of the implicit finite-size artifacts.
Therefore, provided that the interaction between nucleon and Kaon is weak,
which we assume throughout the present analysis,
the naive expectation for the 2nd-lowest NK scattering states 
denoted by the upper line in Fig.~\ref{negativeGSES}
would be able to follow the energies of the 2nd-lowest NK scattering states
even on the $L=8$ lattices in our setup.

\subsection{The volume dependence of the 2nd-lowest state energy}
\label{Negative2-2}

\begin{figure}[h]
\begin{center}
\includegraphics[scale=0.46]{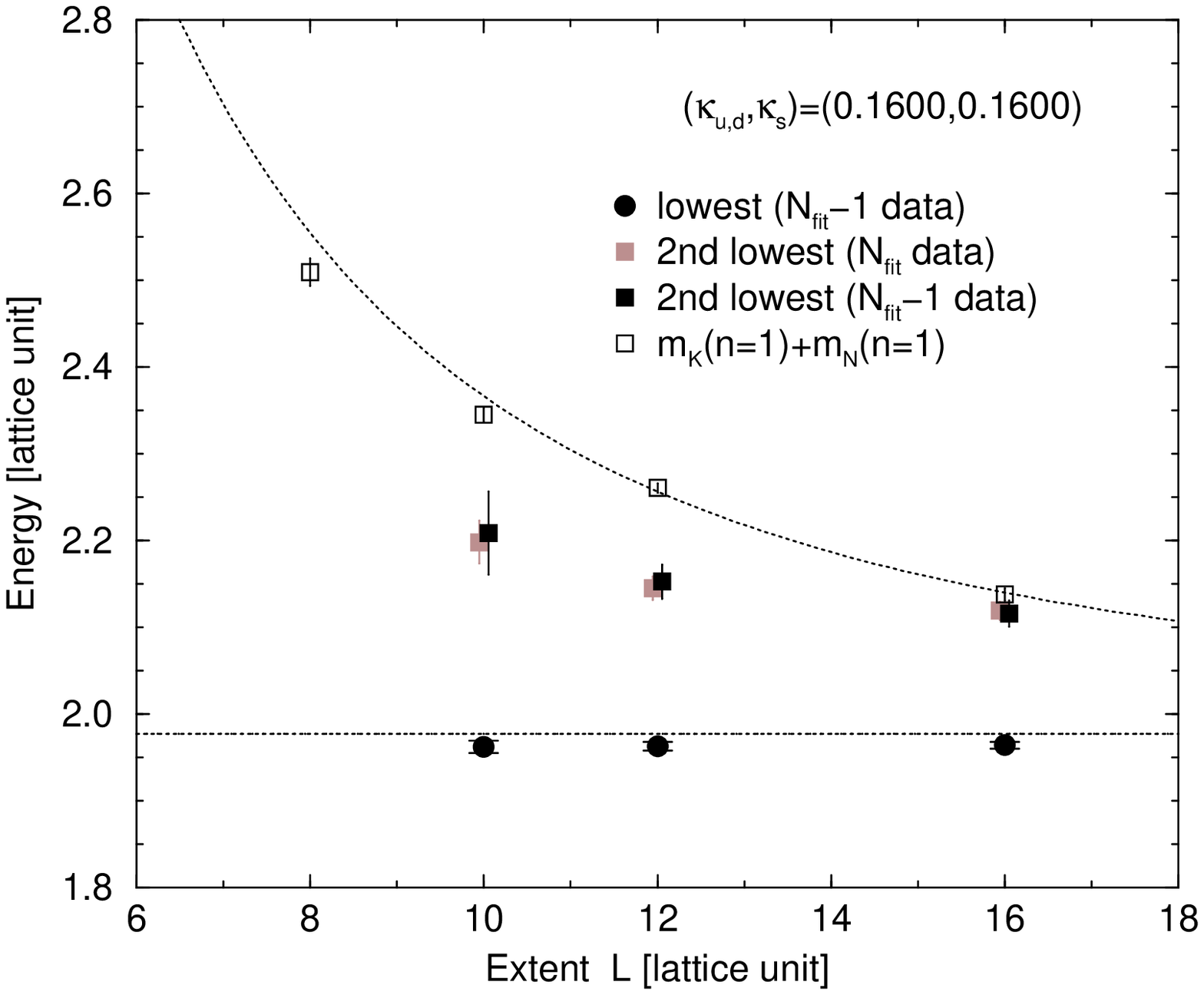}\\
\includegraphics[scale=0.46]{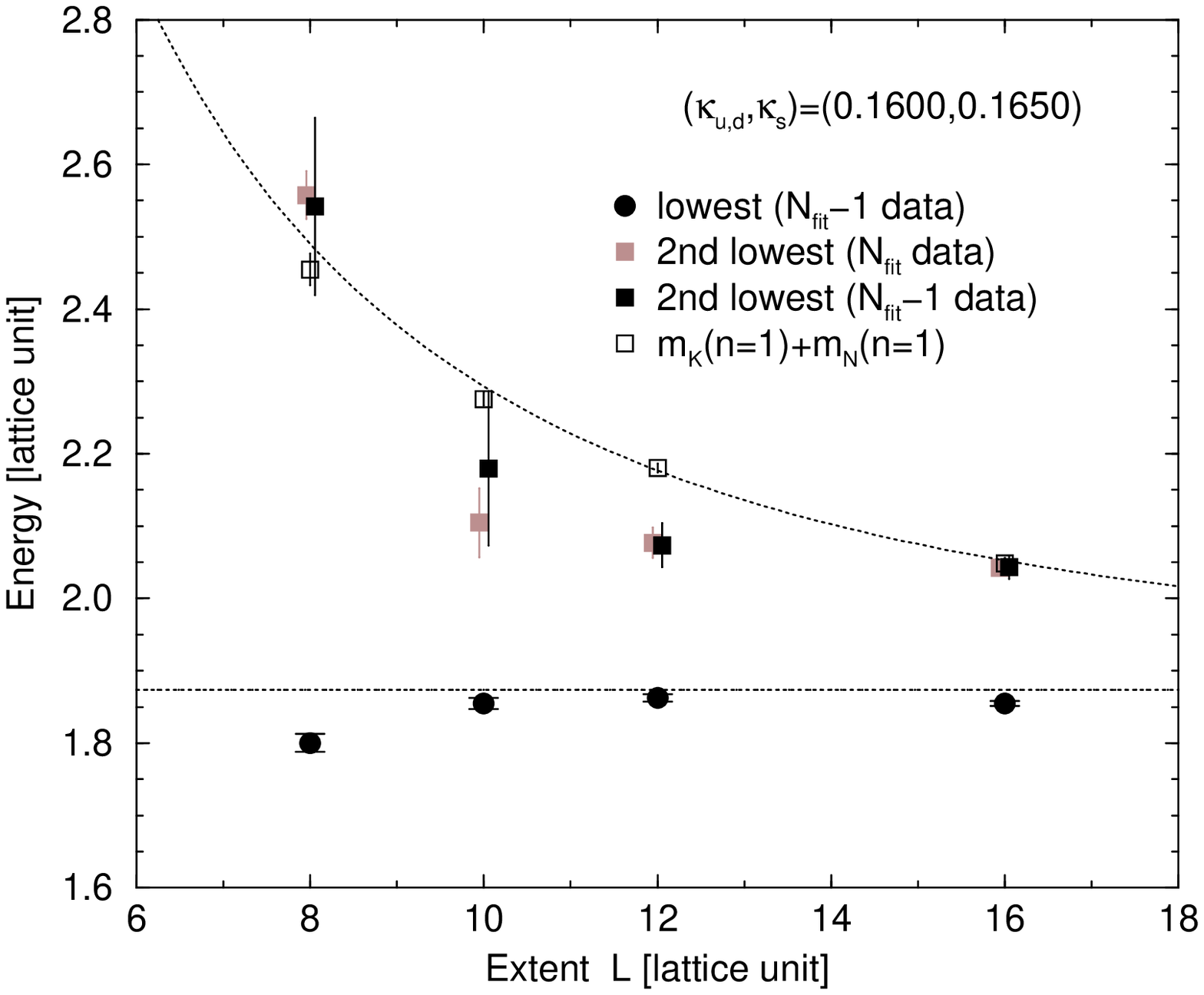}
\includegraphics[scale=0.46]{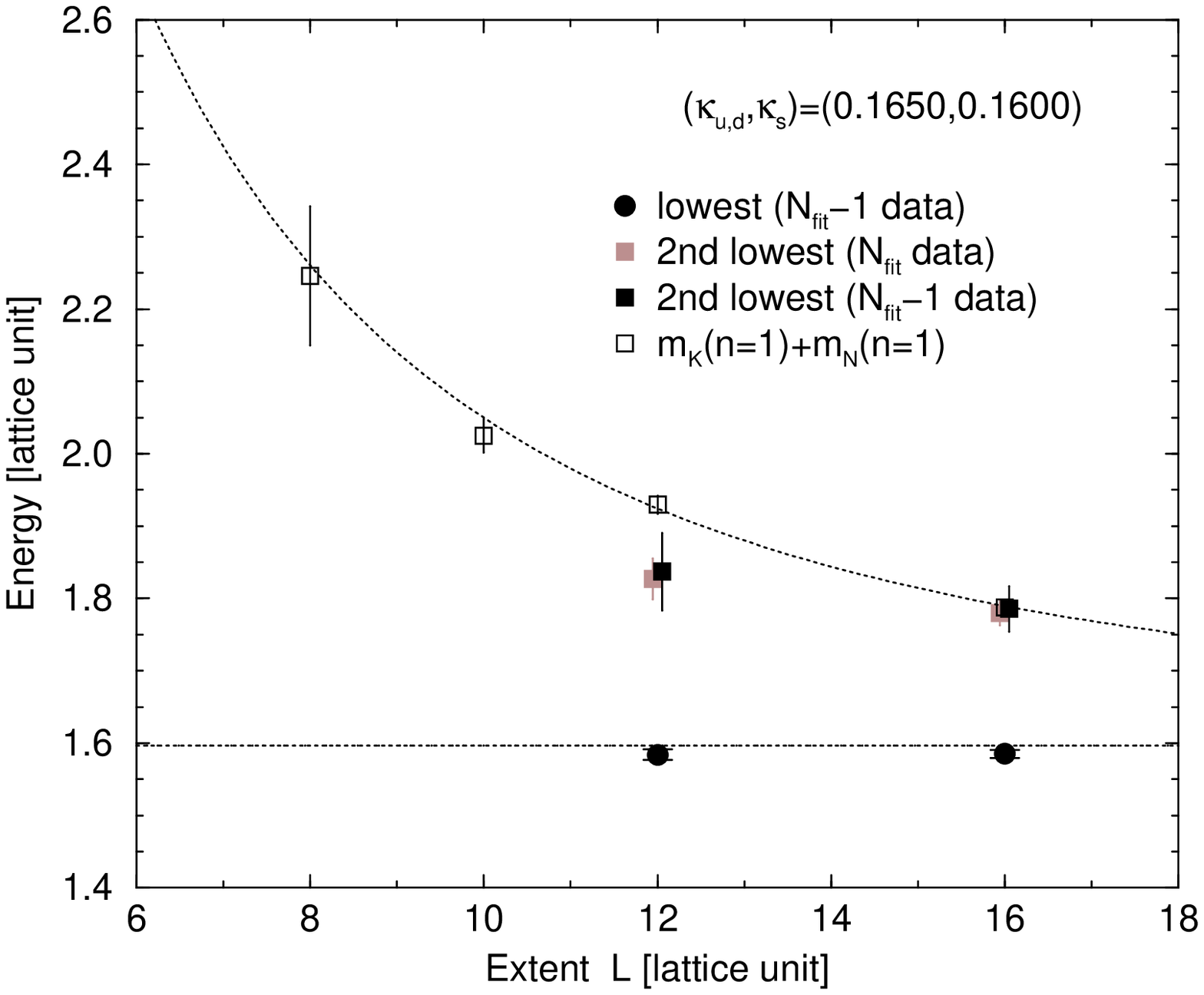}\\
\includegraphics[scale=0.46]{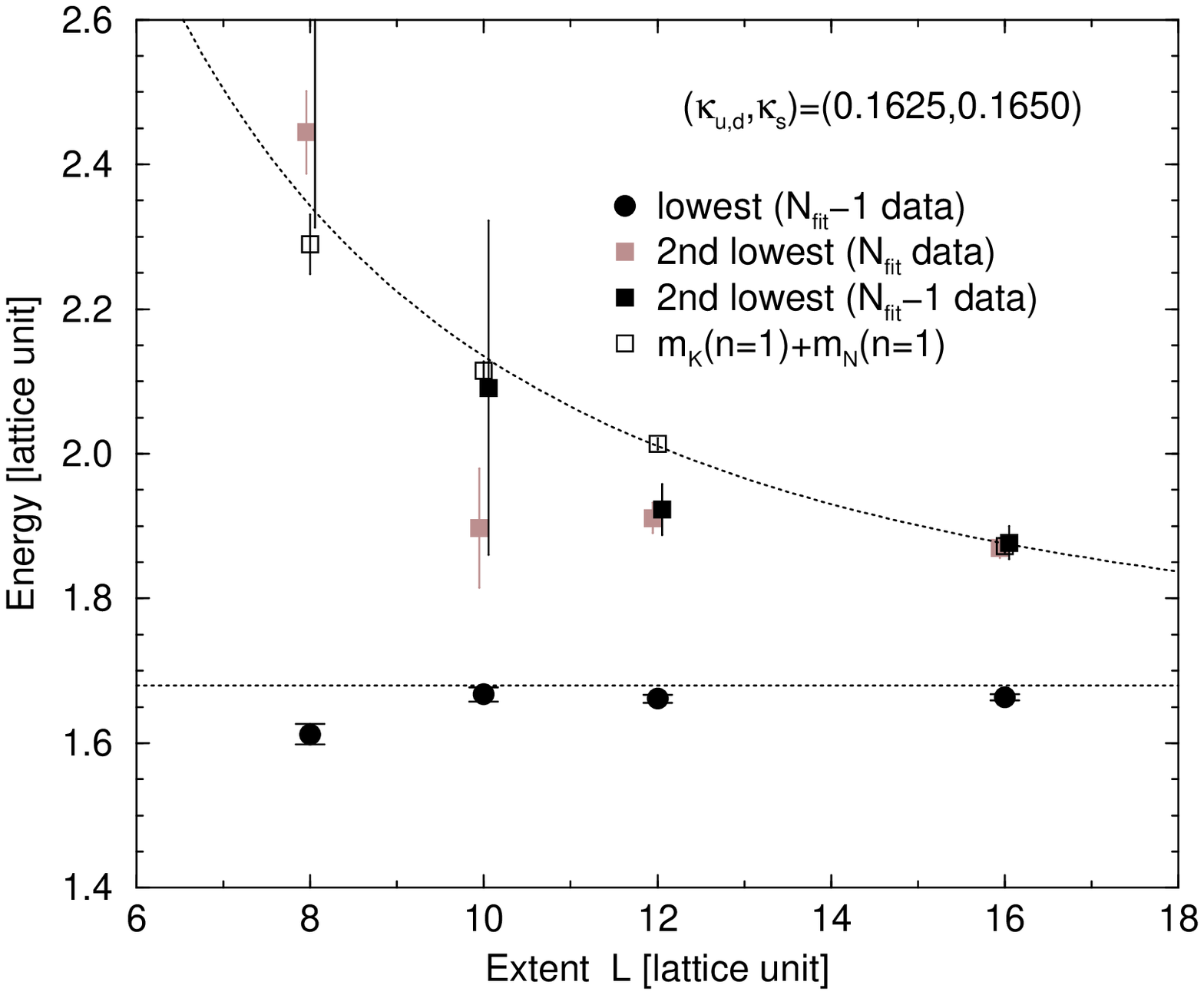}
\includegraphics[scale=0.46]{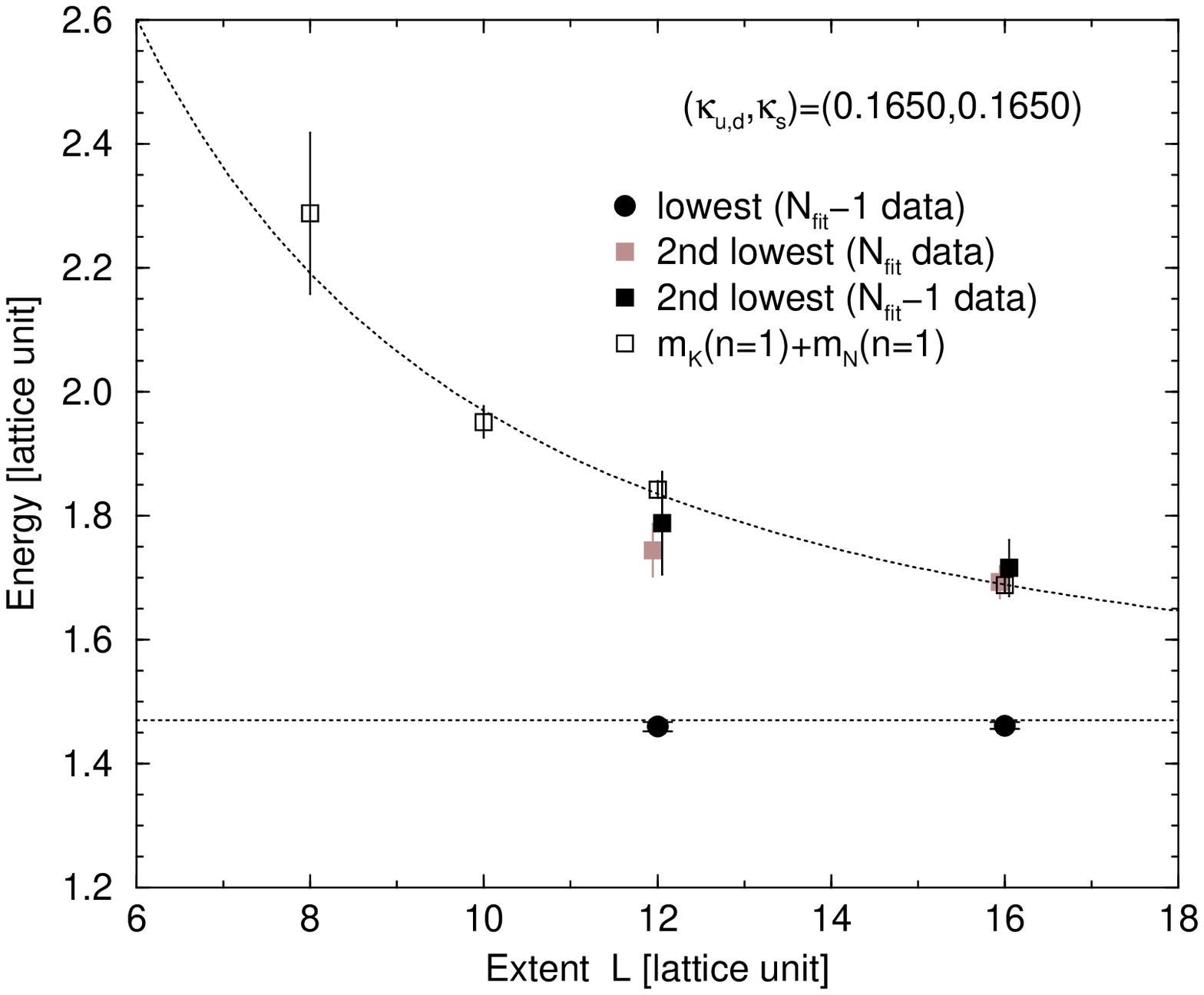}
\end{center}
\caption{\label{negativeGSES}
The black (gray) filled-squares denote
the lattice QCD data of the 2nd-lowest state in
$(I,J^P)=(0,\frac12^-)$ channel
extracted with $N_{\rm fit}-1$ ($N_{\rm fit}$) data 
plotted against the lattice extent $L$.
The filled circles represent 
the lattice QCD data $E_0^-$ of the lowest state in
$(I,J^P)=(0,\frac12^-)$ channel.
The open symbols are the sum 
$E_N^{\rm |\vec{\bf n}|=1} + E_K^{\rm |\vec{\bf n}|=1}$
of energies of nucleon $E_N^{\rm |\vec{\bf n}|=1}$
and Kaon $E_K^{\rm |\vec{\bf n}|=1}$ with the
smallest lattice momentum 
$|\vec {\bf p}|=\frac{2\pi}{L}|\vec{\bf n}|=2\pi/L$.
The upper line represents
$\sqrt{M_N^2+|{\bf p}|^2}+\sqrt{M_K^2+|{\bf p}|^2}$
with $|{\bf p}|=2\pi/L$ the smallest relative momentum
on the lattice.
The lower line represent the simple sum $M_N+M_K$
of the masses of nucleon $M_N$ and Kaon $M_K$.
We adopt the central values of $M_N$ and $M_K$ 
obtained on the largest lattice to draw the two lines.
}
\end{figure}

We compare the lattice data $E_1^-$ with
the expected behaviors $E_{NK}^{\rm |\vec{\bf n}|=1}$
for the 2nd-lowest NK scattering states.
The filled squares in Fig.~\ref{negativeGSES}
denote $E_1^-$, the 2nd-lowest-state energies in this channel.
The black and gray symbols are the lattice data obtained by the fits with 
$N_{\rm fit}-1$ and $N_{\rm fit}$ time slices, respectively 
(see the criterion.3 in the Sec.~\ref{Data}).
The upper line shows 
the expected energy-dependence on $V$
of the 2nd-lowest NK scattering state
$E_{NK}^{\rm |\vec{\bf n}|=1}\equiv
\sqrt{M_N^2+(2\pi/L)^2}+\sqrt{M_K^2+(2\pi/L)^2}$ estimated
with the next-smallest relative momentum between N and K,
and with the masses $M_K$ and $M_N$ extracted on the $L=16$ lattices.
Although the lattice QCD data $E_1^-$ and 
the expected lines $E_{NK}^{\rm|\vec{\bf n}|=1}$
almost coincide with each other on the $L=16$ lattices, 
which one may take as the characteristics of
the 2nd-lowest scattering state,
the data $E_1^-$ 
do not follow $E_{NK}^{\rm|\vec{\bf n}|=1}$
in the smaller lattices.
(At the smallest lattices with $L=8$ (1.4 fm) in the physical unit, 
some results apparently coincide with each other
again. However we consider that the volume with 
$L\sim1.4$ fm is too small for the pentaquarks;
it is difficult to tell which is the origin of the coincidence,
uncontrollable finite volume effects of the pentaquarks
or expected volume dependence of the 2nd-lowest NK scattering state.)
Especially when the quarks are heavy,
composite particles will be rather compact and we expect smaller 
finite volume effects besides those arising from the lattice momenta 
$\vec{\bf p}=\frac{2\pi}{L}\vec{\bf n}$.
Moreover the statistical errors are also well controlled for the 
heavy quarks.
Thus, the significant deviations in $1.5\lesssim L\lesssim 3$ fm 
with the combination of the heavy quarks, 
such as $(\kappa_{u,d},\kappa_s)=(0.1600,0.1600)$,
are reliable and 
the obtained states are difficult to explain as the NK scattering states.

Therefore one can understand this behavior with the view
that this state is a resonance state rather than a scattering state.
In fact, while the data with the lighter quarks have rather strong volume
dependences which can be considered to arise due to the finite size of a
resonance state, the lattice data exhibit almost no volume dependence
with the combination of the heavy quarks
especially in $1.5\lesssim L\lesssim 3$ fm,
which can be regarded as the characteristic of resonance states.

\section{The volume dependence of the spectral weight}
\label{Negative2-3}

For further confirmation, we investigate 
the volume dependence of
the spectral weight~\cite{Metal04}.
As mentioned in Sec.~\ref{Formalism},
the correlation function $\langle O(T)O^\dagger(0)\rangle$
can be expanded as 
$\langle O(T)O^\dagger(0)\rangle 
=\sum W_ie^{-E_iT}$.
The spectral weight of the $i$-th state
is defined as the coefficient $W_i$
corresponding to the overlap of the operator $O(t)$ 
with the $i$-th excited-state.
The normalization conditions of the field $\psi$ and the
states $|i\rangle$ give rise to the volume dependence
of the weight factors $W_i$
in accordance with the types of the operators $O(t)$. 

For example, 
in the case when a correlation function 
is constructed from a point-source
and a zero-momentum point-sink, as
$\sum_{\vec{\bf x}}\langle\Theta(\vec{{\bf x}},T+t_{\rm src})
\overline{\Theta}(\vec{{\bf 0}},t_{\rm src})\rangle$,
the weight factor $W_i$
takes an almost constant value
if $|i\rangle$ is the resonance state where 
the wave function is localized.
If the state $|i\rangle$ is a two-particle state, the 
situation is more complicated. Nevertheless if there is
almost no interaction between the two particles, the 
weight factor is expected to be proportional to $\frac{1}{V}$.
In the case when a source is a wall operator 
$\overline{\Theta}_{\rm wall}(t_{\rm src})$ as taken in this work, 
a definite volume dependence of $W_i$ is not known.
Therefore, we re-examine the lowest state and the 2nd-lowest state
in $(I,J^P)=(0,\frac12^-)$ channel
using the locally-smeared source  
$\overline{\Theta^{2}}_{\rm smear}(t_{\rm src})\equiv 
\sum_{{\bf \vec x}\in \Gamma}\overline{\Theta^2}({\bf \vec x},t_{\rm src})$
with $\Gamma\equiv {(\{0,3\},\{0,3\},\{0,3\})}$,
which we introduce to partially enhance the ground-state overlap.
Since smeared operators, whose typical sizes are much smaller than
the total volume, can be regarded as local operators,
we can discriminate the states using the locally-smeared operators
as in the case of point operators.
(We also investigated the weight factor using the point source.
The results are consistent with those
obtained using the locally-smeared one, but are rather noisy.)
We adopt the hopping parameter $(\kappa_{u,d},\kappa_s)=(0.1600,0.1600)$
and additionally employ $14^3\times 24$ lattice for this aim.
\begin{figure}[h]
\begin{center}
\includegraphics[scale=0.4]{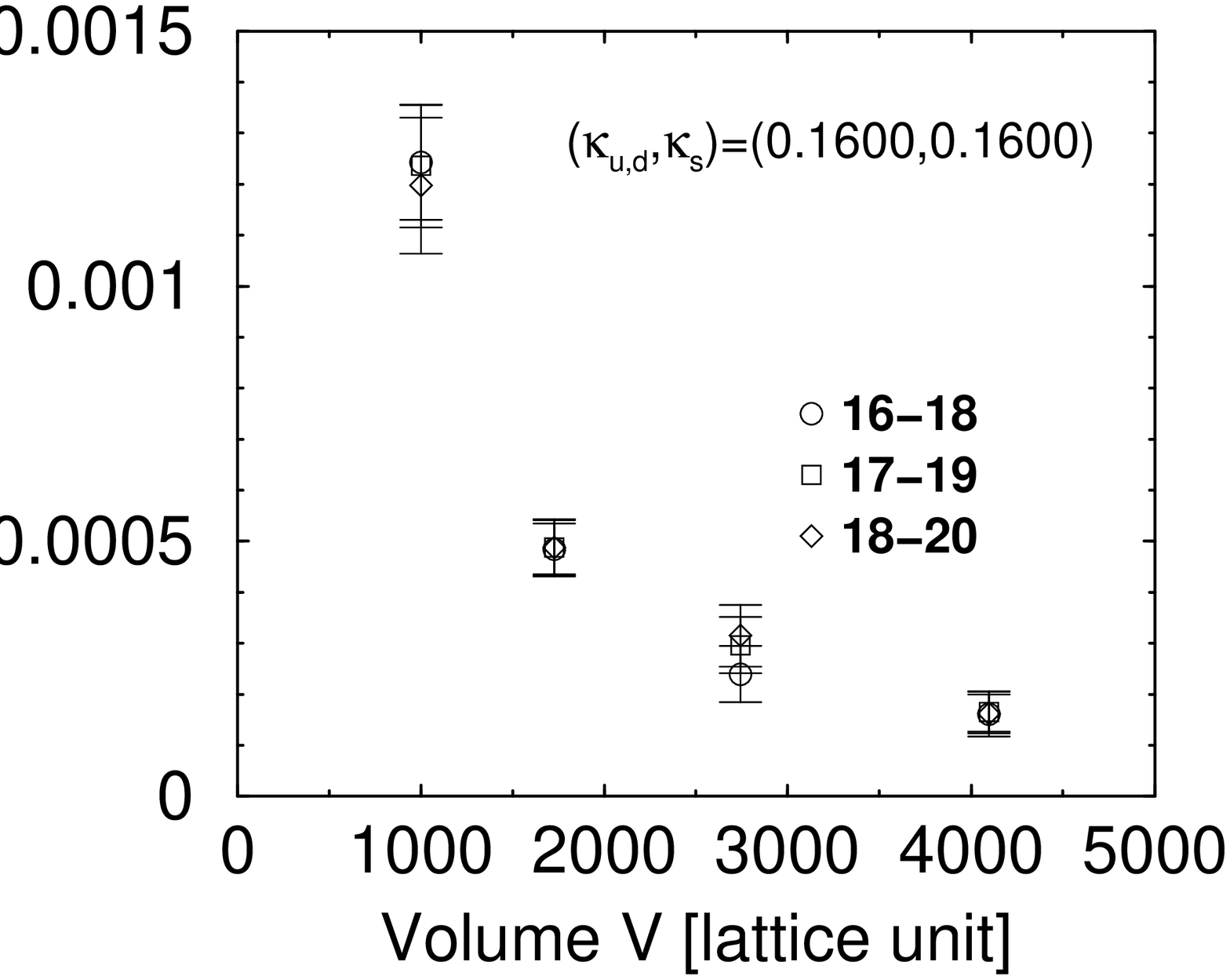}
\hspace{1cm}
\includegraphics[scale=0.4]{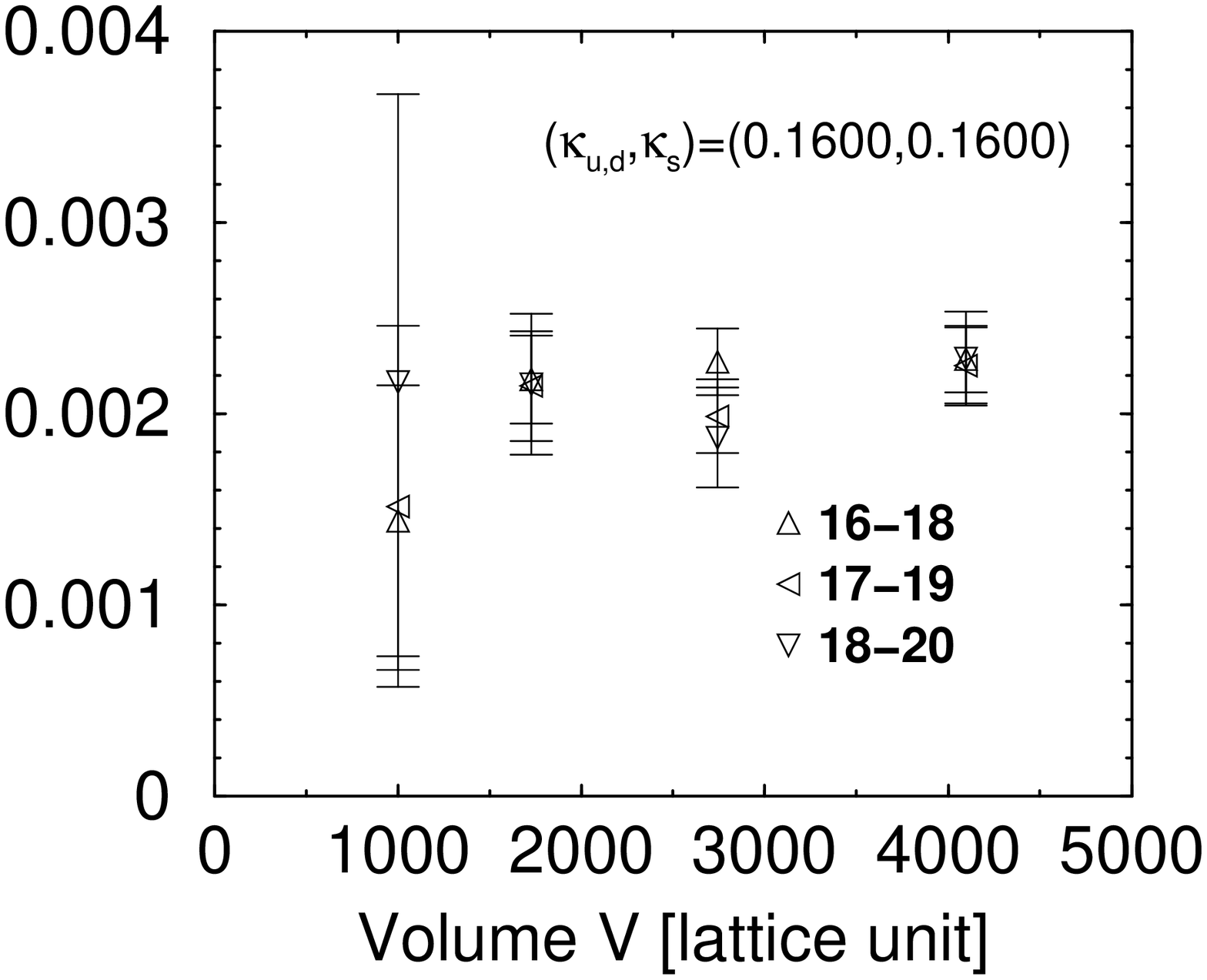}
\end{center}
\caption{\label{weightfactor}
The spectral weight factors defined in Sec.~\ref{Negative2-3} are plotted
against the lattice volume $V$.
The left figure shows $W_0$ for the lowest state in $(I,J^P)=(0,\frac12^-)$
channel and the right figure shows
$W_1$ for the 2nd-lowest state.
In the case when the weight factor $W_i$ 
for the $i$-th state $|i\rangle$
in a point-point correlator
shows no volume dependence, $|i\rangle$ is likely to be a resonance state.
On the contrary, when the $i$-th state $|i\rangle$ is a two-particle
 state,
$W_i$ scales according to $1/V$.
}
\end{figure}

We extract $W_0$ and $W_1$ using the two-exponential fit
as $\sum_{\vec{\bf x}}\langle\Theta^2(\vec{{\bf x}},T+t_{\rm src})
\overline{\Theta^{2}}_{\rm smear}(\vec{{\bf 0}},t_{\rm src})\rangle =W_0e^{-V_0T}+W_1e^{-V_1T}$.
The fit with four free parameters $W_0$, $W_1$, $V_0$ and $V_1$
is however unstable and therefore we
fix the exponents using the obtained values $E_0^-$ and $E_1^-$.
The weight factors $W_0$ and $W_1$ are then obtained
through the two-parameter fit 
as $\sum_{\vec{\bf x}}\langle\Theta^2(\vec{{\bf x}},T+t_{\rm src})
\overline{\Theta^{2}}_{\rm smear}(\vec{{\bf 0}},t_{\rm src})\rangle =W_0e^{-E_0^-T}+W_1e^{-E_1^-T}$
in as large $t$ range ($T_{\rm min}\leq T+t_{\rm src} \leq T_{\rm max}$) as possible
in order to avoid the contaminations of the higher excited-states
than the 2nd-excited state (3rd-lowest state),
which will bring about the instability of the fitted parameters.
The fluctuations of $E_0^-$ and $E_1^-$
are taken into account through the Jackknife error estimation.
Fig.~\ref{weightfactor} includes all the results
with the various fit range as 
($T_{\rm min}$,$T_{\rm max}$)=
(16,18),(17,19),(18,20) to see the fit-range dependence.
Though the results have some fit-range dependences,
the global behaviors are almost the same among the three.

The left figure in Fig.~\ref{weightfactor} shows
the weight factor $W_0$
of the lowest state in $(I,J^P)=(0,\frac12^-)$
channel against the lattice volume $V$.
We find that $W_0$ decreases as $V$ increases
and that the dependence on $V$ is consistent with $\frac{1}{V}$,
which is expected in the case of two-particle states.
It is again confirmed that 
the lowest state in this channel is the NK scattering state
with the relative momentum $|{\bf p}|=0$.
Next, we plot the weight factor $W_1$ of the 2nd
lowest state in the right figure.
In this figure, almost no volume dependence against $V$ is found,
which is the characteristic of the state
in which the relative wave function is localized.
(In Appendix, we try another prescription
to estimate the volume dependences of the spectral weights,
which requires no multi-exponential fit.)
This result can be considered as one of the evidences
of a resonance state lying slightly above the NK threshold.

To summarize this section,
the volume dependence analysis
of the eigenenergies and the weight factors of the 2nd-lowest state
in $(I,J^P)=(0,\frac12^-)$ channel
suggests the existence of a resonance state.
Although there remain the statistical errors and the possible
finite-volume artifacts, the data can be consistently accounted for
assuming the 2nd-lowest state to be different from ordinary 
scattering states. If the 2nd-lowest state were
an ordinary scattering state, one had to assume a large systematic 
errors for heavier quarks which is hard to understand consistently. 

\section{$(I,J^P)=(0,\frac12^+)$ channel}
\label{Positive}

\begin{figure}[htb]
\begin{center}
\includegraphics[scale=0.34]{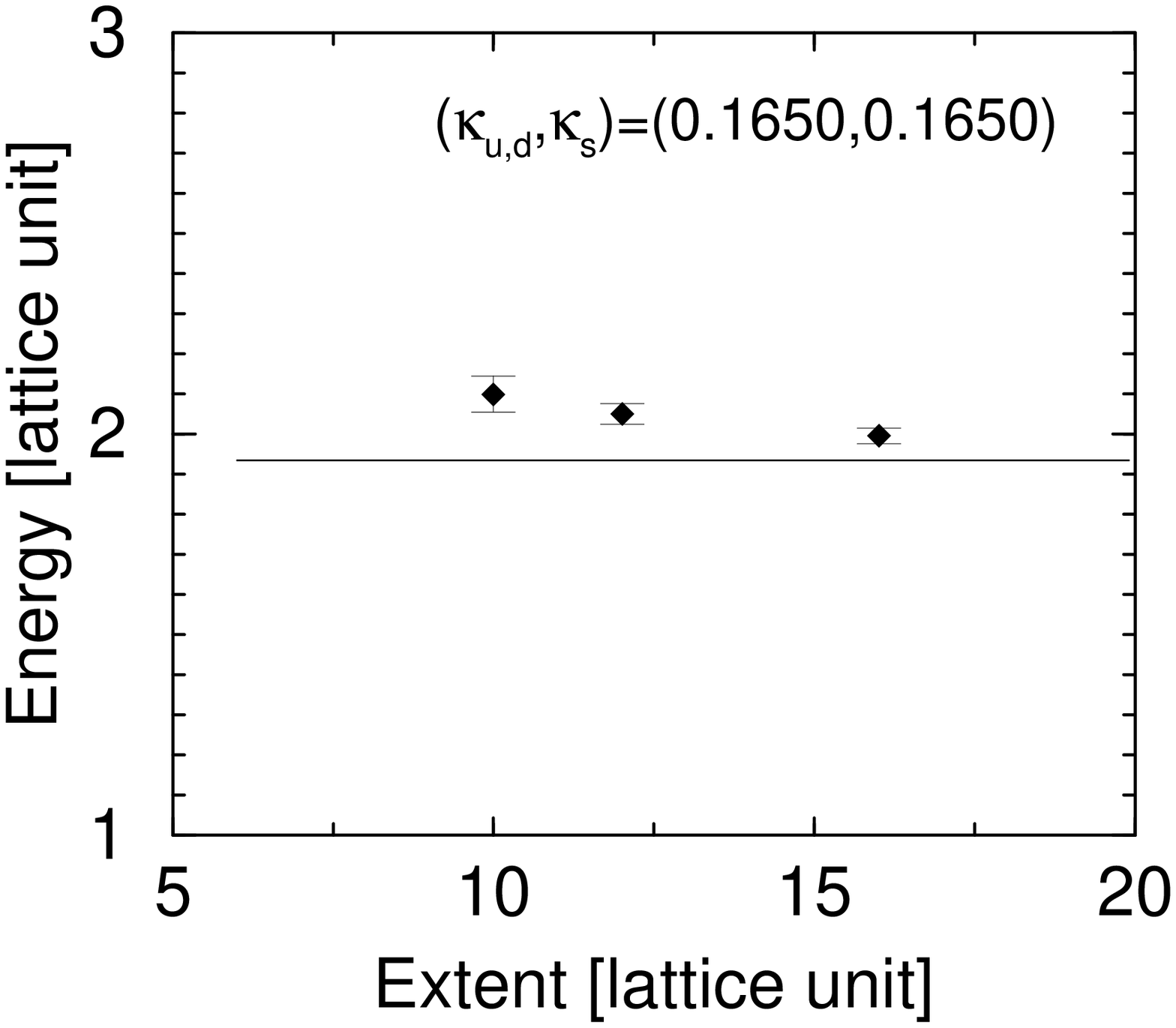}
\includegraphics[scale=0.34]{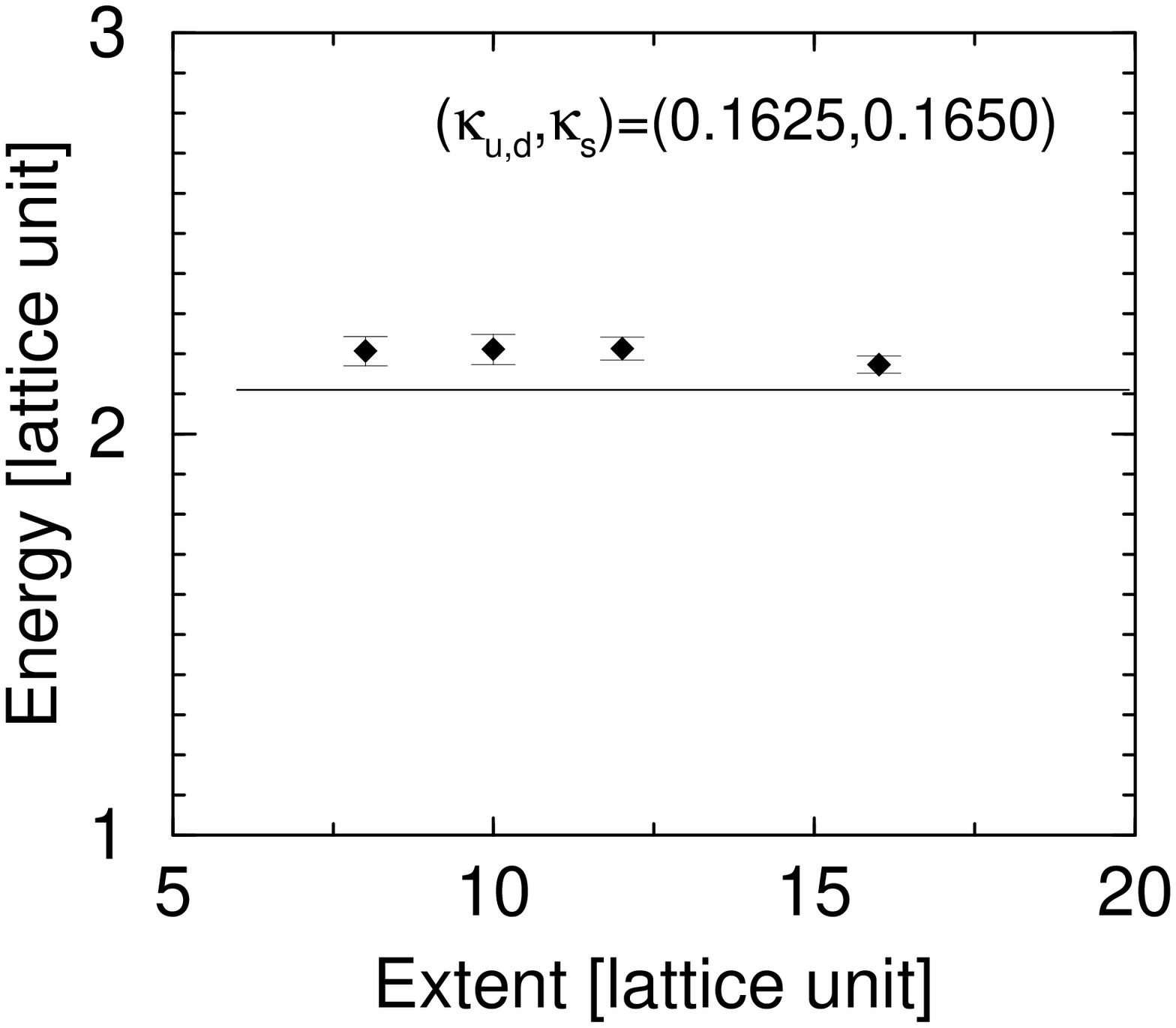}
\includegraphics[scale=0.34]{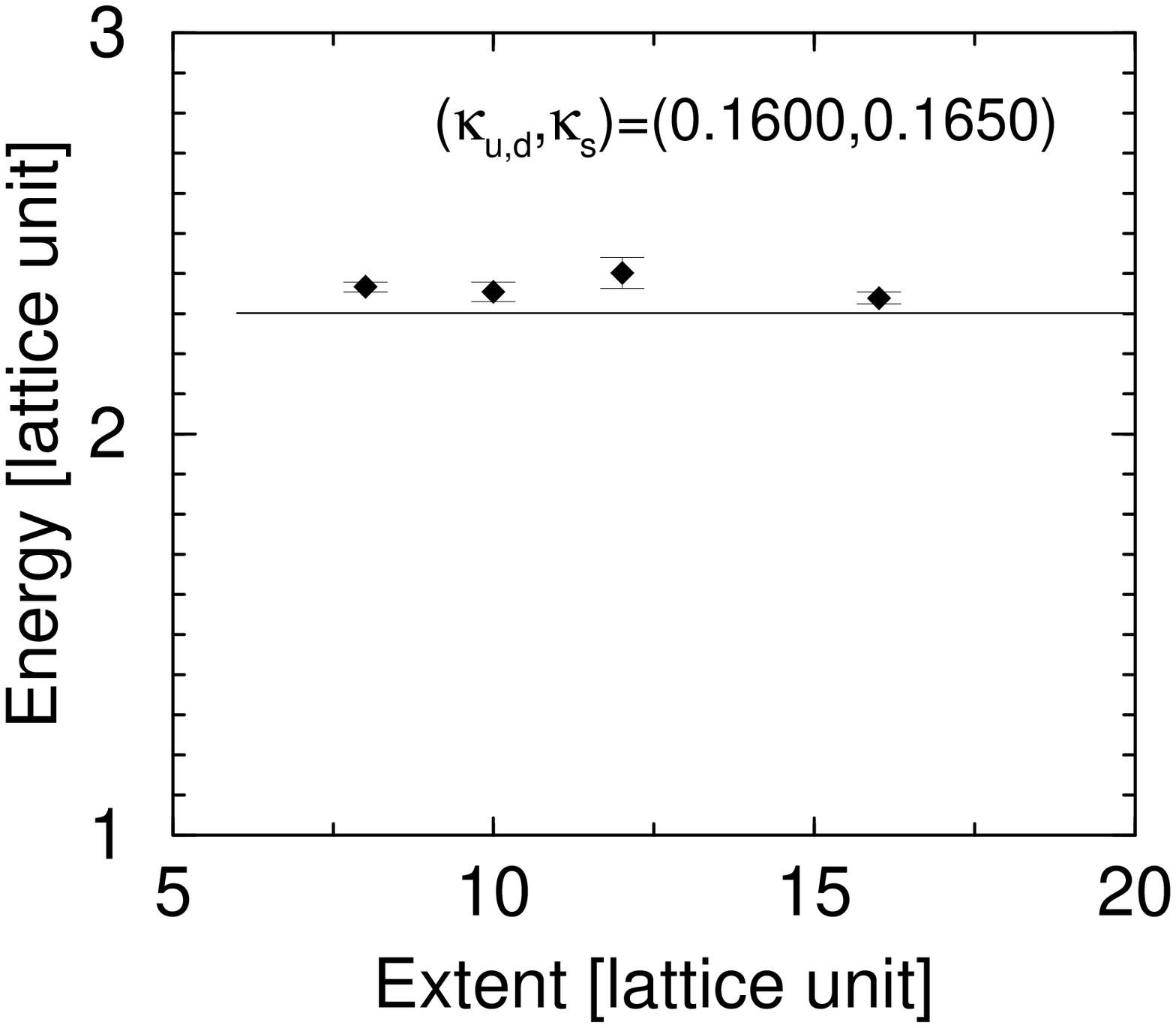}
\includegraphics[scale=0.34]{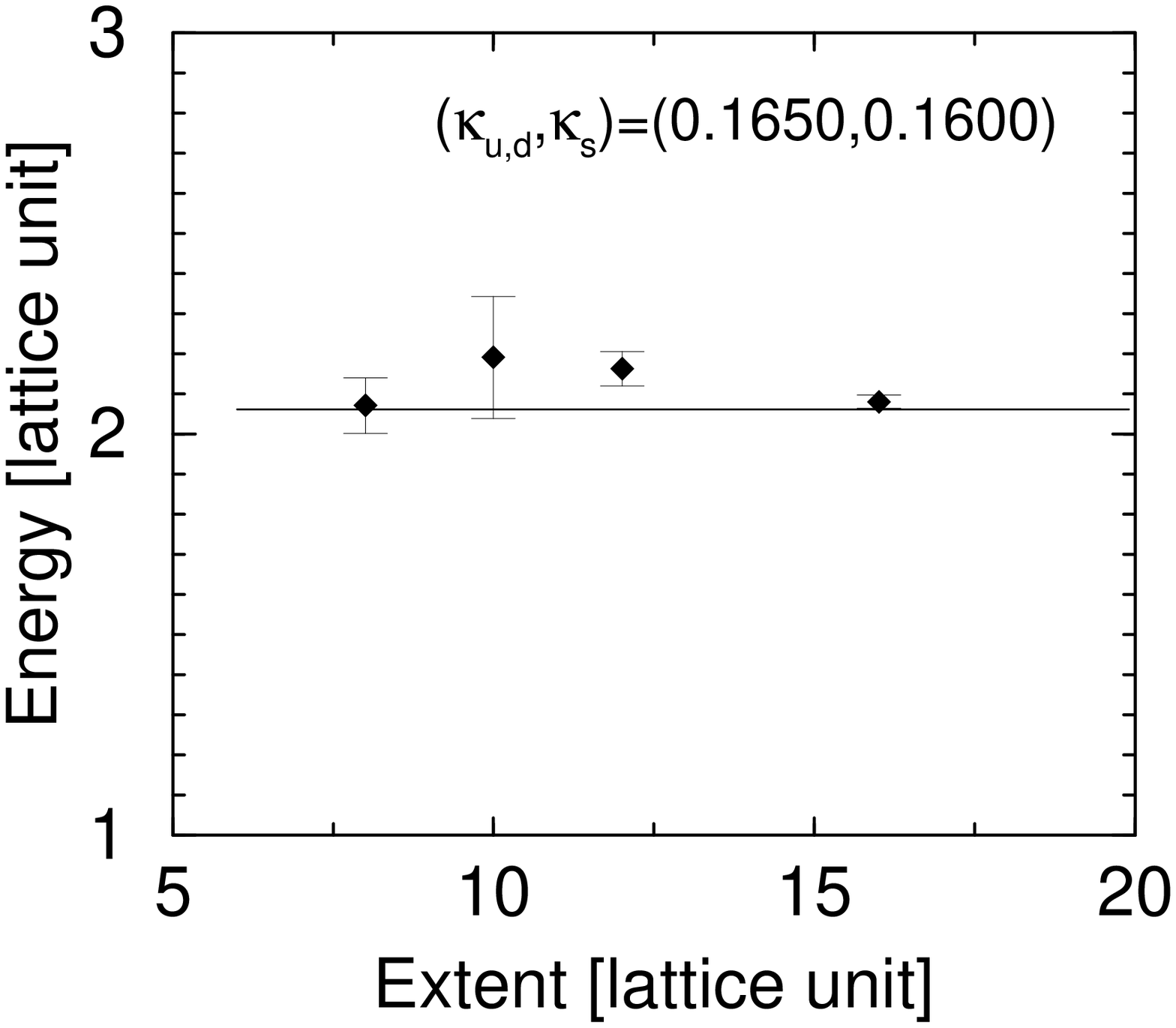}
\includegraphics[scale=0.34]{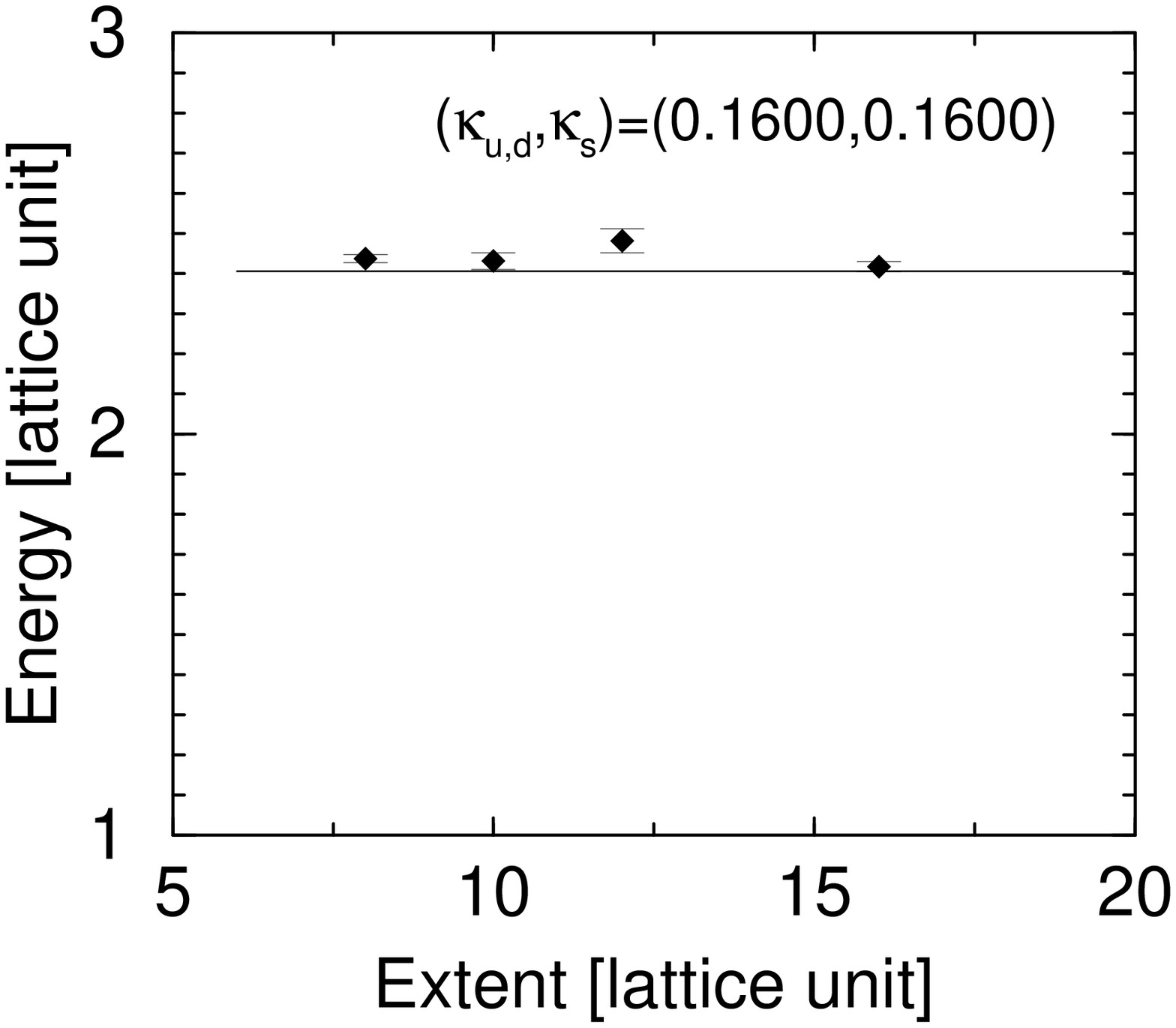}
\end{center}
\caption{\label{positiveSS}
The lattice QCD data in the $(I,J^P)=(0,\frac12^+)$ channel
are plotted against the lattice extent $L$.
The solid line denotes the simple sum $M_{N^*}+M_K$ of
the masses of the lowest-state negative-parity nucleon $M_{N^*}$
and Kaon $M_K$ obtained with the largest lattice.
}
\end{figure}
In the same way as $(I,J^P)=(0,\frac12^-)$ channel,
we have attempted to diagonalize the correlation matrix
in $(I,J^P)=(0,\frac12^+)$ channel
using the wall-sources $\overline{\Theta}_{\rm wall}(t)$
and the zero-momentum point-sinks $\sum_{\vec{\bf x}}\Theta(\vec{\bf x},t)$.
In this channel, the diagonalization is rather unstable and we
find only one state except for tiny contributions of possible other states.
We plot the lattice data in Fig.~\ref{positiveSS}.
One finds that they have almost no volume dependence
and that they coincide with the solid line
which represents the simple sum 
$M_{N^*}+M_K$ of $M_{N^*}$ and $M_K$,
with $M_{N^*}$ the mass of the ground state
of the {\it negative-parity} nucleon.
From this fact,
the state we observe is concluded to be the $N^*$-$K$ scattering state
with the relative momentum $|{\bf p}|=0$.
It may sound strange
because the p-wave state of N and K with the relative
momentum $|{\bf p}|=2\pi/L$
should be lighter than
the $N^*$-$K$ scattering state with the relative momentum $|{\bf p}|=0$;
this lighter state is missing in our analysis.
This failure would be due to the wall-like operator $\Theta_{\rm wall}(t)$.
The fact that the wall operator $\Theta_{\rm wall}(t)$
is constructed by the spatially spread quark fields 
$\sum_{\vec{\bf x}}q({\rm x})$ 
with zero momentum
may leads to the large overlaps 
with the NK scattering state with zero relative momentum.
The relation between operators and overlap coefficients
is an interesting problem and is to be explored in detail
for further studies.
Anyway, the strong dependence on the choice of operators
suggests that it is needed to try 
various types of operators before giving the final conclusion.

\begin{figure}[h]
\begin{center}
\includegraphics[scale=0.4]{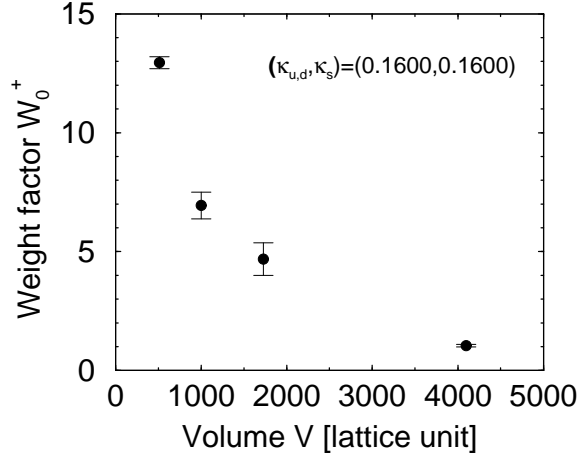}
\end{center}
\caption{\label{weightfactor2}
The spectral weight factor of the extracted state
in $(I,J^P)=(0,\frac12^+)$ channel
with the hopping parameters
$(\kappa_{u,d},\kappa_s)=(0.1600,0.1600)$
is plotted against the lattice volume $V$.
We note here that the 
${1}/{V}$-like volume dependence {\it is not} always
the characteristics of scattering states
when we do not adopt point-point correlators as is the present case.
To conclude from this dependence,
we need to determine the precise volume dependences of the weight factors
in wall-point correlators.
}
\end{figure}

Before closing this section, we show the spectral weight $W_0^+$
obtained by the fit using the form 
$\langle \Theta^1(t+t_{\rm src})\overline{\Theta^1}_{\rm wall}(t_{\rm src})\rangle =W_0^+e^{-E_0^+t}$
in Fig.~\ref{weightfactor2}.
Although 
one sees the $\frac{1}{V}$-like volume dependence in Fig.~\ref{weightfactor2},
one can conclude nothing only from this behavior
unless the precise volume dependences of the weight factors
in wall-point correlators are estimated.

\section{Discussion}\label{Discussion}

\subsection{Operator dependence}

We here mention the operator dependences in $(I,J^P)=(0,\frac12^-)$ channel.
As is seen in Sec.~\ref{Positive},
the overlap factors with states strongly depend on the choice of operators.
We survey the effective masses of the five correlators;
$\sum_{\rm \vec x}\langle \Theta^1({\rm\vec x},T+t_{\rm src}) 
\overline{\Theta^1}_{\rm wall}(t_{\rm src})\rangle$,
$\sum_{\rm \vec x}\langle \Theta^2({\rm\vec x},T+t_{\rm src}) 
\overline{\Theta^2}_{\rm wall}(t_{\rm src})\rangle$,
$\sum_{\rm \vec x}\langle \Theta^1({\rm\vec x},T+t_{\rm src}) 
\overline{\Theta^1}(t_{\rm src})\rangle$,
$\sum_{\rm \vec x}\langle \Theta^2({\rm\vec x},T+t_{\rm src})
\overline{\Theta^2}(t_{\rm src})\rangle$
and
$\sum_{\rm \vec x}\langle \Theta^3({\rm\vec x},T+t_{\rm src}) 
\overline{\Theta^3}(t_{\rm src})\rangle$.
The effective mass E(T) is defined as a ratio
between correlators with the temporal separation $T$ and $T+1$,
\begin{equation}
E(T)\equiv \ln \frac{\langle O(T) O(0)^\dagger\rangle}{\langle O(T+1) O(0)^\dagger\rangle},
\end{equation}
which can be expressed in terms of the eigenenergies and spectral
weights as
\begin{equation}
E(T)
=
\ln \frac{\sum_i W_ie^{-E_iT}}{\sum_i W_ie^{-E_i(T+1)}}
\sim
E_0+\frac{W_1}{W_2}e^{-(E_1-E_0)T}+....
\end{equation}
A plateau in E(T) at $E_0$ implies the ground-state dominance in the correlator.
Effective mass plots $E(T)$ are often used to find the range
where correlators show a single-exponential behavior;
the higher excited-state contributions $W_ie^{-E_iT}(i>0)$ are negligible
in comparison with the ground-state component $W_0e^{-E_0T}$.

Here, $\Theta^3$ is an interpolation operator defined as
\begin{equation}
\Theta^3(x)\equiv
\varepsilon^{\rm abc}\varepsilon^{\rm aef}\varepsilon^{\rm bgh}
[u_e(x)Cd_f(x)][u_g(x)C\gamma_5d_h(x)]C\bar s_c(x)
\end{equation}
which has a di-quark structure 
similar to that
proposed by Jaffe and Wilczek~\cite{JW03},
and is also used in Refs.~\cite{S03,IDIOOS04,CH04}.
\begin{figure}[h]
\begin{center}
\includegraphics[scale=0.4]{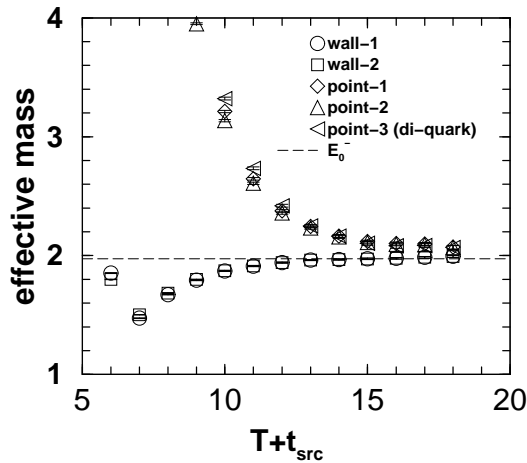}
\end{center}
\caption{\label{variouseffectivemass}
The effective mass plots constructed from
$\sum_{\rm \vec x}\langle \Theta^1({\rm\vec x},T+t_{\rm src}) 
\overline{\Theta^1}_{\rm wall}(t_{\rm src})\rangle$,
$\sum_{\rm \vec x}\langle \Theta^2({\rm\vec x},T+t_{\rm src}) 
\overline{\Theta^2}_{\rm wall}(t_{\rm src})\rangle$,
$\sum_{\rm \vec x}\langle \Theta^1({\rm\vec x},T+t_{\rm src}) 
\overline{\Theta^1}(t_{\rm src})\rangle$,
$\sum_{\rm \vec x}\langle \Theta^2({\rm\vec x},T+t_{\rm src})
\overline{\Theta^2}(t_{\rm src})\rangle$ and
$\sum_{\rm \vec x}\langle \Theta^3({\rm\vec x},T+t_{\rm src}) 
\overline{\Theta^3}(t_{\rm src})\rangle$,
in $(I,J^P)=(0,\frac12^-)$ channel
in $16^3\times 24$ lattice at $\beta =5.7$ employing the hopping parameters 
($\kappa_{u,s},\kappa_s$)=(0.1600,0.1600) are plotted,
along with the dashed line which denotes the lowest-state
energy $E_0^-$.
}
\end{figure}
Fig.~\ref{variouseffectivemass} shows
the effective mass plots constructed from
$\sum_{\rm \vec x}\langle \Theta^1({\rm\vec x},T+t_{\rm src}) 
\overline{\Theta^1}_{\rm wall}(t_{\rm src})\rangle$,
$\sum_{\rm \vec x}\langle \Theta^2({\rm\vec x},T+t_{\rm src}) 
\overline{\Theta^2}_{\rm wall}(t_{\rm src})\rangle$,
$\sum_{\rm \vec x}\langle \Theta^1({\rm\vec x},T+t_{\rm src}) 
\overline{\Theta^1}(t_{\rm src})\rangle$,
$\sum_{\rm \vec x}\langle \Theta^2({\rm\vec x},T+t_{\rm src})
\overline{\Theta^2}(t_{\rm src})\rangle$ and
$\sum_{\rm \vec x}\langle \Theta^3({\rm\vec x},T+t_{\rm src}) 
\overline{\Theta^3}(t_{\rm src})\rangle$.
One can see two typical behaviors in the figure.
One is the line damping from a large value to the energy $E_0^-$ of the lowest NK
scattering state. The other is the one arising upward to $E_0^-$.
Surprisingly,
the differences of the spinor structure or the color structure among the operators
are hardly reflected in the effective mass plots.
The difference is enough to perform the variational method
but seems insufficient for a clear change of the effective mass plots.
Instead, the effective mass plots seem sensitive to 
the spatial distribution of operators.
The upper three symbols are data using the point-point correlators
and the lower two symbols are those from the wall-point correlators.
This means the overlap factor with each state
is controlled mainly by the spatial distribution
rather than the internal structure of operators, except for the overall constant.
The spatially smeared operators seem to have larger overlaps
with the scattering state with the relative momentum $|{\bf p}|=0$.
(One can find that the overlap factor of the wall operator with the observed state
in $(I,J^P)=(0,\frac12^+)$ in Fig.~\ref{weightfactor2}
is 1000 times larger than those of point operators
in Fig.~\ref{weightfactor}.)
One often expects that the overlap with a state
could be enhanced using an operator whose spinor or color structures
resembles the state.
We find however no such tendency in the present analysis.
The insensitivity to the spinor structures
may come from the fact that the KN-type operator ($\Theta^1$)
and the di-quark type operator ($\Theta^3$) are directly related
by a factor of $\gamma_5$ and a Fierz rearrangement~\cite{Metal04}.
Though we have no idea about the mechanism of the insensitivity to the color structure
at present, this insensitivity would have some connection with
the internal color-structure of $\Theta^+$.

The upper three data 
slowly damp and do not reach the lowest energy $E_0^-$ in this $T$ range,
which can be explained in terms of the spectral weight.
As is seen in Fig.~\ref{weightfactor},
$W_0$ is ten-times 
smaller than $W_1$ in the case of the point-point correlator.
Then, the term
$\frac{W_1}{W_0}e^{-(E_1^--E_0^-)t}$ in the effective mass
survives at relatively large $T$.
Hence the effective mass needs larger $T$ to show a plateau at $E_0^-$.
The insensitivity of the overlaps to the internal
structure of operators could be helpful for us:
We have adopted 
two operators whose color and spinor structures are different
from each other.
Although the difference is enough in $(I,J^P)=(0,\frac12^-)$ channel,
it may be insufficient in $(I,J^P)=(0,\frac12^+)$ channel,
which leads to the failure in the diagonalization.
If we use operators 
with spatial distributions different from each other,
it would be more effective in the diagonalization method.

\subsection{Comment on other works}

Here we comment on other works previously published,
especially for the pioneering works  
by Csikor {\it et al.}~\cite{CFKK03} and Sasaki~\cite{S03}.
The simulation condition for the former is rather similar to ours.

Csikor {\it et al.} first 
reported the possible pentaquark state slightly above the NK threshold
in $(I,J^P)=(0,\frac12^-)$ channel in~\cite{CFKK03}.
In Ref.~\cite{CFKK03}, they tried chiral extrapolations and taking the
continuum limit at the quenched level for the possible pentaquark
state. However they used the single-exponential fit analysis for the 
non-lowest state, namely the possible pentaquark state, for the
main results.
It is difficult to justify their result unless 
the coupling of the operators to the lowest NK state
is extremely small.

Sasaki found a double-plateau in the effective mass plot
and identified the 2nd-lowest plateau as the signal of $\Theta^+$.
Unfortunately, we does not find a double-plateau in the present analysis.
The double-plateau-like behavior 
in effective mass plots
can appear only under the 
extreme condition that $W_1$ is much larger than $W_0$.
$W_1$ which is ten times larger than $W_0$ seen in Fig.~\ref{weightfactor}
and the statistical fluctuations
may cause the deviation of the effective mass plot
from the single monotonous line.
In fact, the effective mass plot very slowly approaches $E_0^-$
as $T$ increases in Fig.~\ref{variouseffectivemass}.
He extracted the mass of the next-lowest state
with a single and double exponential fits.
The result do not contradict with ours.

Ref.~\cite{Metal04} reports a lattice QCD study
which adopted the overlap-fermions with the exact chiral symmetry.
The hybrid-boundary method was suggested in Ref.~\cite{IDIOOS04}
and the authors tried to single out the possible resonance state.
In these two studies,
the absence of resonance states with a mass
a few hundred MeV above the NK threshold was concluded.
We have not found the resonance state which coincides 
just with the mass of $\Theta^+$ in the chiral limit.
In this sense, the results in Refs.~\cite{Metal04,IDIOOS04}
are not inconsistent with ours.

\subsection{chiral extrapolation}\label{chiralextrapolation}

We perform chiral extrapolations 
for Kaon, nucleon, NK threshold
(a simple sum of a Kaon mass and a nucleon mass)
and the 2nd-lowest state
in the $(I,J^P)=(0,\frac12^-)$ channel.
We adopt the lattice data with $16^3\times 24$ lattice, the largest
lattice in our analysis.
One can find in Fig.~\ref{negativeGSES}
that the 2nd-lowest state,
which is expected to be a resonance state,
is already affected by the finite volume effects for $L\leq 12$
with the lightest combination of quarks,
and we therefore adopt the largest-lattice data for safety.
We can expect from this fact that
the typical diameter of this resonance is about 2 fm or longer
and that it is desirable
to use larger lattices than, $(2.5\ {\rm fm})^3$
for the analysis of $\Theta^+$.

\begin{figure}[h]
\begin{center}
\includegraphics[scale=0.5]{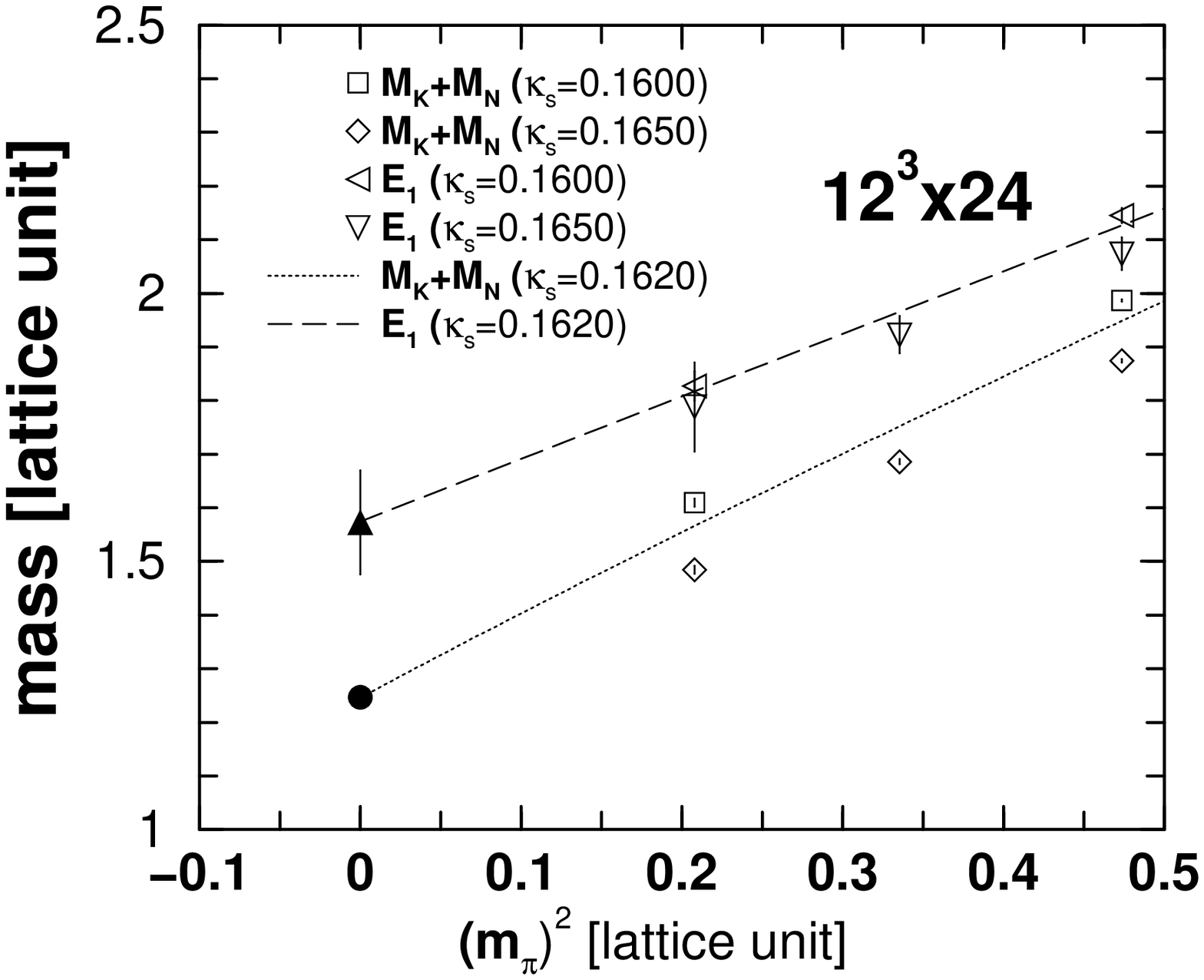}
\includegraphics[scale=0.5]{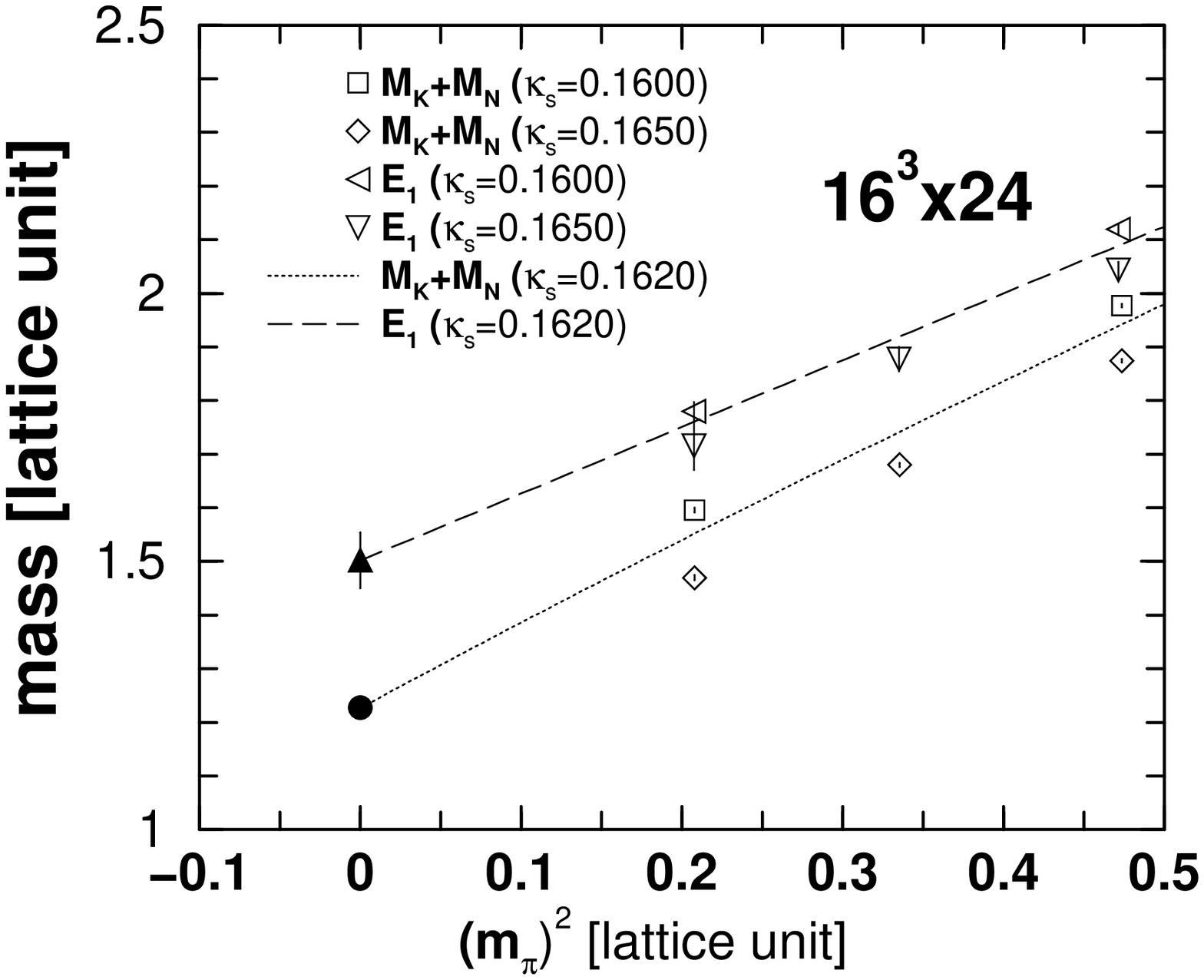}
\end{center}
\caption{\label{chiralextrapolations}
A comparison of the chiral extrapolations in $(I,J^P)=(0,\frac12^-)$ channel
on $12^3\times 24$ (left)  and  $16^3\times 24$ (right) lattices for the 
NK threshold energy $M_K+M_N$ and
the 2nd-lowest state energy $E_1^-$.
$M_K+M_N$ and $E_1^-$ are plotted against $M_{\pi}^2$. 
The filled circles (triangles) denote the energies 
of the NK threshold (2nd-lowest state) in the chiral limit.
$\kappa_s$ is fixed to be $\kappa_s \sim 0.1620$ 
so that the physical Kaon mass
is reproduced in the chiral limit.
}
\end{figure}

In Fig.~\ref{chiralextrapolations},
$M_K+M_N$ and $E_1^-$ 
obtained with each combination of quark masses for $12^3\times 24$ and 
$16^3\times 24$ lattices are plotted against $m_\pi^2$. 
We assume the linear function of quark masses,
$E_{\rm B}(m_{\rm u,d}, m_{\rm s})=
b_{00}+b_{10}m_{\rm u,d}+b_{01}m_{\rm s}$,
for nucleon and the 2nd-lowest state
with $b_{ij}$ free parameters fitted using the five lattice data.
We determine the critical $\kappa$ ($\kappa_c$) by $M_\pi^2$
and fix the $\kappa_s$ 
so that the physical Kaon mass is reproduced in the chiral limit,
using the form for pseudo scalar mesons
$E^2_{\rm PS}(m_{\rm u,d}, m_{\rm s})=
a_{10}m_{\rm u,d}+a_{01}m_{\rm s}$.
The chiral-extrapolated values of
$M_K$, $M_N$, $M_K+M_N$ and $E_1^-$ for $16^3\times 24$
lattice are 0.4274(12), 0.7996(60), 1.227(6) and 1.500(52)
in the lattice unit
and 0.5001(14), 0.9355(70), 1.436(7) and 1.755(61)
in the unit of GeV, respectively. We find that the results for 
$12^3\times 24$ lattice are consistent within errors as shown in 
Fig.~\ref{chiralextrapolations}.

The value of $E_1^-$=1755(61) MeV in the chiral limit is significantly 
larger than the mass of $\Theta^+(1540)$ in the real world.
How can we interpret this deviation? One possibility is the systematic 
errors from the discretization, the chiral extrapolation, or quenching.
Another possibility is that the observed 2nd-lowest state might be a
signal of a resonance state lying higher than $\Theta^+$.
Unfortunately there is no clear explanation at this point. 
Obviously more extensive studies on finer lattices with lighter quark
masses in unquenched QCD are required. However, we can at least 
conclude that {\it our quenched lattice calculations suggest the 
existence of a resonance-like state slightly above the NK threshold 
for the parameter region we have investigated.}

\section{Summary and Future works}\label{Summary}

We have performed the lattice QCD study of the 
$(S,I,J)=(+1,0,\frac12)$ states
on $8^3\times 24$, $10^3\times 24$, $12^3\times 24$
and $16^3\times 24$ lattices at $\beta$=5.7 at the quenched level
with the standard plaquette gauge action and Wilson quark action.
To avoid the possible contaminations originating from the (anti)periodic
boundary condition,
which are peculiar to the pentaquark
and have not been properly noticed in previous studies,
we have adopted the Dirichlet boundary condition in the temporal
direction for the quark field.
With the aim to separate states clearly,
we have adopted two independent operators with $I=0$ and $J^P=\frac12$
so that we can construct a $2\times 2$ correlation matrix.

From the correlation matrix of the operators,
we have successfully obtained the energies of the lowest state and 
the 2nd-lowest state in the $(I,J^P)=(0,\frac12^-)$ channel.
The volume dependence of the energies and spectral weight factors
show that 
the 2nd-lowest state in this channel is likely to be
a resonance state located slightly above the NK threshold
and that
the lowest state is the NK scattering state with the relative momentum
$|{\bf p}|=0$.
As for the $(I,J^P)=(0,\frac12^+)$ channel,
we have observed only one state in the present analysis,
which is likely to be a $N^*K$ scattering state
of the ground state of the negative-parity nucleon $N^*$ and Kaon
with the relative momentum $|{\bf p}|=0$.

We have also investigated the overlaps using five independent operators.
As a result, we have found that
the overlaps seem to be insensitive to the spinor and color
structure of operators
while
the overlaps are mainly controlled by the spatial distributions
of operators, at least for a few low-lying state in this analysis.
For the diagonalization method,
it may be more effective to vary the spatial distributions rather
than the internal structures.

The volume dependence of $E_1^-$ suggests that
this resonance-like state in the $(I,J^P)=(0,\frac12^-)$ channel
is a rather spread object with the radius of about 1 fm or more.
The possibility of a resonance state lying in
$(I,J^P)=(0,\frac12^-)$ channel is desired to be confirmed
by other theoretical studies,
such as quark models, QCD sum rules, string models 
and so on~\cite{O04,J04,P03,EMN04,SDO04,TS04-2,HH05}.
Unfortunately, four quarks {\it uudd} and one antiquark $\bar s$
in $J=\frac12^-$ state
can hardly reproduce the unusually narrow width of $\Theta^+$ 
so far, while the obtained mass in $J^P=\frac12^-$ channel
could be assigned to the observed resonance state~\cite{TS04-2,HH05}.
Hence, $J^P=\frac32^-$ or $J^P=\frac12^+$ states
are favored to reproduce the width in terms of the quark model.
However, there are many unknown problems left so far
such as the internal structure of multi-quark 
hadrons~\cite{J04,JW03,TMNS01,BKST04,OST04,AK04}
or the dynamics of the string/flux-tubes~\cite{BKST04,OST04}.
The discovery of $\Theta^+$
gives us many challenges in the hadron physics
and more detailed theoretical study including the lattice QCD studies 
are awaited.

For further analyses,
a variational analysis using the $3\times 3$ correlation matrix or
larger matrices will be desirable.
The observation of wave functions will be also useful
to distinguish a resonance state from scattering states
and to investigate the internal structures of hadrons.
We can use the lattice QCD calculations in order to estimate
the decay width~\cite{DDDJFMP98} and to study the flux-tube 
dynamics~\cite{BKST04,OST04,C04},
which should give useful inputs for model calculations.

\section*{Acknowledgments}

We acknowledge the Yukawa Institute for Theoretical Physics at Kyoto 
University, where this work was initiated from the discussions during 
the YITP-W-03-21 workshop on ``Multi-quark Hadrons: four, five and more?''. 
T.~T.~T. thanks Dr.~F.~X.~Lee for the useful advice.
T.~U. and T.~O. thank Dr. T.~Yamazaki for the fruitful discussion.
T.~T.~T. and T.~U. were supported by the Japan Society for the
Promotion of Science (JSPS) for Young Scientists. 
T.~O. and T.~K. are supported by Grant-in-Aid for Scientific research 
from the Ministry of Education, Culture, Sports, Science and
Technology of Japan (Nos. 13135213,16028210, 16540243)
and (Nos. 14540263), respectively. 
 This work is also partially supported by the 21st Century for Center 
of Excellence program.  
The lattice QCD Monte Carlo calculations have been performed
on NEC-SX5 at Osaka University and on HITACHI-SR8000 at KEK.

\section*{NOTE ADDED}
After the completion of this paper,
Refs.~\cite{Letal05,CFKKT05,AT05} 
which also study the pentaquark state with lattice QCD
have appeared on the preprint server.

\begin{table}[h]

($\kappa_{u,d}$,$\kappa_s$)=(0.1650,0.1650)

\begin{tabular}{c c c c c c c c c}
\hline
size & $M_\pi$ & $M_K$ & $M_N$ & $M_{N^*}$ & 
$E_0^-$ & $E_1^-$ & $E_{NK}$ & $E_0^+$ \\ \hline\hline
$8^3\times 24 $ & 0.4378(49) & 0.4378(49) & 1.0463(89) & 1.4706(338) &
 --- & --- & 1.3841(202) & ---\\
$10^3\times 24$ & 0.4543(17) & 0.4543(17) & 1.0591(91) & 1.4313(292) &
 --- & --- & 1.4704(202) & 2.0987(45) \\
$12^3\times 24$ & 0.4563(13) & 0.4563(13) & 1.0281(74) & 1.4760(309) &
1.4601( 75) & 1.7881(829) & 1.4715(107) & 2.0502(26) \\
$16^3\times 24$ & 0.4556(11) & 0.4556(11) & 1.0143(44) & 1.4791(278) &
1.4616( 56) & 1.7157(452) & 1.4743( 74) & 1.9951(19) \\
\hline\hline
\end{tabular}
\vspace{.5cm}

($\kappa_{u,d}$,$\kappa_s$)=(0.1625,0.1650)

\begin{tabular}{c c c c c c c c c}
\hline
size & $M_\pi$ & $M_K$ & $M_N$ & $M_{N^*}$ & 
$E_0^-$ & $E_1^-$ & $E_{NK}$ & $E_0^+$ \\ \hline\hline
$8^3\times 24 $ & 0.5747(21) & 0.5130(31) & 1.2112(133) & 1.5973(220) &
1.6123(139) & 2.6447(3322) & 1.6119(195) & 2.2065(368) \\
$10^3\times 24$ & 0.5785(11) & 0.5199(14) & 1.1814( 55) & 1.5712( 93) &
1.6673( 96) & 2.0912(2305) & 1.6687(127) & 2.2120(393) \\
$12^3\times 24$ & 0.5792(10) & 0.5205(10) & 1.1655( 47) & 1.5930(175) &
1.6612( 55) & 1.9228(348) & 1.6763( 70) & 2.2145(293) \\
$16^3\times 24$ & 0.5789( 9) & 0.5209(10) & 1.1590( 37) & 1.5888(169) &
1.6636( 42) & 1.8769(225) & 1.6745( 50) & 2.1734(223) \\
\hline\hline
\end{tabular}
\vspace{.5cm}

($\kappa_{u,d}$,$\kappa_s$)=(0.1600,0.1650)

\begin{tabular}{c c c c c c c c c}
\hline
size & $M_\pi$ & $M_K$ & $M_N$ & $M_{N^*}$ & 
$E_0^-$ & $E_1^-$ & $E_{NK}$ & $E_0^+$ \\ \hline\hline
$8^3\times 24 $ & 0.6839(18) & 0.5761(25) & 1.3270(110) & 1.6956( 54) &
1.8002(121)  & 2.5420(1226) & 1.8109(259) & 2.3666(122) \\
$10^3\times 24$ & 0.6873(10) & 0.5819(12) & 1.3070( 48) & 1.6810(131) & 
1.8549( 78) & 2.1797(1067) & 1.8574( 90) & 2.3547(253) \\
$12^3\times 24$ & 0.6883( 9) & 0.5816(11) & 1.3003( 48) & 1.7198(221) &
1.8627( 51) & 2.0736(306) & 1.8680( 57) & 2.4004(396) \\
$16^3\times 24$ & 0.6867( 9) & 0.5818( 9) & 1.2921( 30) & 1.7193(222) &
1.8546( 37) & 2.0429(156) & 1.8668( 42) & 2.3403(162) \\
\hline\hline
\end{tabular}
\vspace{.5cm}

($\kappa_{u,d}$,$\kappa_s$)=(0.1650,0.1600)

\begin{tabular}{c c c c c c c c c}
\hline
size & $M_\pi$ & $M_K$ & $M_N$ & $M_{N^*}$ & 
$E_0^-$ & $E_1^-$ & $E_{NK}$ & $E_0^+$ \\ \hline\hline
$8^3\times 24 $ & 0.4378(49) & 0.5761(25) & 1.0463( 89) & 1.4705(338) &
 --- & --- & 1.5150(259) & 2.0510(732) \\
$10^3\times 24$ & 0.4543(17) & 0.5823(17) & 1.0774(128) & 1.4313(292) &
 --- & --- & 1.6088(173) & 2.1897(1549)\\
$12^3\times 24$ & 0.4563(13) & 0.5816(11) & 1.0281( 74) & 1.4760(309) &
1.5838( 72) & 1.8368(538) & 1.5951(100) & 2.1645(440) \\
$16^3\times 24$ & 0.4556(11) & 0.5818( 9) & 1.0143( 44) & 1.4791(278) &
1.5852( 55) & 1.7855(313) & 1.5987( 73) & 2.0823(176) \\
\hline\hline
\end{tabular}
\vspace{.5cm}

($\kappa_{u,d}$,$\kappa_s$)=(0.1600,0.1600)

\begin{tabular}{c c c c c c c c c}
\hline
size & $M_\pi$ & $M_K$ & $M_N$ & $M_{N^*}$ & 
$E_0^-$ & $E_1^-$ & $E_{NK}$ & $E_0^+$ \\ \hline\hline
$8^3\times 24 $ & 0.6839(18) & 0.6839(18) & 1.3270(110) & 1.6956( 54) &
 --- & --- & 1.9239(226) & 2.4376(105) \\
$10^3\times 24$ & 0.6873(10) & 0.6873(10) & 1.3070( 48) & 1.6810(131) &
1.9622( 73) & 2.2085(478) & 1.9617( 87) & 2.4319(215) \\
$12^3\times 24$ & 0.6883( 9) & 0.6883( 9) & 1.2987( 42) & 1.7198(221) &
1.9632( 51) & 2.1528(195) & 1.9705( 56) & 2.4820(312) \\
$16^3\times 24$ & 0.6867( 9) & 0.6867( 9) & 1.2906( 30) & 1.7193(222) &
1.9641( 40) & 2.1158(153) & 1.9723( 41) & 2.4180(138) \\
\hline\hline
\end{tabular}
\caption{\label{results}
The pion masses $M_\pi$, Kaon masses $M_K$, nucleon masses $M_N$,
masses of the ground state of negative-parity nucleon $M_{N^*}$,
energies of the lowest state $E_0^-$,
energies of the 2nd-lowest state $E_1^-$,
energies of the lowest state $E_{NK}$ (obtained by single-exponential fit)
in the $(I,J^P)=(0,\frac12^-)$ channel are listed.
The energies of the obtained state $E_0^+$
in the $(I,J^P)=(0,\frac12^+)$ channel are also listed.
$\kappa_{u,d}$ and $\kappa_s$ are the hopping parameters
for $u,d$ quarks and $s$ quark respectively.
}
\end{table}

\appendix
\section{additional estimations of weight factors}
\label{app:wf}

In this appendix, we make another trial to estimate
volume dependences of weight factors in $(I,J^P)=(0,\frac12^-)$ channel.
As seen in Sec.~\ref{Negative2-3},
we have extracted the weight factors using double-exponential fit,
which is however rather unstable and we have therefore fixed the exponents.
We here discuss the possibility of methods
without any multi-exponential fits.
Let us again consider $N\times N$ correlation matrices
constructed by $\gamma$-sink $\alpha$-source and $\gamma$-sink
$\beta$-source correlators.
Here $\alpha,\beta,\gamma$ denote the types of operators,
such as ``point'' or ``wall'' or ``smear'' and so on.
The notations are the same as those in Sec.~\ref{Formalism}.
The $\gamma -\alpha$ and $\gamma -\beta$ correlation matrices are
described as
\begin{eqnarray}
{\cal C}_{IJ}^{\rm \gamma\alpha}(T)\equiv
\langle {\Theta}^I_{\gamma}(T+t_{\rm src}) 
\overline{{\Theta}^{J}_{\alpha}}(t_{\rm src})\rangle
=(C^{\gamma\dagger}\Lambda(T)C^\alpha)_{IJ}+d_{IJ}e^{-E_NT}+...
\\
{\cal C}_{IJ}^{\rm \gamma\beta}(T)\equiv
\langle {\Theta}^I_{\gamma}(T+t_{\rm src}) 
\overline{{\Theta}^{J}_{\beta}}(t_{\rm src})\rangle
=(C^{\gamma\dagger}\Lambda(T)C^\beta)_{IJ}+d'_{IJ}e^{-E_NT}+...
\end{eqnarray}
with $N\times N$ matrices ($d_{IJ}e^{-E_NT}+...$) and 
($d'_{IJ}e^{-E_NT}+...$) being
possible higher excited-state contaminations.
We hereby consider two quantities;
$\left[{\cal C}_{IJ}^{\rm \gamma\alpha}(T){\cal C}_{IJ}^{\rm
\gamma\alpha}(T+1)^{-1}\right]^T
{\cal C}_{IJ}^{\rm \gamma\alpha}(T)$
defined using one type of the correlation matrix and 
$[{\cal C}_{IJ}^{\rm \gamma\alpha}(T)]^{-1}{\cal C}_{IJ}^{\rm \gamma\beta}(T)$,
which with large $T$ lead to
\begin{equation}
\left[{\cal C}_{IJ}^{\rm \gamma\alpha}(T){\cal C}_{IJ}^{\rm
\gamma\alpha}(T+1)^{-1}\right]^T
{\cal C}_{IJ}^{\rm \gamma\alpha}(T)
= C^{\gamma\dagger} C^\alpha
+{\cal F}(D(T))+... .
\end{equation}
and
\begin{equation}
[{\cal C}_{IJ}^{\rm \gamma\alpha}(T)]^{-1}{\cal C}_{IJ}^{\rm
 \gamma\beta}(T)
=
(C^{\alpha})^{-1}C^\beta
+{\cal F}'(D(T))+... ,
\end{equation}
respectively.
Here ${\cal F}(D(T))$ and ${\cal F}'(D(T))$ are
terms including $N\times N$ diagonal matrix
$D(T)\equiv {\rm diag}(e^{-(E_N-E_{N-1})T},...,
e^{-(E_N-E_0)T})$.
Then, each component of 
$\left[{\cal C}_{IJ}^{\rm \gamma\alpha}(T){\cal C}_{IJ}^{\rm
\gamma\alpha}(T+1)^{-1}\right]^T
{\cal C}_{IJ}^{\rm \gamma\alpha}(T)$
and
$[{\cal C}_{IJ}^{\rm \gamma\alpha}(T)]^{-1}{\cal C}_{IJ}^{\rm
\gamma\beta}(T)$
gets stable and shows a plateau in large $T$ region,
where ${\cal F}(D(T))$ and ${\cal F}'(D(T))$ are negligible.

Next, we relate these quantities to spectral weights.
For this aim, we simply take the determinants.
In the case when the correlation matrices are $2\times 2$ matrices,
the determinant $\det \left((C^\alpha)^{-1}C^\beta\right)$ is explicitly 
written as
$\det C^\beta/\det C^\alpha=
\varepsilon^{IJ}C^\beta_{0I}C^\beta_{1J}/
\varepsilon^{I'J'}C^\alpha_{0I'}C^\alpha_{1J'}$, and
the determinant $\det \left(C^{\gamma\dagger} C^\beta\right)$ is
expressed as
$(\det C^{\gamma\dagger})\times (\det C^\beta)
=(\varepsilon^{IJ}C^{\gamma\dagger}_{I0}C^{\gamma\dagger}_{J1})
\times
(\varepsilon^{I'j'}C^\beta_{0I'}C^\beta_{1J'})
=\varepsilon^{IJ}\varepsilon^{I'J'}
C^{\gamma\dagger}_{I0}C^\beta_{0I'}
C^{\gamma\dagger}_{J1}C^\beta_{1J'}$.
The term $C^\beta_{0I}C^\beta_{1J}$ 
($C^\alpha_{0I}C^\alpha_{1J}$)
denotes the product of the overlaps of the $\beta$($\alpha$)-type
operator with the lowest state and the 2nd-lowest state.
On the other hand,
$C^{\gamma\dagger}_{I0}C^\beta_{0I'}$
($C^{\gamma\dagger}_{J1}C^\beta_{1J'}$)
corresponds to the spectral weight
for the lowest (2nd-lowest) state in the $\beta$-$\gamma$ correlator
in terms of a volume dependence.

\begin{figure}
\begin{center}
\includegraphics[scale=0.35]{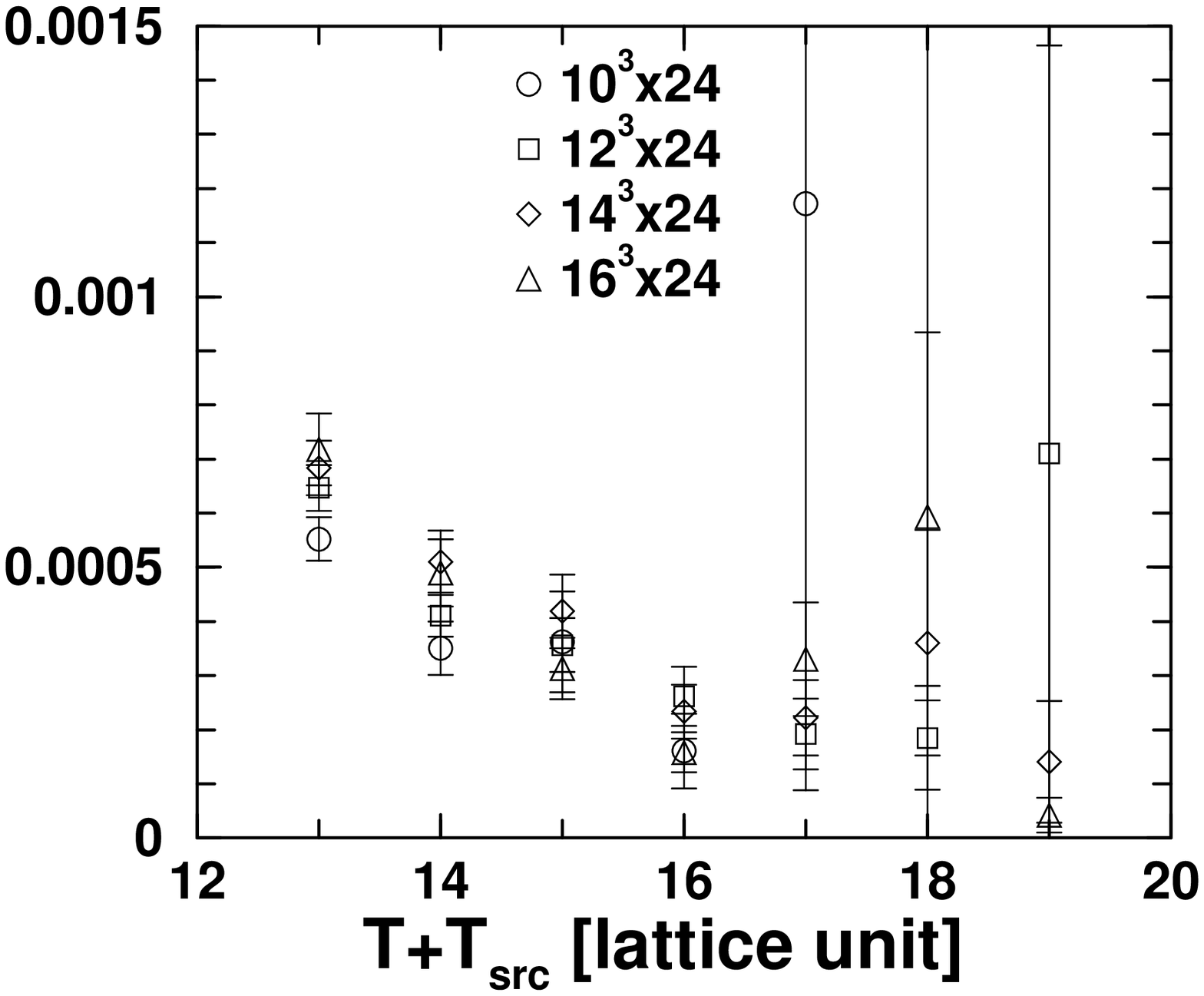}
\includegraphics[scale=0.35]{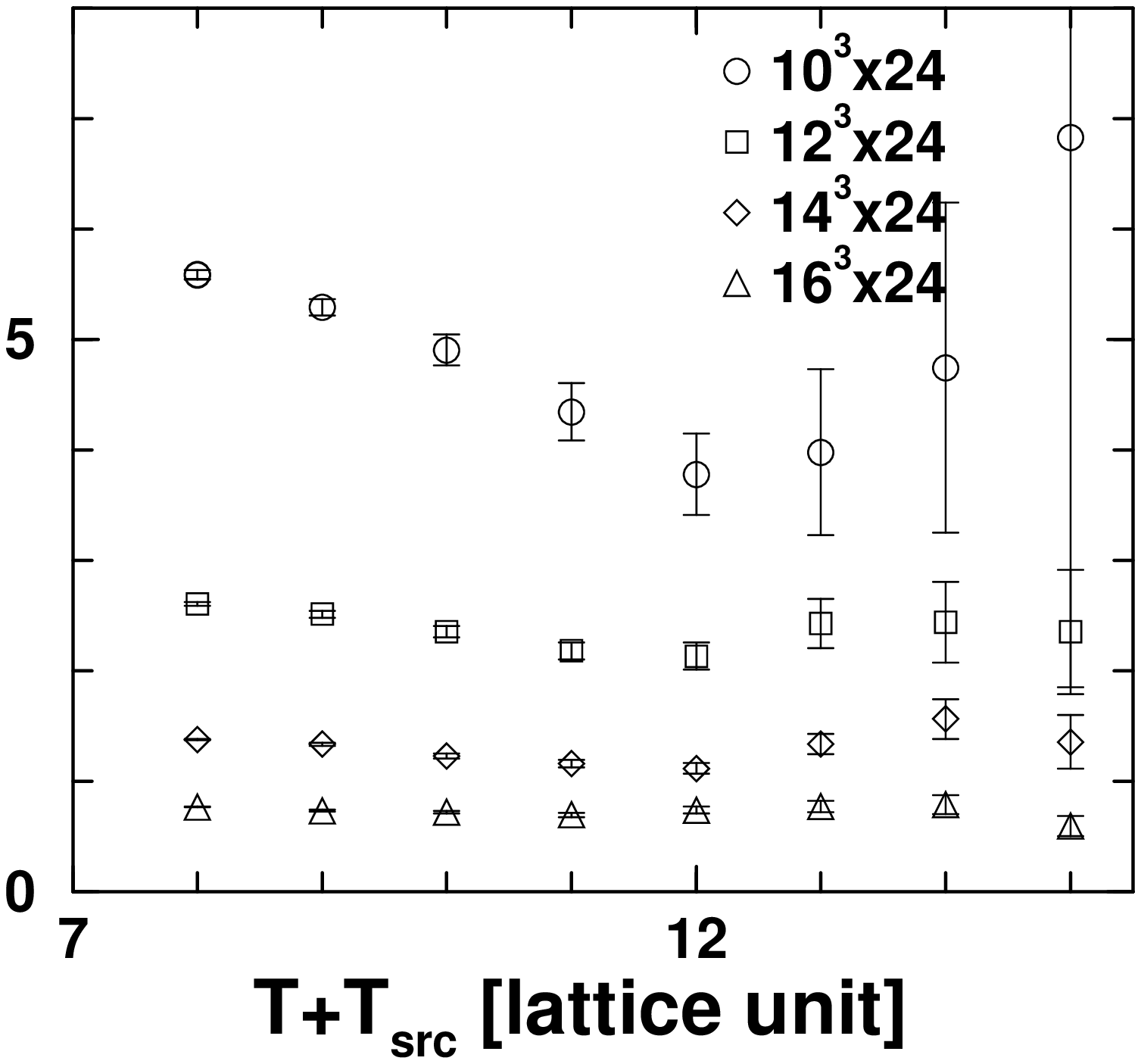}
\includegraphics[scale=0.35]{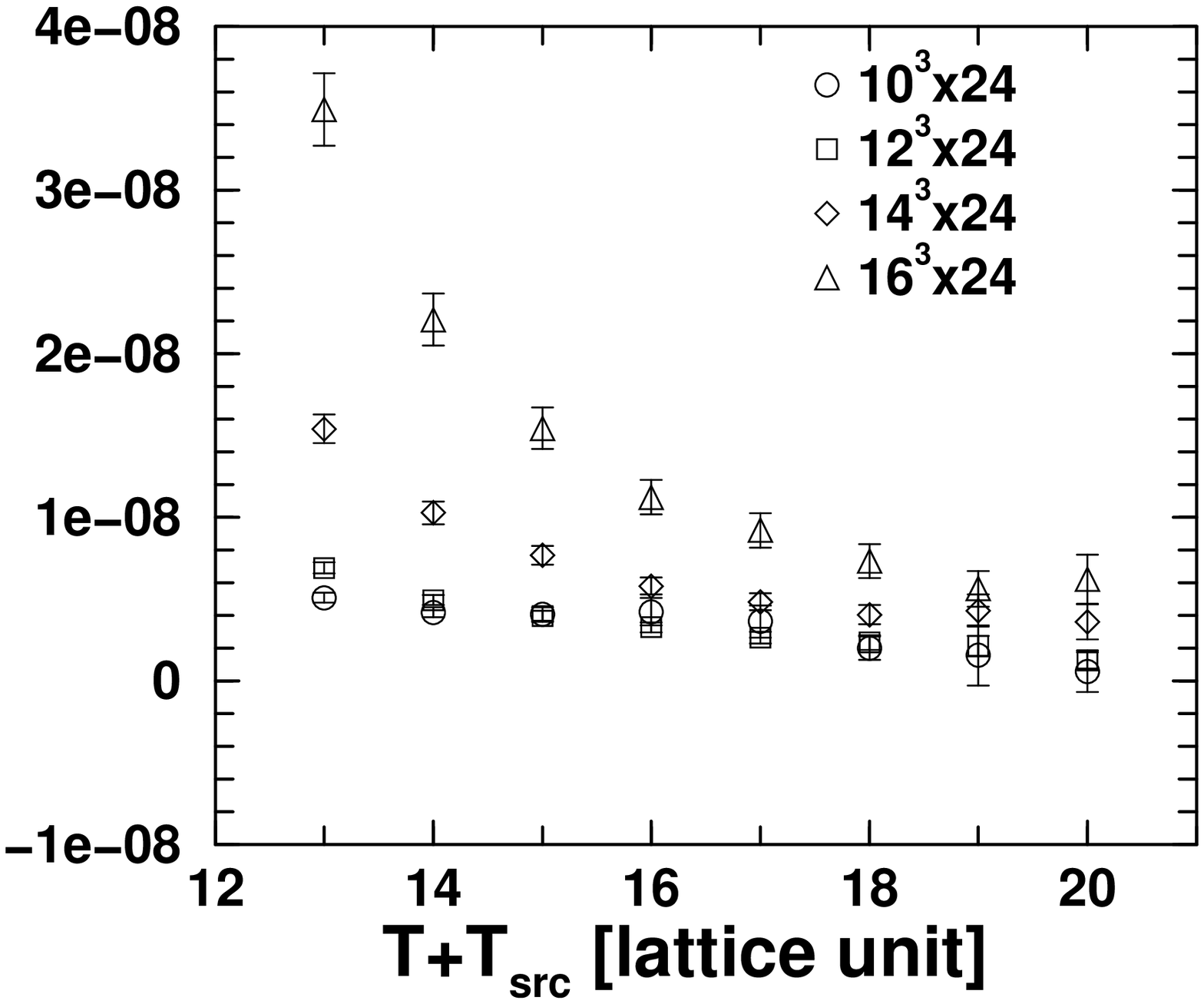}
\end{center}
\caption{
The left figure shows
$\det\left(\left[{\cal C}_{IJ}^{\rm PS}(T){\cal C}_{IJ}^{\rm PS}(T+1)^{-1}\right]^T
{\cal C}_{IJ}^{\rm PS}(T)\right)$ 
as the function of $T$ on each volume.
The middle figure is the plot of 
$\det\left(\left[{\cal C}_{IJ}^{\rm PW}(T){\cal C}_{IJ}^{\rm PW}(T+1)^{-1}\right]^T
{\cal C}_{IJ}^{\rm PW}(T)\right)$ against $T$.
${\rm Det}\left([{\cal C}_{IJ}^{\rm PW}(T)]^{-1}{\cal C}_{IJ}^{\rm PS}(T)\right)$
is plotted in the right figure.
${\rm Det}\left(\left[{\cal C}_{IJ}^{\rm PS}(T){\cal C}_{IJ}^{\rm PS}(T+1)^{-1}\right]^T
{\cal C}_{IJ}^{\rm PS}(T)\right)$,
$\det\left(\left[{\cal C}_{IJ}^{\rm PW}(T){\cal C}_{IJ}^{\rm PW}(T+1)^{-1}\right]^T
{\cal C}_{IJ}^{\rm PW}(T)\right)$ and
$\det\left([{\cal C}_{IJ}^{\rm PW}(T)]^{-1}{\cal C}_{IJ}^{\rm PS}(T)\right)$
show plateaus in the large $T$ region
and coincide with
$\det (C^{{\rm P}\dagger} C^{\rm S})$,
$\det (C^{{\rm P}\dagger} C^{\rm W})$ and $\det ((C^{\rm W})^{-1}C^{\rm S})$
respectively.
\label{app1}}
\end{figure}

\begin{figure}
\begin{center}
\includegraphics[scale=0.35]{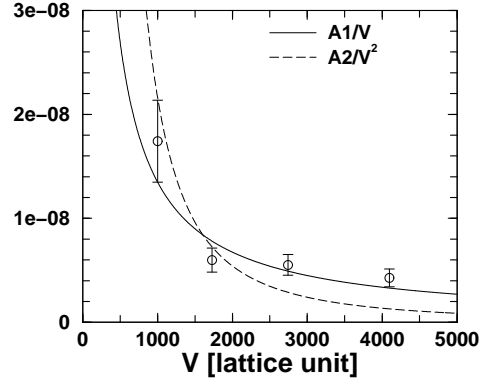}
\end{center}
\caption{
${\rm Det} (C^{{\rm P}\dagger} C^{\rm S})$ is plotted as the function of the
lattice volume $V$.
The solid line denotes the best-fit curve by $A_1/V$
and the dashed line does the best-fit curve by $A_2/V^2$.
The best fit parameters are $A_1=1.35$ and $A_2=2.17$,
and $\chi^2/N_{\rm df}$ is 1.72 and 7.13 respectively.
These data behaves consistently in accordance with $1/V$.
\label{app2}}
\end{figure}

Let us consider the several cases 
when ($\alpha,\beta,\gamma$)=\{W(wall), S(smeared), P(point)\}.
The term $\det (C^{{\rm P}\dagger} C^{\rm S})$ behaves
showing the same volume dependence
as the product of the spectral weights for
the lowest and the 2nd-lowest state in the smeared-point correlator,
which {\it should be} $\sim\frac{1}{V}\times 1=\frac{1}{V}$
{\it if} the lowest state is a scattering state and the 2nd-lowest state
is a resonance state.
The left panel in Fig.~\ref{app1} represents
$\det\left(\left[{\cal C}_{IJ}^{\rm PS}(T){\cal C}_{IJ}^{\rm PS}(T+1)^{-1}\right]^T
{\cal C}_{IJ}^{\rm PS}(T)\right)$ on each volume.
However, 
$\det\left(\left[{\cal C}_{IJ}^{\rm PS}(T){\cal C}_{IJ}^{\rm PS}(T+1)^{-1}\right]^T
{\cal C}_{IJ}^{\rm PS}(T)\right)$ on each volume,
which approaches $\det (C^{{\rm P}\dagger} C^{\rm S})$ with large $T$,
has relatively large errors and fluctuations
with no clear plateau and we fail to extract 
$\det (C^{{\rm P}\dagger} C^{\rm S})$.
This would be due to the smallness
of the signals in smeared-point correlators.
Meanwhile,
$\det\left(\left[{\cal C}_{IJ}^{\rm PW}(T){\cal C}_{IJ}^{\rm PW}(T+1)^{-1}\right]^T
{\cal C}_{IJ}^{\rm PW}(T)\right)$ shown in the middle panel in Fig.~\ref{app1}
and
$\det\left([{\cal C}_{IJ}^{\rm PW}(T)]^{-1}{\cal C}_{IJ}^{\rm PS}(T)\right)$
shown in the right panel in Fig.~\ref{app1},
which approach $\det (C^{{\rm P}\dagger} C^{\rm W})$ and 
$\det ((C^{\rm W})^{-1}C^{\rm S})$
respectively, show  relatively clear plateaus.
Therefore we extract 
$\det (C^{{\rm P}\dagger} C^{\rm W})$ and $\det ((C^{\rm W})^{-1}C^{\rm S})$
by the fits
$\det (C^{{\rm P}\dagger} C^{\rm W})=
\det\left(\left[{\cal C}_{IJ}^{\rm PW}(T){\cal C}_{IJ}^{\rm PW}(T+1)^{-1}\right]^T
{\cal C}_{IJ}^{\rm PW}(T)\right)$
and 
$\det ((C^{\rm W})^{-1}C^{\rm S})=
\det\left([{\cal C}_{IJ}^{\rm PW}(T)]^{-1}{\cal C}_{IJ}^{\rm PS}(T)\right)$
in the $T$ range where they show plateaus
and we finally obtain 
$\det (C^{{\rm P}\dagger} C^{\rm S})$ as
$\det (C^{{\rm P}\dagger} C^{\rm S})=
\det (C^{{\rm P}\dagger} C^{\rm W})\times 
\det ((C^{\rm W})^{-1}C^{\rm S})$.
In Fig.~\ref{app2},
we show $\det (C^{{\rm P}\dagger} C^{\rm S})$ obtained by the prescription shown above.
The solid line denotes the best-fit curve by $A_1/V$
and the dashed line does the best-fit curve by $A_2/V^2$.
The best fit parameters are $A_1=1.35$ and $A_2=2.17$,
and $\chi^2/N_{\rm df}$ is 1.72 and 7.13 respectively.
The volume dependence of $\det (C^{{\rm P}\dagger} C^{\rm S})$
seems not to be inconsistent with $\frac{1}{V}$.
If we know the precise volume dependence of overlaps of wall operators,
we may be discriminate the states
using $\det (C^{{\rm P}\dagger} C^{\rm W})$ or 
$\det ((C^{\rm W})^{-1}C^{\rm S})$.

\end{document}